\documentclass[11pt,preprint2]{aastex}

\slugcomment{submitted to \apjs}

\begin{document}

\title{Hubble  Space Telescope  Near-Infrared Snapshot  Survey  of 3CR
radio source counterparts at low redshift\altaffilmark{1}}


\author{
Juan P. Madrid\altaffilmark{2},
Marco Chiaberge\altaffilmark{2,3}, 
David Floyd\altaffilmark{2},  
William B. Sparks\altaffilmark{2},    
Duccio~Macchetto\altaffilmark{2,4},
George~K.~Miley\altaffilmark{5},
David Axon \altaffilmark{6},
Alessandro Capetti \altaffilmark{7},
Christopher~P.~O'Dea \altaffilmark{6},
Stefi Baum \altaffilmark{6},
Eric~Perlman \altaffilmark{9}, 
Alice Quillen  \altaffilmark{8} 
}

\altaffiltext{1}{Based on  observations made with  the NASA/ESA Hubble
Space Telescope,  obtained at  the Space Telescope  Science Institute,
which is operated  by the Association of Universities  for Research in
Astronomy, Inc., under NASA  contract NAS 5-26555.  These observations
are associated with program 10173.}

\altaffiltext{2}{ Space  Telescope Science Institute,  3700 San Martin
Drive, Baltimore, MD 21218.}

\altaffiltext{3}{On leave from INAF, Instituto di Radioastronomia, Via
P.  Gobetti 101, Bologna, Italy, 40126-I.}

\altaffiltext{4}{Affiliated with  the Space Telescope  Division of the
European   Space   Agency,    ESTEC,   Noordwijk,   The   Netherlands.}

\altaffiltext{5}{Leiden Observatory, P.O. Box 9513, NL-2300 RA Leiden,
The Netherlands.}

\altaffiltext{6}{Department   of  Physics,   Rochester   Institute  of
Technology, 85 Lomb Memorial Drive, Rochester, NY 14623.}

\altaffiltext{7}{INAF  - Osservatorio  Astronomico  di Torino,  Strada
Osservatorio 20, 10025 Pino Torinese, Italy.}

\altaffiltext{8}{Department  of Physics  and Astronomy,  University of
Rochester, Bausch \& Lomb Hall, P.O. Box 270171, 600 Wilson Boulevard,
Rochester, NY 14627.}

\altaffiltext{9}{Joint Center for Astrophysics, Department of Physics,
University  of  Maryland,   Baltimore  County,  1000  Hilltop  Circle,
Baltimore, MD 21250.}


\begin{abstract}

We present  newly acquired images of the  near-infrared counterpart of
3CR radio sources.   All the sources were selected  to have a redshift
of  less  than  0.3  to   allow  us  to  obtain  the  highest  spatial
resolution. The  observations were carried  out as a  snapshot program
using the  Near-Infrared Camera and  Multiobject Spectrograph (NICMOS)
on-board the Hubble  Space Telescope (HST). In this  paper we describe
69 radio galaxies  observed for the first time  with NICMOS during HST
cycle  13. All  the objects  presented here  are  elliptical galaxies.
However,  each  of  them  has  unique characteristics  such  as  close
companions, dust lanes, unresolved nuclei, arc-like features, globular
clusters and jets clearly visible from the images or with basic galaxy
subtraction.

\end{abstract}



\section{Introduction} 

Extragalactic  radio   sources  are  often   associated  with  massive
elliptical galaxies.   These so called  radio galaxies are one  of the
most extraordinary  astrophysical phenomena, powered,  it is generally
believed,  by supermassive  black holes  in the  galaxy  nuclei. Radio
galaxies  are found in  a variety  of environments  and across  a wide
range  of  redshifts.  

The Revised Third Cambridge  Catalogue (3CR) (Bennett 1962a, 1962b) is
the best studied  sample of radio loud galaxies  and quasars.  Spinrad
et al.   (1985) confirmed 298  extragalactic radio sources of  the 3CR
catalogue. This  catalogue was made  based on the radio  properties of
the  sources,   and  therefore  its  selection   criteria  are  mostly
independent with respect to  orientation and HST wavelengths.  The 3CR
catalogue has, thus, excellent attributes for a survey.

Here we describe all the  observations taken with our snapshot program
in  the HST  Cycle 13.   We are  using NICMOS  to obtain  high spatial
resolution, deep images  in the near infrared $H$-band  of 3CR sources
at  low redshift, $z<0.3$.   The primary  goal of  this project  is to
characterize the radio galaxy hosts free from the obscuring effects of
dust, or at  least, substantially reduced relative to  the optical and
UV.  We also  want to establish how radio galaxy  hosts compare to the
hosts  of the  most  powerful high-z  AGN,  to QSO,  and to  quiescent
ellipticals.   The  high  linear  resolution  of  HST  grants  us  the
possibility  to observe  details of  the galaxies  such  as point-like
nuclei,  jets, and  hot spots.   These near-infrared  observations are
aimed at providing a  zero redshift comparison sample for observations
at high redshift.

This NICMOS snapshot program is  a major enhancement to the dataset of
HST  observations of  the 3CR  sources.  Successful  snapshot programs
have been  carried out during  past cycles in  the optical and  in the
ultraviolet by Sparks and  collaborators.  The WFPC2 observations were
presented by  Martel et  al.  (1999)  for $z<0.1$, by  de Koff  et al.
(1996) for $0.1<z<0.5$,  and by McCarthy  et al.  (1997)  for $z>0.5$.
UV  observations with  STIS were  published  by Allen  et al.  (2002).
These successive HST programs have  generated a database with a sample
completeness  comparable, for  some range  of redshift,  to  the radio
catalogue.   This  database   provides  an  excellent  foundation  for
statistical studies as a vast  number of observations of these sources
are  available for  comparison  at other  wavelengths  and with  other
instruments.

The  content of  this paper  is as  follows: Section  2  describes the
observation strategy,  Section 3 discusses the different  steps of the
data reduction, Section 4 highlights results obtained from this survey
and gives near-infrared photometry,  Section 5 presents three galaxies
with  jets visible  with  galaxy  subtraction, in  Section  6 we  make
comments on the image of each individual galaxy.

We   use  throughout   this   paper  H$_0$=71   km.s$^{-1}$Mpc$^{-1}$,
$\Omega_M=0.27$, and $\Omega_\Lambda=0.73$.


\section{Sample selection and observations}

There is a total of 115  objects in the 3CR catalogue with $z<0.3$, 18
of them have been observed  with NICMOS during previous cycles as part
of other  HST programs.   We present 69  sources that have  never been
observed before, this corresponds to  71$\%$ of the sample.  These new
observations  are associated  with  program 10173,  PI: Sparks.   This
NICMOS survey  uses snapshot exposures, a capability  developed by the
STScI to  maximize the observing efficiency of  HST.  The observations
obtained during this survey are taken at irregular intervals that fill
the scheduling gaps between other accepted GO programs.  The remaining
28 sources  were not scheduled  for observation during cycle  13.  The
observation log  is presented in  Table 1. Since the  observed targets
are randomly chosen based on the constraints of the observing schedule
of HST,  the 69 galaxies  we present here  are not biased  towards any
particular  characteristic. Thus,  they are  suitable  for statistical
analysis.

The radio properties  of the observed sample are given  in Table 2. We
present  the  flux density  and  radio power  at  178  MHz, the  radio
spectral index, largest angular size, position angle, and Fanaroff and
Riley   (1974)   morphological   classification.   The   sources   are
representative  of large,  steep spectrum,  high power  radio sources.
The sources span three decades in radio power from log$_{10} \sim $ 25
to 28. The edge-darkened, lower luminosity FRI sources make up 26\% of
the sample.

All  the observations  presented  here were  carried  out with  NICMOS
Camera 2 (NIC2)  on MULTIACCUM observing mode.  We  use the MULTIACCUM
sequence  STEP32, which  consists  of  rapid reads  up  to 32  seconds
followed by 32 second steps.  NIC2 has a field of view of $19.2\arcsec
\times 19.2\arcsec$ and a projected pixel size of $0.076\arcsec \times
0.075\arcsec$. The measured FWHM for a bright source is 0.14\arcsec.

The field  size allows us to  detect both the host  galaxy and nuclear
regions  at the redshifts  we aim.   The cooresponding  physical scale
over  our redshift  range  varies from  0.335  kpc/arcsecond to  4.376
kpc/arcsecond. 

We use the  F160W filter, the analog of the $H$  band.  This filter is
centered at  1.6037$\mu$m, covering a wavelength  range from 1.4$\mu$m
to  1.8$\mu$m, and  includes  the Paschen  $\beta$  emission line  for
objects at $z<0.1$.  For further details the reader is referred to the
NICMOS Instrument Handbook (Noll et al. 2004).

All images  have the same total  exposure time of  1152 seconds, split
into  four  exposures of  288s.   We  perform  sub-pixel dithering  to
improve  the PSF sampling  and remove  bad pixels.   We use  a squared
dither pattern, with 7.4 pixels offsets.


\section{Data processing}

We obtain  the data from  the Multimission Archive at  Space Telescope
(MAST).   The  data  are  processed  by  the  standard  ``on-the-fly''
reprocessing  calibration pipeline.  The  pipeline separates  the data
into science  and engineering data,  creates fits files  and populates
its  headers. The  pipeline also  calibrates the  raw data  using {\sc
calnica} (Noll  et al.  2004). We do  the subsequent reduction  of the
data with the NOAO  image processing software Image Reduction Analysis
Facility (IRAF).

The most important anomalies that we encounter in NIC2 images are: the
pedestal effect,  the amplifier  glow, the coronographic  spot, cosmic
rays        and       bad       pixels        \footnote{see       also
http://www.stsci.edu/hst/nicmos/}.   In this  section we  describe the
steps we  follow to get  rid of these  anomalies on the  NICMOS images
using different IRAF tasks.

NIC2 is divided into four  quadrants of $128\times128$ pixels each. On
visual inspection  the raw images  appear to have one  quadrant darker
than the  others.  This is  particularly apparent for images  with low
background, e.g. small galaxies. This  is known as the pedestal effect
and is part  of the instrumental signature for  NICMOS.  The origin of
this  effect is an  additional offset  introduced during  the detector
reset.  The pedestal  effect is  stochastic and  time  variable, which
makes it  impossible to remove  with the standard pipeline  offered by
the STScI (Noll et al. 2004).

We  remove the pedestal  effect with  the tasks  {\sc pedsky}  or {\sc
pedsub} from the NICMOS package under the HST calibration of the Space
Telescope Science Data Analysis  System (STSDAS).  

These two tasks determine the  pedestal offset of each quadrant with a
similar technique.  Both {\sc pedsky} and {\sc pedsub} assume that the
image  pixel value is  the sum  of the  astronomical signal  (source +
sky), and the pedestal effect modulated by the flatfield.  These tasks
use the flatfield  as input and loop over a range  of trial values for
the pedestal and  find the best fit to the  science image, for greater
details see Bushouse et al. (2000).

{\sc Pedsky} does the fit of  the pedestal only with those pixels near
the background  level.  This  task dismisses pixels  containing signal
from a  source during  the fitting process.  {\sc Pedsky}  removes the
pedestal effect  and also  subtracts the sky  background. We  use {\sc
pedsky}  for  galaxies  that  cover  only  a  small  fraction  of  the
detector. This task  is more effective when a large  area of the image
is  free from  sources in  order  to allow  a better  estimate of  the
background.

For sources covering  a large portion of the detector  we use the task
{\sc pedsub} to remove the pedestal effect.  {\sc Pedsub} has a series
of   filters    to   remove   unwanted   features    such   as   large
galaxies. Contrary to  {\sc pedsky}, {\sc pedsub} does  not remove the
background. In  the case where the  background is not  removed, we use
{\sc msky2}, a task written by Mark Dickinson (private communication),
that can  fit the  sky interactively, and  subtract the sky  mode from
each image to obtain a zero level background.

Given the small  projected size of the detector,  a few galaxies cover
the whole chip, for these images we adopt the median background of the
other frames as the sky level. This is determined by measuring the sky
level on  152 single  exposures of the  dither pattern.  In  these 152
exposures  the source covers  only a  small fraction  of the  field of
view. We  find a median  background of 0.046  counts per pixel  with a
standard deviation of  0.01806. For the galaxies covering  most of the
detector, the  subtracted sky level  amounts, on average, to  1.3\% of
the flux inside an aperture of 1kpc radius.

Some images  have visible  noise in the  corners due to  the amplifier
glow of  NIC2. This glow is  caused by the  readout amplifier situated
close to each  corner of the detector. Each time  the detector is read
out  the amplifier  warms  up  and emits  infrared  radiation that  is
detected by the chip (Noll et  al. 2004). This anomaly does not affect
the overall appearance of the image.
 
In order  to obtain  images of a  better quality we  create individual
masks for  each object  to cover the  small blemishes of  the detector
such as residual cosmic rays  and bad pixels.  Furthermore, NIC2 has a
hole  through the  camera to  allow coronographic  observations, which
creates a circular spot on the images of about ten pixels in diameter.
We treat this area with a  mask to conceal the coronographic spot.  We
also create cosmic rays masks, with the task {\sc drizzle$_-$cr}, that
we incorporate into the individual  masks for each object.  The pixels
of  column 128  in NIC2  contains the  first pixels  read out  in each
quadrant  and have  an incorrect  bias subtracted  from them.  This is
called the "photometrically challenged" column. We applied a bad pixel
mask to the  pixels of this column.  We  recover, thus, information on
the image.

The  second extension  of the  NICMOS  fits files  contains the  error
image.  We  convert these  error files into  weightings which  are the
inverse square of  the noise per pixel. The  mask files are multiplied
into  the weight  files described  above,  giving zero  weight to  bad
pixels in the individual files but leaving the others unchanged.
 
We combine  the four calibrated,  background subtracted images  of the
dither  pattern with  the task  {\sc drizzle}  under the  package {\sc
dither} (Fruchter \&  Hook 2002) .  The final  weight files are called
when dithering the four exposures  of each object. The projected pixel
size in the final image is $0.038\arcsec$.

{\sc Drizzle}  is also  used to apply  the necessary rotations  to the
images to obtain a north up, east left orientation. These final images
are presented in Figures 1-77.


\section{Analysis} 

\subsection{Companions and morphology}

The high spatial resolution provided by HST allows us to obtain images
of the 3CR radio galaxies  with unprecedented details.  

All galaxies  imaged during this  program are ellipticals but  each of
them  has its own  peculiar characteristics  and environment.   In our
sample of 69  galaxies 36 of them (52$\%$)  have companions present in
the field  of view of NIC2. More  than one third of  this sample, i.e.
25 galaxies,  have an unresolved  nucleus and nine galaxies  have dust
lanes.   Nineteen galaxies,  primarily the  closest, show  off nuclear
point-like sources likely to be globular clusters.  Four galaxies have
intriguing  arc-like  structures, which  are  perhaps  lensed arcs  or
merger remnants.

\subsection{Aperture photometry}

We use the IRAF task {\sc radprof} to measure the flux of the galaxies
at five different radii. The  radii of the apertures we use correspond
to: 1 kpc, 5 kpc, 10 kpc, 15 kpc, 20 kpc.  Given the small size of the
detector  we cannot  always measure  the  flux at  all five  different
radii.

The count rate ({\sc cr}) is converted into flux using {\sc photflam},
the  inverse  sensitivity, in  the  header  of  the calibrated  NICMOS
images:

\begin{mathletters}
{\sc photflam}=1.74779.10$^{-19}$ erg cm$^{-2}$ \AA DN$^{-1}$
\end{mathletters}

We determine  the magnitudes in  the Space Telescope (ST)  system with
the expression:

\begin{mathletters}
m$_{ST}$  =  - 2.5log({\sc photflam $\times$ cr}) + {\sc photzpt}
\end{mathletters} 

where {\sc  photzpt} is the ST  magnitude zero point  i.e.  -21.1, see
the  HST  Data Handbook  for  NICMOS  (Dickinson  et al.  2002).   The
limiting magnitude  for the  images presented here  is about  25.3. We
present the results  of the near-infrared photometry in  Table 3. Note
that these magnitudes correspond to the total observed flux inside the
radii set above.  We did  not perform photometry for the nuclei, which
will be analyzed in detail in a forthcoming paper (Chiaberge et al. in
preparation)


\section{Ellipse residuals}

We    have   used   the    {\sc   ellipse}    task   in    IRAF   {\sc
stsdas.analysis.isophote},  based  on  the algorithm  of  Jedrzejewski
(1987),  to  fit elliptical  isophotes  to  the  NICMOS images.   Each
isophote   is  fitted   at  a   pre-defined,  fixed   semi-major  axis
length. Starting  from a  first guess, the  image is sampled  along an
elliptical  path  (defined by  central  coordinates $(x_{c},  y_{c})$,
position  angle   $\theta$,  and  ellipticity  $b/a$)   to  produce  a
1-dimensional intensity distribution as a function of $\theta$.

Modeling with {\sc ellipse} allows  us to track radial changes in the
ellipticity and  position angle of  the isophotes of  galaxies. Models
based on  the best-fitting ellipse parameters are  constructed in {\sc
iraf}  using the  task  {\sc bmodel}.  

A complete  sample of model-subtracted residual images  of this survey
will be presented by Floyd et  al. (in preparation).  Here we show two
examples where jets are  clearly visible with galaxy subtraction. This
is  the case  for 3C66B,  3C133 Figures  10,23.  Figure  62  shows the
residual image for 3C401, its jet  is also visible in the NICMOS image
(Chiaberge at al. 2005).


\section{Notes on individual sources}

We provide here a short  description of the most prominent features of
the NICMOS  images for each object.

\begin{description}

\item[3C20]  The near-infrared  image reveals  a circular  galaxy. One
source  is  visible approx.   $4.7\arcsec$  to  the  northwest of  the
nucleus.  A  bright unresolved source  is present $5.2\arcsec$  to the
southeast.  This galaxy has a  known radio hotspot detected by de Koff
et  al.  (1996)  with WFPC2.  This hotspot  falls outside  the smaller
field of view of NIC2.

\item[3C28]  3C28  is an  elliptical  elongated  on  the northwest  to
southeast direction.   At least two  other sources are present  in the
field of view: one is on the edge of the chip to the northwest and the
second one is  $10.3\arcsec$ to the southwest.  This  galaxy is the cD
of the X-ray cluster Abell 115 (McCarthy et al.  1995).

\item[3C29]  A round,  undisturbed  elliptical with  a bright  central
unresolved component (nucleus).  Several globular clusters are present
in the field  of view. This galaxy has a faint  compact nucleus in the
optical and in  the UV. The UV nucleus was first  detected by Allen et
al. (2002).

\item[3C31] This  galaxy covers  most of the  area of the  detector. A
face  on dust  ring around  the galaxy's  nucleus can  be seen  on the
infrared image.  The WFPC2 image shows that this dust ring is composed
of several dust  strands with a larger absorption  on the southwestern
side  (Martel et  al. (1999)  and de  Koff et  al. (2000)). 

\item[3C33.1] Elliptical  galaxy with  no obvious disturbance  with an
unresolved nucleus  and a single  close companion about  $4\arcsec$ to
the south-southwest.  A bright  unresolved source lies  $12\arcsec$ to
the southwest,  three other faint unresolved sources  are also visible
as well as another double system $12\arcsec$ to the northwest.

\item[3C35] This flattened galaxy is elongated northwest to southeast.
The  WFPC2  image of  this  galaxy  shows  an unremarkable  elliptical
(Martel et  al.  1999) while the  STIS image exhibits  a faint compact
nucleus (Allen at al. 2002).

\item[3C52] The  near-infrared image reveals an elongated  galaxy in a
crowded field.  One unresolved  source is present $6.5.\arcsec$ to the
northwest and  a second unresolved  source is located  $9.7\arcsec$ to
the  north-northeast of the  center of  the galaxy.  The near-infrared
image does  not show the dust  lane, perpendicular to  the radio axis,
visible in the optical (de Koff et al. (1996, 2000)).

\item[3C61.1] The galaxy lies in the middle of the field of view which
shows three  prominent resolved  companions. 3C61.1 is  elliptical and
compact. We  do not see  the tails of  emission to the south  and east
detected by de Koff et al. (1996) with WFPC2. One unresolved source is
present to the north of the galaxy.

\item[3C66B]  This NICMOS  image shows  a round  galaxy with  a bright
unresolved  nucleus,  several  globular  clusters and  two  unresolved
sources to the south.  We notice  a small asymmetry in the shape of
this  galaxy with  its northwest  side more  sharply defined  than the
southeast one.  Zirbel \& Baum (1998) claim that this slight asymmetry
may be  caused by a  neighbor to the  southeast.  The jet of  3C66B is
clearly visible in this NICMOS image after basic galaxy subtraction as
shown in section 5.

\item[3C75N] 3C75N is slightly  elongated on the east-southeast to the
west-northwest direction.  This is  the northern component of a system
of  two elliptical  galaxies, the  edge of  the southern  companion is
visible at the  bottom of the image. Some  faint globular clusters are
also seen.

\item[3C76.1] 3C76.1 is an elliptical  with a very elongated galaxy to
the north,  it is unclear from  this image if they  are interacting or
not.

\item[3C79]   A   galaxy   with   elliptical   morphology,   elongated
north-south.  The  image shows  two companions: one  bright unresolved
source less than one arcsecond south of the nucleus, and the other one
to the north.  3C79 shows a complex morphology in the optical, with no
compact nucleus (de Koff et al. 1996)

\item[3C83.1] The  near-infrared  image shows an  elliptical elongated
north  to south and  covering most  of the  detector area.   The image
reveals a thin dust lane  wrapping around the nucleus. A bright nearby
star generates  diffraction spikes crossing the  detector.  Some faint
globular clusters are also visible.

\item[3C88]   Elliptical  elongated   northwest   to  southeast,   the
radio-axis is perpendicular to the elongation of the galaxy.  Globular
clusters are visible on the image  and there is no sign of disturbance
or merger.
 
\item[3C105] Highly  flattened elliptical, faintly  disturbed. A faint
unresolved source is visible  4$\arcsec$ to the west-southwest off the
nucleus.

\item[3C111]  This galaxy  has a  very  bright nucleus  with the  host
clearly visible. 3C111 is an elliptical elongated north to south.

\item[3C123] This elliptical galaxy is a member of a cluster. At least
three diffuse companions and two point-like sources are present in the
field of view. The optical image of de Koff et al. (1996) shows also a
faint diffuse source.

\item[3C129]  An  elongated   arc-shaped  source  is  clearly  visible
$3.4\arcsec$ east of the nucleus  of the galaxy.  Globular clusters in
this galaxy  are also  detected.  Several unresolved  foreground stars
are present in the field of view.

\item[3C129.1] This very round elliptical  galaxy is found in a region
of  high foreground  stellar  density. It  is  the cD  of the  cluster
4U0446+44. Globular clusters are also possibly present in the image.

\item[3C130]   This  radiogalaxy   is  a   round   elliptical  without
peculiarities. It covers a large portion of the detector and lies in a
region  of high  foreground stellar  density. Thus  a large  number of
point sources are visible. Some of them may be globular clusters.

\item[3C133] We  have discovered  a new optical-IR  jet in  3C133, see
Floyd  et al.   (2006) for  further  discussion. The  jet and  eastern
hotspot  are  well  resolved,  and  visible at  both  optical  and  IR
wavelengths, in spite of the  low galactic latitude.  The infrared jet
follows the morphology of the inner  part of the radio jet, with three
distinct, aligned,  bright knots east  of the unresolved  nucleus. The
host is a round elliptical.

\item[3C135] This FRII  radio galaxy is a cluster  member (McCarthy et
al. 1995) with a close companion to the southwest.

\item[3C165] The  near infrared image reveals  an elliptical elongated
north  to  south in  a  crowded field.   A  source  is present  approx
$7\arcsec$ to the northeast. 3C165 is a cluster member.

\item[3C171] This  galaxy of slightly  disturbed elliptical morphology
has particularly strong extended emission line regions. These emission
line regions are aligned with the radio emission (Heckman et al. 1984,
Blundell  1996).  The  NICMOS  image shows  faint  tails, most  likely
Pa$\beta$  emission, to  the  east and  west  overlapping the  optical
emission line regions, see zoom image.

\item[3C173.1]   Slightly  boxy   elliptical   with  faint   companion
$3\arcsec$ south.  An asymmetric, low surface brightness component, is
visible to the northwest,  perhaps evidence of recent merger activity.
3C173.1 is a member of a group.

\item[3C180] Our NICMOS image shows  a boxy giant elliptical. 3C180 is
a  member of  a cluster  according to  McCarthy et  al.  (1995).  Five
unresolved sources are present in the field of view.

\item[3C184.1]  This galaxy  is a  cluster member  with  several faint
companions. It  is an  elongated elliptical with  noticeable disturbed
isophotes.   An arc-shaped  structure  is present  $\thicksim2\arcsec$
northeast of the unresolved nucleus.

\item[3C192] This  object appears  as a round,  undisturbed elliptical
galaxy  in the  near-infrared,  while  it shows  a  compact core  with
diffuse emission  structures in  the UV (Allen  et al.  2002). Several
globular clusters are also detected. 

\item[3C196.1]  The  near-infrared image  shows  an elliptical  galaxy
elongated northeast to  southwest, which is the same  direction of the
radio emission. The same morphology is seen in the optical (de Koff et
al. 1996 and Baum et al. 1988 ). Five other sources populate the field
of view.

\item[3C197.1] This galaxy has  a round, slightly disturbed morphology
and an unresolved nucleus.

\item[3C198]  The NICMOS  image shows  an elongated  elliptical galaxy
with a  bright unresolved  nucleus. The circumnuclear  compact sources
seen  in the  UV  image  (Allen et  al.  2002) do  not  appear in  the
near-infrared.

\item[3C213.1] This radio  galaxy has an extended area  of emission to
the southeast. The near-infrared counterpart of the northern radio hot
spot is clearly visible. The southern  hot spot is also visible but is
much dimmer.  The two hotspots are also present on the WFPC2 image (de
Koff et al.  1996, de Vries et al. 1997).

\item[3C219] A member of a cluster of galaxies, this source shows five
different  companions  of  different  types and  a  bright  unresolved
nucleus.   As noted  by de  Koff et  al.  (1996)  3C219 appears  to be
interacting with the large companion to the southeast, however, McLure
et al.  (1999)  modeled the WFPC2 image and found  no evidence of such
interaction.

\item[3C223]  Our  near infrared  image  shows  an asymmetric  galaxy,
slightly elongated northeast to southwest.

\item[3C223.1]  The  image  shows  a  galaxy  with  a  very  elongated
elliptical shape and a thin dust disk to the southeast, see zoom image
with different  intensity scale.   The dust disk  is best seen  in the
optical (de Koff et al.  2000).   A tail of emission is present to the
northeast.

\item[3C227] This is a broad line radio galaxy. We can clearly see the
host galaxy and its bright  nucleus on this near-infrared image.  Both
the  optical and  UV  images  show a  very  bright unresolved  nucleus
(Martel et al.   1999, Allen et al.  2000). A  close companion is also
visible 5$\arcsec$ southeast off the nucleus.

\item[3C236] Elliptical elongated from the northeast to the southwest.
A dust lane is notable on  the WFPC2 images and was detected by Martel
et al.  (1999).  The absorption map  of de Koff et al.  (2000) clearly
shows this  dust disk too.  O'dea  at al. (2001) found  four very blue
regions of star formation on the edge of the dust lane.

\item[3C277.3] The  host galaxy  of Coma A  is a round  elliptical.  A
source  elongated  northeast  to  southwest is  present  approximately
$5.8\arcsec$ to  the southeast.  This source  is the radio  knot K1 as
defined by  Miley et al.  (1981)  and van Breugel et  al.  (1985). The
knot  is also  visible  in the  optical  (Capetti et  al.  2000).   An
unresolved  source, likely  to be  associated to  knot K2,  is visible
$1.8\arcsec$ southwest of K1.

\item[3C285] Irregular clumps of emission linked to star formation are
detected in the ultraviolet image  of this galaxy (Allen et al. 2002).
The near-infrared  image shows a flattened  galaxy elongated southeast
to northwest.   The dust lane  perpendicular to the elongation  of the
galaxy  is visible  in  both the  WFPC2  and NICMOS  images.  A  faint
arc-like source is present approx.  $5.1\arcsec$ to the southeast, see
zoom image.

\item[3C287.1] This  broad line radio  galaxy has a  bright unresolved
nucleus  with visible  diffraction spikes,  the host  is  an elongated
elliptical. A companion is present $5.1\arcsec$ to the northeast.

\item[3C288]  An E0 elliptical  galaxy with  round isophotes,  a faint
unresolved  nucleus,  and  two  compact nearby  companions.  One  more
extended companion is visible to the west.  Two fainter companions are
also present, one to the northwest and the other one to the southwest.

\item[3C303] This  galaxy is characterized by a  strong X-ray emission
(Hardcastle \&  Worrall 1999) and  has a bright unresolved  nucleus in
the  near-infrared   image.   A  companion   is  visible  $5.6\arcsec$
northeast of the nucleus.

\item[3C310] As noted  by Martel et al. (1999)  for the optical image,
this  galaxy  is  flattened  east-west,  on  an  almost  perpendicular
direction to the  radio jet axis. The edge of a  companion to the east
is visible  on the corner of  the chip.  There is  a bright foreground
star to the south of the galaxy. Globular clusters and a faint diffuse
companion $7\arcsec$ northwest of the nucleus are also seen.

\item[3C314.1] The  NICMOS image shows an  elliptical galaxy elongated
east-northeast to west-southwest.  The  isophotes are elongated in the
near-infrared as well as in the optical (de Koff et al. 1996).

\item[3C315]  This  is  a  very  elongated  galaxy  associated  to  an
``X''-shaped radio source.  A faint arc like feature is visible at the
southern end  of the galaxy, see  zoom image.  3C315 has  a very round
large companion galaxy to the  south.  A small faint source is visible
to the southwest of the chip.

\item[3C319]  The host  galaxy associated  to  3C319 is  close to  the
southwest  edge of  the  image.  It appears  as  a slightly  elongated
elliptical. Three companions are also visible in the image.

\item[3C321]  This galaxy  shows a  clearly visible  dust lane  on our
near-infrared image.  A companion in  the process of merging (Roche \&
Eales, 2000) is located $3.5\arcsec$ to the northwest and is elongated
towards  the central  source along  the  radio axis.  Bright knots  of
emission are present along the northern border of the dust lane on the
UV image (Allen et al. 2002).

\item[3C346] The northwestern galaxy  of the double system corresponds
with the  radio core and has a  very bright nucleus. The  radio jet of
this  galaxy bends at  a very  bright knot,  unresolved in  the NICMOS
image.  This knot is located  $\thicksim2.2\arcsec$ to the east of the
nucleus.   The jet  and its  knots are  visible in  this near-infrared
image and  in the  optical as well  (de Koff  et al. 1996).   The main
companion  galaxy  is  highly  asymmetric and  possibly  merging  with
3C346. A  galaxy, likely to  be an edge-on  spiral, is visible  to the
southwest of the image.

\item[3C348] This  source, also known as  Hercules A, is  a cluster cD
galaxy.  3C348  is a  double-galaxy system in  the process  of merging
(Sadun \& Hayes, 1993). The radio core coincides with the southeastern
galaxy.   This galaxy has  faint dust  rings around  the core  in this
near-infrared image.  These  rings are discussed in detail  by Baum et
al. (1996) based on WFPC2 observations.

\item[3C349]  The  NICMOS  image  shows a  boxy,  elongated,  slightly
asymmetric elliptical galaxy with an unresolved nucleus.

\item[3C353] A  round giant elliptical  with no signs  of disturbance.
Martel    et    al.     (1999)    find   very    circular    isophotes
($e\thickapprox0.04$) on the optical image.  A few small sources, most
likely globular clusters, are visible around this galaxy.

\item[3C371]  This object  is  often classified  as  a BL  Lac in  the
literature. In our image it appears  as a round and smooth galaxy with
a very bright nucleus that causes marked diffraction spikes.

\item[3C379.1]  Slightly  elongated  elliptical  galaxy. We  notice  a
curious arc-like feature $5\arcsec$ to the west. This arc like feature
might be a lensed arc or a merger remnant, see zoom image.

\item[3C381] Smooth elliptical with a close projected companion galaxy
to the  east.  Roche \& Eales  (2000), with a U-band  image taken with
the Wide Field Camera on  the Isaac Newton Telescope, find an apparent
tidal distortion that confirms  that the two galaxies are interacting.
There  is an  additional faint  unresolved  source $\thicksim2\arcsec$
south.

\item[3C382]  This galaxy has  a very  bright unresolved  nucleus with
marked diffraction spikes.   This galaxy lies on a  field populated by
unresolved  sources. 

\item[3C386]  This elliptical  galaxy  fills a  large  portion of  the
detector.  New spectroscopic evidence shows  that what appears to be a
bright  nucleus is a  star superimposed  on the  center of  the galaxy
(Marchesini et  al.  private  communication; Chiaberge et  al.  2002).
Several point  sources are discernible on  the field of  view, some of
them are possibly globular clusters. 

\item[3C388]  This  is the  cD  galaxy of  an  Abell  class 2  cluster
(Prestage \& Peacock 1988) with  a close bright and round companion to
the southwest.  Two possible nearby companions are  visible within the
halo of the galaxy, about 1'' north of the nucleus. These two possible
companions were also detected by  Martel et al.  (1999) in their WFPC2
image. In the near-infrared, they appear unresolved.

\item[3C390.3] The host galaxy of  this BLRG is clearly visible in our
NICMOS    image.    A    faint   arc-like    component    is   visible
$\thicksim3\arcsec$ southeast of the nucleus.

\item[3C401] This  elliptical shows an unresolved  nucleus.  The image
shows a faint jet about 6$\arcsec$ southwest of the nucleus identified
by  Chiaberge  et al.   (2005)  as  the  infrared counterpart  of  the
brightest region  of the radio jet.   Chiaberge et al.   find that the
infrared emission  dominates the spectral energy  distribution of this
jet. An  elliptical companion  is visible $4\arcsec$  to the  north. A
bright  unresolved  source  is   present  $7.5\arcsec$  north  of  the
core. This  object lies on  the radio jet  axis but is  not associated
with any  of the radio  features. In the  optical image of de  Koff et
al. (1996)  it appears as a  diffuse source. It is  possible that this
object is an infrared-bright background galaxy.

\item[3C402] A large, elongated,  smooth elliptical with several point
like sources which might be globular clusters.

\item[3C403]  Elliptical galaxy  with  a companion  to the  southeast.
This  galaxy appears  to be  in a  region of  high  foreground stellar
density.

\item[3C430]  A  dust  lane  around  the  nucleus  of  this  elongated
elliptical is weakly visible  on this image.  Several foreground stars
are present in the field of  view. Some of this compact sources may be
globular clusters.

\item[3C433] This cluster member (McCarthy, 1995) has a peculiar radio
jet oriented north-south. We see an elongated elliptical galaxy with a
bright  nucleus. The  nucleus is  totally absent  in the  optical band
(Chiaberge et  al. 1999). A patch  of emission detected by  de Koff et
al.  (1996)  in the optical  is also present  to the northwest  of the
nucleus  on our  NICMOS  image.   Two companion  galaxies  lie to  the
north-northeast. A  bright foreground star is visible  on the northern
corner of the image.

\item[3C436] The galaxy has  an elongated shape, somewhat aligned with
the radio image. An elongated companion is present $8.7\arcsec$ to the
east-northeast.

\item[3C438] This  cluster member has  at least three  companions, the
brightest  of which lies  approx. $4\arcsec$  to the  northeast.  This
galaxy appears  to be in a  region of high  foreground stellar density
given the presence of eight unresolved sources on the HST image.

\item[3C445]  This BLRG  appears as  a  round elliptical  with a  very
bright unresolved nucleus and clearly visible diffraction spikes. Some
faint globular clusters are also visible.

\item[3C449]  The NICMOS  image of  this  galaxy shows  the dust  lane
detected  with WFPC2  by  Martel  et al.   (1999)  although with  less
dramatic features, see zoom image. The dust lane has been modeled as a
warped disk  by Tremblay et al.  (2005).  We also  detect the abundant
globular  clusters  seen in  the  optical.   The  galaxy is  elongated
north-south and occupies a large portion of the NIC2 chip.

\item[3C452] An elliptical elongated east  to west showing no signs of
disturbance. A faint compact source is visible $1\arcsec$ southwest of
the nucleus. 

\item[3C465] The NICMOS image shows an elongated galaxy with its close
companion  visible  to  the  north.  Globular  clusters  surround  the
galaxy. This galaxy,  the brightest galaxy in the  cluster Abell 2634,
shows  a faint  unresolved nucleus  from the  near-infrared to  the UV
(Allen et al.  2002, Chiaberge et al. 1999) and a visible dust lane in
the optical (Martel et al.  1999).

\end{description}


\section {Conclusion}

We  have   presented  69  HST  NICMOS  images   of  the  near-infrared
counterparts of  radio sources of  the 3CR catalogue at  low redshift.
These  NICMOS images  allow us  to charaterize  the host  galaxies and
their environment. We also detect  special features such as jets, dust
disks,  nuclei,   and  hot  spots.   This   database  of  high-quality
near-infrared  images presented  in  this paper  provide an  excellent
foundation for future statistical studies of radio galaxies and galaxy
evolution.


\acknowledgments

This  research has  made  use  of the  NASA  Astrophysics Data  System
Bibliographic services.  This  research also made use made  use of the
NASA/IPAC Extragalactic  Database (NED) which  is operated by  the Jet
Propulsion  Laboratory,  California  Institute  of  Technology,  under
contract with  the National Aeronautics and  Space Administration.  We
are grateful  to Eddie  Bergeron and Santiago  Arribas for  helping us
with the NICMOS data reduction.




\clearpage

\begin{deluxetable}{llcccc}  
\tablecaption{Observation Log\label{tbl-1}}   \tablewidth{0pt}
\tablehead{\colhead{Source} & \colhead{Date (UT)} & \colhead{$\alpha$}& \colhead{$\delta$} &  \colhead{Gal Lat}
\\
\colhead{(1)}& \colhead{(2)}& \colhead{(3)}& \colhead{(4)}& \colhead{(5)}
}

\startdata

3C20         &  2005 Feb 27 &  00 43 09.27   &   +52 03 36.66  &  -10.79  \\
3C28         &  2005 Jun 13 &  00 55 50.65   &   +26 24 36.93  &  -36.45  \\
3C29	     &  2004 Dec  4 &  00 57 34.88   &   -01 23 27.55  &  -64.22  \\
3C31	     &  2005 Jun 17 &  01 07 24.99   &   +32 24 45.02  &  -30.34  \\
3C33.1       &  2004 Aug 15 &  01 09 44.27   &   +73 11 57.2   &  +10.38  \\
3C35	     &  2005 Mar 16 &  01 12 02.29   &   +49 28 35.33  &  -13.26  \\
3C52	     &  2005 Mar 11 &  01 48 28.90   &   +53 32 27.9   &   -8.38  \\
3C61.1       &  2004 Aug  9 &  02 22 36.00   &   +86 19 08.0   &  +23.73  \\
3C66B	     &  2004 Nov  5 &  02 23 11.46   &   +42 59 31.34  &  -16.77  \\
3C75N	     &  2004 Nov 11 &  02 57 41.55   &   +06 01 36.58  &  -44.93  \\
3C76.1       &  2005 Feb  6 &  03 03 15.0    &   +16 26 19.85  &  -35.96  \\
3C79	     &  2004 Oct 30 &  03 10 00.1    &   +17 05 58.91  &  -34.46  \\
3C83.1       &  2005 Mar 12 &  03 18 15.8    &   +41 51 28.0   &  -13.13  \\
3C88	     &  2004 Nov  6 &  03 27 54.17   &   +02 33 41.82  &  -42.02  \\
3C105	     &  2004 Oct 26 &  04 07 16.46   &   +03 42 25.68  &  -33.62  \\
3C111	     &  2004 Dec  8 &  04 18 21.05   &   +38 01 35.77  &   -8.82  \\
3C123	     &  2004 Dec  7 &  04 37 04.4    &   +29 40 13.2   &  -11.66  \\
3C129	     &  2004 Dec  8 &  04 49 09.07   &   +45 00 39.0   &   +0.14  \\
3C129.1      &  2004 Nov 22 &  04 50 06.7    &   +45 03 06.0   &   +0.30  \\
3C130	     &  2005 Apr 19 &  04 52 52.78   &   +52 04 47.53  &   +5.12  \\
3C133	     &  2004 Dec 13 &  05 02 58.4    &   +25 16 28.0   &   -9.91  \\
3C135	     &  2005 Apr  8 &  05 14 08.3    &   +00 56 32.0   &  -21.04  \\
3C165	     &  2005 Apr 26 &  06 43 06.6    &   +23 19 03.0   &   +8.67  \\
3C171	     &  2004 Nov 14 &  06 55 14.72   &   +54 08 58.27  &  +22.23  \\
3C173.1      &  2004 Nov 22 &  07 09 24.34   &   +74 49 15.19  &  +27.27  \\
3C180	     &  2005 Feb 20 &  07 27 04.77   &   -02 04 30.97  &  +06.96  \\
3C184.1      &  2004 Nov 26 &  07 43 01.28   &   +80 26 26.3   &  +28.86  \\
3C192	     &  2005 Jan  8 &  08 05 35.0    &   +24 09 50.0   &  +26.40  \\
3C196.1      &  2005 Feb  1 &  08 15 27.73   &   -03 08 26.99  &  +17.07  \\
3C197.1      &  2005 Apr 19 &  08 21 33.7    &   +47 02 37.0   &  +34.48  \\
3C198	     &  2005 May  3 &  08 22 31.9    &   +05 57 7.0    &  +22.95  \\
3C213.1      &  2005 Feb 12 &  09 01 05.3    &   +29 01 46.0   &  +39.67  \\
3C219	     &  2004 Sep 14 &  09 21 8.64    &   +45 38 56.49  &  +44.76  \\
3C223	     &  2005 Feb 10 &  09 39 52.76   &   +35 53 59.12  &  +48.66  \\
3C223.1      &  2005 Jan 18 &  09 41 24.04   &   +39 44 42.39  &  +48.92  \\
3C227	     &  2005 Mar 28 &  09 47 45.14   &   +07 25 20.33  &  +42.29  \\
3C236	     &  2004 Nov  2 &  10 06 01.7    &   +34 54 10.0   &  +53.98  \\
3C277.3      &  2005 Mar 24 &  12 54 12.06   &   +27 37 32.66  &  +89.21  \\
3C285	     &  2004 Dec  5 &  13 21 17.8    &   +42 35 15.0   &  +73.39  \\
3C287.1      &  2005 Jul 16 &  13 32 53.27   &   +02 00 44.73  &  +62.99  \\
3C288	     &  2004 Oct 31 &  13 38 50.0    &   +38 51 10.7   &  +74.66  \\
3C303	     &  2004 Dec 26 &  14 43 02.74   &   +52 01 37.5   &  +57.50  \\
3C310	     &  2004 Aug 13 &  15 04 57.18   &   +26 00 56.87  &  +60.21  \\
3C314.1      &  2005 Feb 24 &  15 10 23.12   &   +70 45 53.4   &  +42.18  \\
3C315	     &  2004 Dec 30 &  15 13 40.0    &   +26 07 27.0   &  +58.30  \\
3C319	     &  2004 Dec 29 &  15 24 05.5    &   +54 28 14.6   &  +51.05  \\
3C321	     &  2004 Dec 27 &  15 31 43.4    &   +24 04 19.0   &  +53.88  \\
3C346	     &  2005 May 19 &  16 43 48.69   &   +17 15 48.09  &  +35.77  \\ 
3C348	     &  2005 May  9 &  16 51 08.16   &   +04 59 33.84  &  +28.95  \\
3C349	     &  2005 Mar 23 &  16 59 28.84   &   +47 02 56.8   &  +38.20  \\
3C353	     &  2004 Sep  9 &  17 20 28.16   &   -00 58 47.06  &  +19.65  \\ 
3C371	     &  2005 Jan 29 &  18 06 50.6    &   +69 49 28.0   &  +29.17  \\
3C379.1      &  2004 Nov  5 &  18 24 32.53   &   +74 20 58.64  &  +27.85  \\
3C381	     &  2004 Nov 11 &  18 33 46.29   &   +47 27 02.9   &  +22.48  \\
3C382	     &  2005 Jun 22 &  18 35 03.45   &   +32 41 46.18  &  +17.45  \\
3C386	     &  2005 Jun 15 &  18 38 26.27   &   +17 11 49.57  &  +10.55  \\
3C388	     &  2004 Oct 19 &  18 44 02.4    &   +45 33 30.0   &  +20.22  \\
3C390.3      &  2004 Sep 17 &  18 42 09.0    &   +79 46 17.0   &  +27.07  \\
3C401	     &  2004 Aug 11 &  19 40 25.14   &   +60 41 36.85  &  +17.77  \\
3C402	     &  2004 Dec 10 &  19 41 46.0    &   +50 35 44.9   &  +13.27  \\
3C403	     &  2004 Nov  6 &  19 52 15.81   &   +02 30 24.4   &  -12.31  \\
3C430	     &  2005 Jan 27 &  21 18 19.15   &   +60 48 06.88  &   +7.96  \\
3C433	     &  2004 Aug 18 &  21 23 44.6    &   +25 04 28.5   &  -17.69  \\ 
3C436	     &  2004 Nov  9 &  21 44 11.74   &   +28 10 18.67  &  -18.77  \\
3C438	     &  2004 Nov 18 &  21 55 52.3    &   +38 00 30.0   &  -12.98  \\
3C445	     &  2005 Jun 24 &  22 23 49.57   &   -02 06 13.08  &  -46.71  \\
3C449	     &  2004 Nov 11 &  22 31 20.63   &   +39 21 30.07  &  -15.92  \\
3C452	     &  2004 Nov 28 &  22 45 48.9    &   +39 41 14.47  &  -17.06  \\
3C465	     &  2004 Sep 28 &  23 38 29.41   &   +27 01 53.03   &  -33.07 \\

\enddata

\tablecomments{Col.   (1),  3CR  number;  col. (2)  observation  date;
col.  (3) right  ascension for  epoch 2000;  col. (4)  declination for
epoch 2000; col. (5) galactic latitude.}

\end{deluxetable}



\clearpage

\begin{deluxetable}{llcccccc}  
\tablecaption{Radio Properties of the NICMOS Snapshot Survey\label{tbl-2}} 
\tablewidth{0pt} \tablehead{\colhead{Source} & \colhead{z} & \colhead{S(178)(Jy)}
& \colhead{log$_{10}$ P$_{178}$}&  \colhead{$\alpha$} & \colhead{LAS} &\colhead{PA}&\colhead{FR} 
\\
\colhead{(1)}& \colhead{(2)}& \colhead{(3)}& \colhead{(4)}& \colhead{(5)}&
\colhead{(6)}& \colhead{(7)}& \colhead{(8)}}

\startdata

3C20	     &  0.174	&  42.9 &  27.43  & 0.67  &   51 & 101 & II \\ 
3C28	     &  0.1952  &  16.3 &  27.11  & 1.06  &   30 & 166 & II \\
3C29	     &  0.04481 &  15.1 &  25.8   & 0.50  &  139 & 160 &  I \\
3C31	     &  0.0167  &  16.8 &  25.0   & 0.57  & 1833 & 159 &  I \\  
3C33.1       &  0.1809  &  13.0 &  26.95  & 0.62  &  216 &  45 & II \\
3C35	     &  0.0670  &  10.5 &  26.0   & 0.77  &  704 &  12 & II \\
3C52	     &  0.2854  &  13.5 &  27.37  & 0.62  &   51 &  20 & II \\
3C61.1       &  0.184	&  31.2 &  27.35  & 0.77  &  186 &   2 & II \\
3C66B	     &  0.0215  &  24.6 &  25.4   & 0.62  &  330 &  54 &  I \\
3C75N	     &  0.02315 &  25.8 &  25.5   & 0.71  &  692 & 111 &  I \\
3C76.1       &  0.0324  &  13.3 &  24.33  & 0.77  &  200 &     &  I \\
3C79	     &  0.25595 &  30.5 &  27.63  & 0.92  &   86 & 105 & II \\
3C83.1       &  0.0255  &  26.0 &  25.6   & 0.64  &  680 &  96 &  I \\
3C88	     &  0.03022 &  15.3 &  25.5   & 0.52  &  259 &  60 & II \\
3C105	     &  0.089	&  17.8 &  25.74  & 0.58  &	 & 335 & II \\
3C111	     &  0.0485  &  64.6 &  26.5   & 0.73  &  220 &  62 & II \\
3C123	     &  0.2177  & 189.0 &  28.28  & 0.70  &   23 & 115 & II \\
3C129	     &  0.0208  &  46.9 &  25.66  & 0.92  &	 &     &  I \\
3C129.1      &  0.0222  &  10.5 &  25.07  & 0.89  &	 &     &  I \\
3C130	     &  0.1090  &  15.5	&  26.67  & 0.89  &	 &     &  I \\
3C133	     &  0.2775  &  22.3 &  27.57  & 0.70  &   12 & 107 & II \\
3C135	     &  0.1253  &  17.3 &  26.75  & 0.92  &  130 &     & II \\
3C165	     &  0.2957  &  13.5 &  27.41  & 0.71  &	 & 159 &    \\
3C171	     &  0.2384  &  19.5 &  27.37  & 0.87  &   30 & 100 & II \\
3C173.1      &  0.2921  &  15.4 &  27.45  & 0.88  &   58 &  17 & II \\
3C180	     &  0.22	&  15.1 &  27.19  & 0.84  &	 &   7 &    \\
3C184.1      &  0.1182  &  13.0 &  26.56  & 0.68  &  167 & 157 & II \\
3C192	     &  0.0598  &  23.0 &  25.10  & 0.79  &  200 &     & II \\
3C196.1      &  0.198	&  18.6 &  27.18  & 1.16  &    4 &  43 & II \\
3C197.1      &  0.1301  &   8.1 &  26.44  & 0.69  &   14 &   2 & II \\
3C198	     &  0.0815  &   9.7 &  26.1   & 0.69  &  338 &  38 & II \\  
3C213.1      &  0.194	&   6.6 &  26.71  & 0.55  &   43 & 162 & II \\
3C219	     &  0.1744  &  41.2 &  27.41  & 0.81  &  184 &  40 & II \\
3C223	     &  0.1368  &  14.7 &  26.75  & 0.74  &  300 & 164 & II \\
3C223.1      &  0.107	&   6.0 &  26.14  & 0.56  &  117 &  15 & II \\
3C227	     &  0.0861  &  30.4 &  26.7   & 0.67  &  246 &  86 & II \\ 
3C236	     &  0.10050 &  14.4 &  26.5   & 0.51  & 2514 & 122 & II \\
3C277.3      &  0.0857  &   9.0 &  26.1   & 0.58  &   50 & 158 & II \\ 
3C285	     &  0.0794  &  11.3 &  25.46  & 0.95  &  184 &     & II \\
3C287.1      &  0.2159  &   8.2 &  26.9   & 0.52  &  112 &  91 & II \\
3C288	     &  0.246	&  18.9 &  27.39  & 0.85  &   16 & 146 &  I \\
3C303	     &  0.141	&  11.2 &  26.66  & 0.76  &   38 &  97 & II \\
3C310	     &  0.0535  &  55.1 &  26.5   & 0.92  &  323 & 165 &  I \\ 
3C314.1      &  0.1197  &  10.6 &  26.49  & 0.95  &  201 & 144 &  I \\
3C315	     &  0.1083  &  17.8 &  26.62  & 0.72  &	 &  11 &  I \\
3C319	     &  0.192	&  15.3 &  27.07  & 0.90  &   93 &  49 & II \\
3C321	     &  0.096	&  13.5 &  26.4   & 0.60  &  309 & 121 & II \\
3C346	     &  0.161	&  10.9 &  26.76  & 0.52  &   13 &  71 & II \\
3C348	     &  0.154	& 351.0 &  28.23  & 1.00  &  191 & 101 & I  \\
3C349	     &  0.205	&  13.3 &  27.07  & 0.74  &   82 & 142 & II \\
3C353	     &  0.03043 & 236.0 &  26.7   & 0.71  &  275 &  85 & II \\ 
3C371	     &  0.0500  &   3.7 &  25.3   & 0.30  &  277 &  59 & II \\
3C379.1      &  0.256	&   7.4 &  27.01  & 0.68  &   76 & 161 & II \\
3C381	     &  0.1605  &  16.6 &  26.94  & 0.81  &   69 &   4 & II \\
3C382	     &  0.0578  &  19.9 &  26.2   & 0.59  &  179 &  50 & II \\
3C386	     &  0.0170  &  23.9 &  25.2   & 0.59  &  288 &  17 &  I \\
3C388	     &  0.091	&  24.6 &  26.6   & 0.70  &   32 &  63 & II \\
3C390.3      &  0.0561  &  47.5 &  26.5   & 0.75  &  231 & 145 & II \\
3C401	     &  0.20104 &  20.9 &  27.25  & 0.71  &   19 &  24 & II \\
3C402	     &  0.0239  &  10.1 &  25.1   & 0.56  &  528 & 163 & II \\
3C403	     &  0.0590  &  17.8 &  26.1   & 0.45  &  230 &  79 & II \\
3C430	     &  0.0541  &  33.7 &  26.3   & 0.72  &   99 &  35 & II \\ 
3C433	     &  0.1016  &  56.2 &  27.06  & 0.75  &   58 & 172 & II \\
3C436	     &  0.2145  &  17.8 &  27.23  & 0.86  &  105 & 173 & II \\
3C438	     &  0.290	&  44.7 &  27.91  & 0.88  &   19 & 136 & I  \\
3C445	     &  0.0562  &  24.8 &  26.2   & 0.85  &  576 & 171 & II \\
3C449	     &  0.0171  &  11.5 &  24.9   & 0.58  & 1742 &  10 & I  \\
3C452	     &  0.0811  &  54.4 &  26.9   & 0.78  &  277 &  79 & II \\
3C465	     &  0.0303  &  37.8 &  25.8   & 0.75  &  375 & 122 & I  \\

 \enddata

\tablecomments{\small Col.  (1), 3CR  number; col.  (2), redshift; col
(3), radio flux  density in Janskys at 178  MHz; col.  (4), log$_{10}$
of  the radio  power  at 178  MHz  in W  Hz$^{-1}$;  col.  (5),  radio
spectral index  determined between 178  and 408 MHz; col.   (6), radio
source largest angular size  in arcseconds; col.  (7), radio structure
position angle measured north  through east; col.  (8), Fanaroff-Riley
type.}

\tablerefs{\small de Koff  et al.  (1996)  and references therein,  Martel et
al.  (1999) and  references therein,  Spinrad et  al.  (1985), updated
values were taken from NED.}

\end{deluxetable}



\clearpage

\begin{deluxetable}{lcccccl}  
\rotate
\tablecaption{HST NICMOS F160W Properties of the 3CR Snapshot Survey\label{tbl-3}} 
\tablewidth{0pt} 

\tablehead{\colhead{Source} & \colhead{m$_{1kpc}$} & \colhead{m$_{5kpc}$}
& \colhead{m$_{10kpc}$} & \colhead{m$_{15kpc}$} & \colhead{m$_{20kpc}$}& \colhead{Comments}
\\
\colhead{(1)}& \colhead{(2)}& \colhead{(3)}& \colhead{(4)}& \colhead{(5)}&
\colhead{(6)}& \colhead{(7)}}

\startdata

3C20	&  20.606 & 19.270  &  19.016 & 18.883  & 18.733 & Close companion \\
3C28	&  21.486 & 19.382  &  18.709 & 18.475  & 18.393 & Close companion \\		
3C29	&  17.966 & 15.741  &	      & 	&	 & Unresolved nucleus, globular clusters\\  
3C31	&  14.795 &	    &	      & 	&	 & Dust lane, globular clusters, unresolved nucleus\\	  
3C33.1  &  20.308 & 19.365  &  18.923 & 18.707  & 18.678 & Close companion, unresolved nucleus\\	
3C35	&  18.410 & 16.908  &  16.365 & 	&	 & \\	
3C52	&  21.348 & 19.436  &  19.006 & 18.842  & 18.779 & Close companion\\				
3C61.1  &  21.873 & 20.613  &  20.325 & 20.270  &	 & Close companion\\  
3C66B	&  15.827 &	    &	      & 	&	 & Globular clusters, unresolved nucleus, jet\\	  
3C75N	&  15.469 &	    &	      & 	&	 & Close companion, globular clusters\\	 
3C76.1  &  17.038 & 15.995  &	      & 	&	 & Close companion, globular clusters\\			
3C79	&  20.942 & 19.470  &  19.108 &  18.962 & 18.898 & Close companion\\			 
3C83.1  &  15.457 &	    &	      & 	&	 & Dust lane, globular clusters\\	  
3C88	&  17.212 & 15.545  &	      & 	&	 & Globular clusters\\			
3C105	&  19.411 & 18.344  &  18.175 & 	&	 & \\			  
3C111	&  17.053 & 16.572  &	      & 	&	 & Unresolved nucleus\\			
3C123	&  21.694 & 19.851  &  19.337 & 19.117  & 19.010 & Close companion\\		  
3C129	&  16.313 &	    &	      & 	&	 & Globular clusters, arc?\\  
3C129.1 &  16.561 &         &         &         &	 & Globular clusters\\	    
3C130	&  18.271 & 16.297  &  15.704 & 15.373  &	 & Globular clusters\\		
3C133	&  20.754 & 19.868  &  19.593 & 19.447  & 19.323 & Close companion, unresolved nucleus, jet\\
3C135	&  19.882 & 18.690  &  18.273 & 18.224  &	 & Close companion\\			  
3C165	&  21.910 & 20.187  &  19.722 & 19.595  & 19.556 & Close companion\\				
3C171	&  21.562 & 19.997  &  19.657 & 19.487  & 19.445 & Close companion, emission line tails\\	
3C173.1 &  21.119 & 19.425  &  18.998 & 18.837  & 18.759 & Close companion\\				
3C180	&  21.534 & 19.572  &  19.009 & 18.790  & 18.716 & \\					  
3C184.1 &  19.261 & 18.432  &  18.284 & 18.275  &	 & Close companion, arc?, unresolved nucleus\\	
3C192	&  17.988 & 16.942  &  16.893 & 	&	 & Globular clusters\\			  
3C196.1 &  21.766 & 19.535  &  18.789 & 18.462  & 18.352 & Close companion, globular clusters?\\	
3C197.1 &  19.914 & 18.668  &  18.367 & 18.295  &	 & Unresolved nucleus\\			  
3C198	&  19.367 & 18.390  &  18.213 & 	&	 & Unresolved nucleus\\			  
3C213.1 &  20.879 & 19.523  &  19.190 & 19.109  & 19.066 & Hot spots\\				  
3C219	&  20.110 & 18.767  &  18.396 & 18.284  & 18.250 & Close companion, unresolved nucleus\\	
3C223	&  20.118 & 18.761  &  18.416 & 18.351  &	 & \\			  
3C223.1 &  18.841 & 17.660  &  17.386 & 17.327  &	 & Unresolved nucleus, dust lane\\		
3C227	&  17.944 & 17.415  &  17.313 & 17.308  &	 & Close companion, unresolved nucleus\\
3C236	&  18.857 & 17.502  &  17.210 & 17.140  &	 & Dust lane\\			  
3C277.3 &  19.120 & 17.626  &  17.327 & 17.257  &	 & Close companion, radio knots\\	
3C285	&  19.167 & 17.624  &  17.248 & 17.163  &	 & Dust lane, arc\\	  
3C287.1 &  19.731 & 18.971  &  18.717 & 18.626  & 18.576 & Close companion, unresolved nucleus\\
3C288	&  22.113 & 19.476  &  18.657 & 18.367  & 18.227 & Close companion, unresolved nucleus\\
3C303	&  19.230 & 18.243  &  17.961 & 17.868  & 17.854 & Close companion, unresolved nucleus\\
3C310	&  18.483 & 17.120  &  16.927 & 	&	 & Close companion, globular clusters\\	
3C314.1 &  20.281 &  18.915 &  18.658 &  18.624 & 18.571 & \\		
3C315	&  19.363 & 18.424  &  18.185 & 	&	 & Close companion\\			
3C319	&  21.093 &	    &	      & 	&	 & Close companion\\
3C321	&  19.047 & 17.266  &  16.597 & 	&	 & Close companion, dust lane\\		
3C346	&  20.162 & 19.060  &  18.345 &  18.019 & 17.932 & Close companion, unresolved nucleus, jet\\	
3C348	&  22.272 & 19.328  &  18.343 &  18.098 & 18.021 & Close companion, dust lane\\		
3C349	&  20.859 & 19.657  &  19.347 &  19.275 & 19.259 & Unresolved nucleus\\	  
3C353	&  16.860 & 15.537  &	      & 	&	 & Globular clusters\\		
3C371	&  16.007 & 15.555  &         & 	&	 & Unresolved nucleus\\
3C379.1 &  21.227 & 19.407  &  18.891 & 18.717  &	 & Arc\\		       
3C381	&  20.072 & 18.819  &  18.545 & 18.413  & 18.290 & Close companion, unresolved nucleus\\
3C382	&  15.931 & 15.366  &  15.143 & 	&	 & Unresolved nucleus\\		
3C386	&  15.429 &	    &	      & 	&	 & Globular clusters\\	
3C388	&  19.534 & 17.101  &  16.337 & 15.884  &	 & Close companion\\		
3C390.3 &  16.806 & 16.249  &  16.187 & 	&	 & Unresolved nucleus, arc?\\
3C401	&  21.874 & 19.853  &  19.160 & 18.671  & 18.590 & Close companion, unresolved nucleus, jet\\
3C402	&  15.915 &	    &	      & 	&	 & Globular clusters\\
3C403	&  17.848 & 16.485  &	      & 	&	 & Close companion\\
3C430	&  17.653 & 16.400  &         & 	&	 & Dust lane, globular clusters\\
3C433	&  18.922 & 17.500  &  17.057 &  16.615 &	 & Close companion, unresolved nucleus\\
3C436	&  20.920 & 19.217  &  18.780 &  18.614 & 18.585 & Close companion\\
3C438	&  21.851 & 19.807  &  19.240 &  19.015 & 18.839 & Close companion\\
3C445	&  16.802 & 16.499  &	      & 	&	 & Unresolved nucleus, globular clusters\\
3C449	&  15.418 &         &	      & 	&	 & Dust lane, globular clusters\\
3C452	&  18.564 & 17.191  &  16.643 & 	&	 & \\
3C465	&  16.105 &         &	      & 	&	 & Close companion, globular clusters, unresolved nucleus\\

 \enddata 

\tablecomments{Col.  (1), 3CR number;  Col. (2) through (6) HST NICMOS
F160W magnitude  inside a circular aperture, centered  on the nucleus,
with  a radius  of 1,  5, 10,  15 and  20 kpc  respectively;  Col. (7)
remarks on special features.}

\end{deluxetable}


\clearpage 



\begin{figure}
\plotone{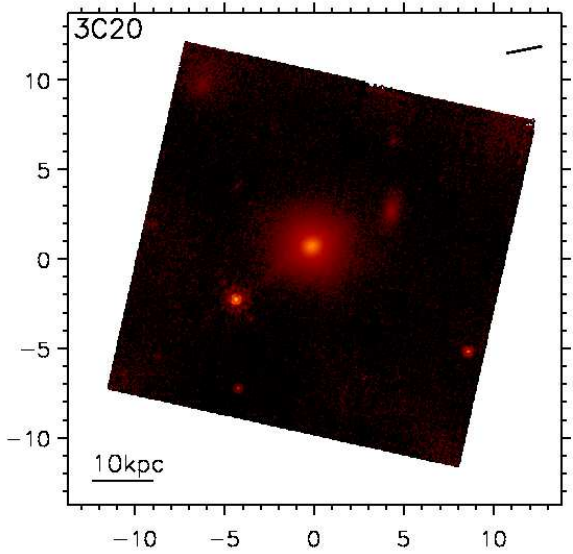}
\caption{ All images  are rotated so they are north  up and east left.
Each figure is plotted within  a box with vertical and horizontal axes
in arcseconds.   The intensity  scale is logarithmic.  A scale  bar is
plotted on  the lower left  corner in kpc,  a second bar on  the upper
right corner  of the  image indicates the  direction of the  radio jet
axis. This is HST/NICMOS F160W image of 3C20}
\end{figure}


\begin{figure}
\plotone{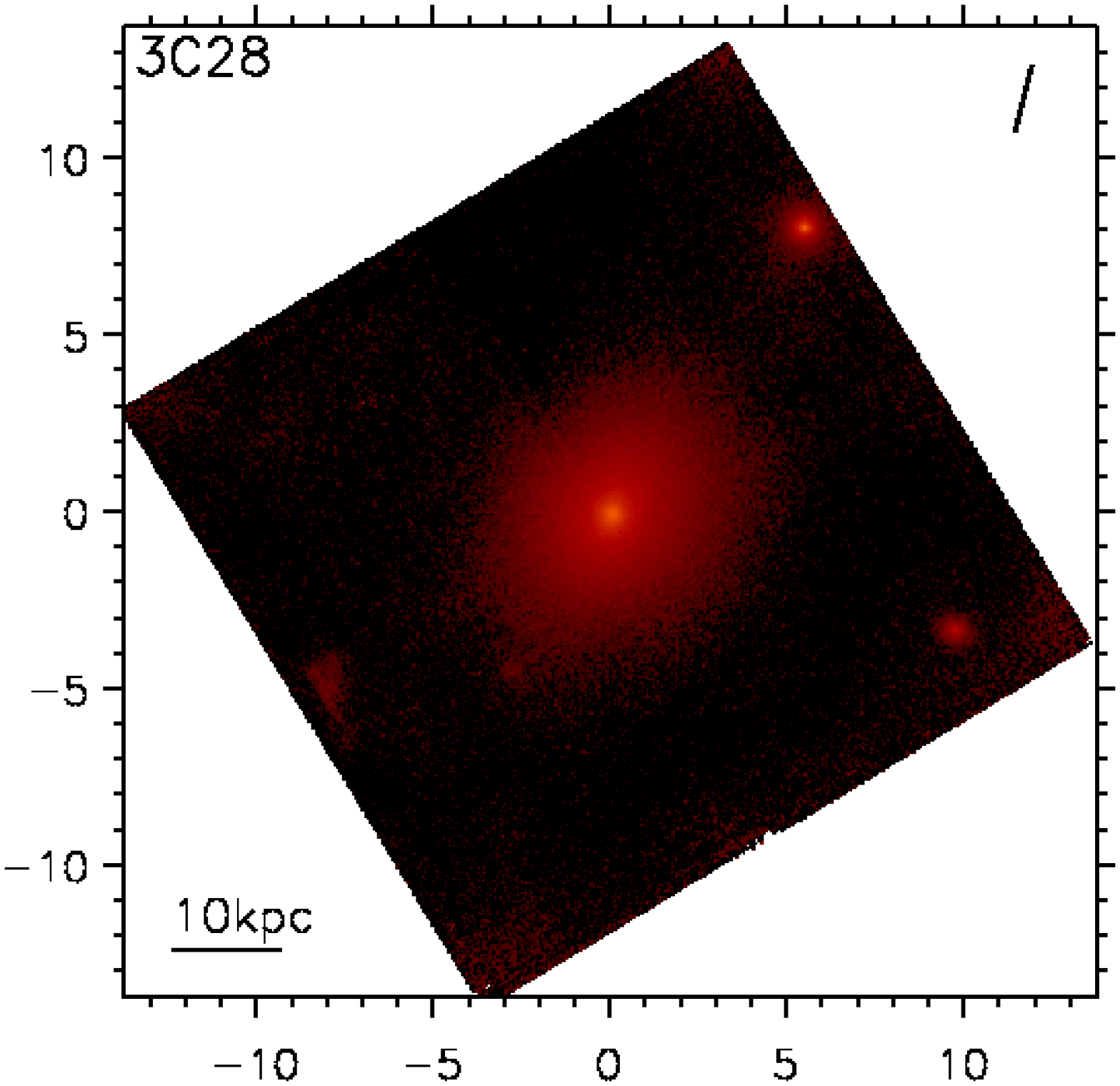}
\caption{HST/NICMOS F160W image of 3C28}
\end{figure}


\begin{figure}
\plotone{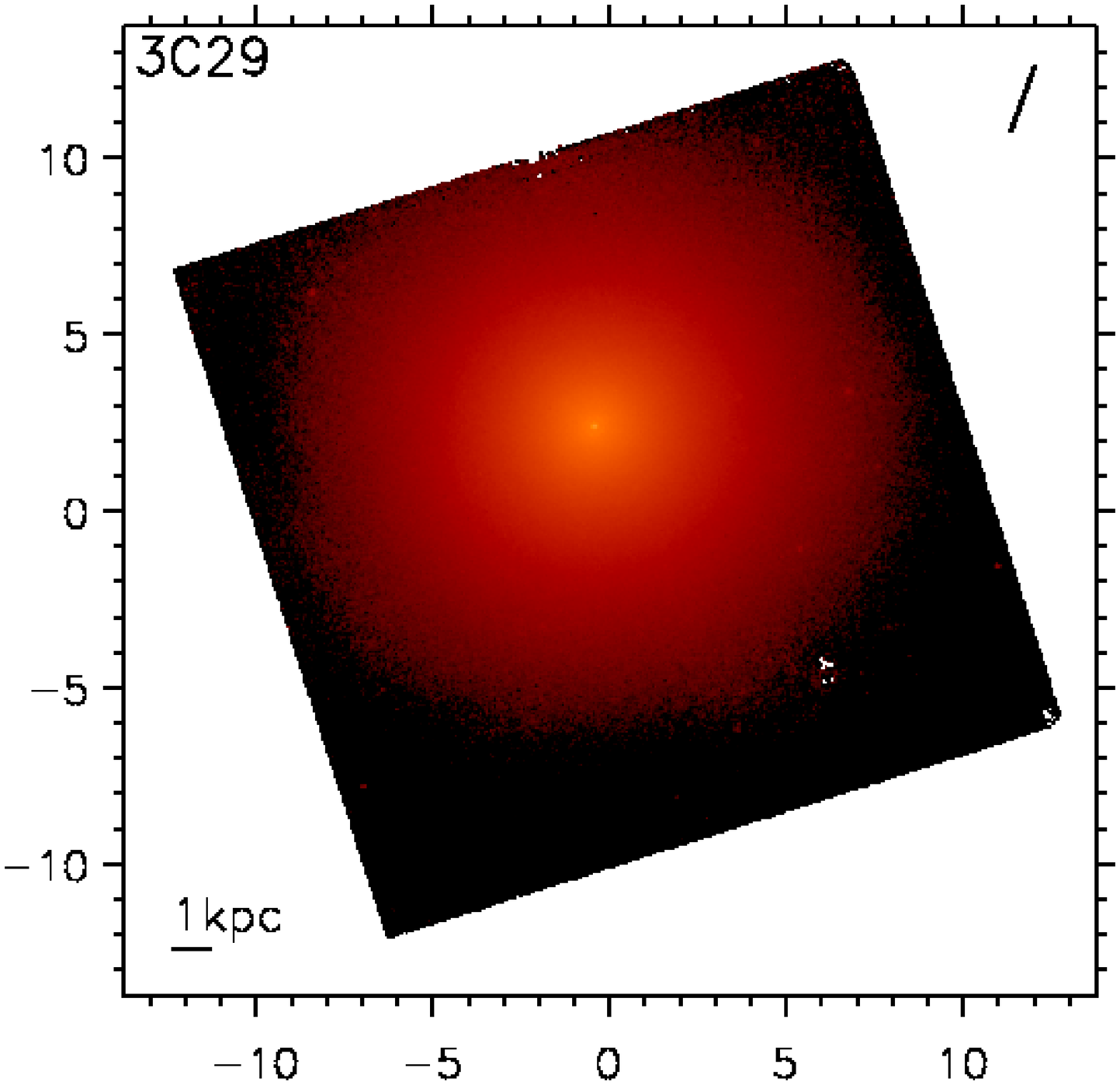}
\caption{HST/NICMOS F160W image of 3C29}
\end{figure}


\begin{figure}
\plotone{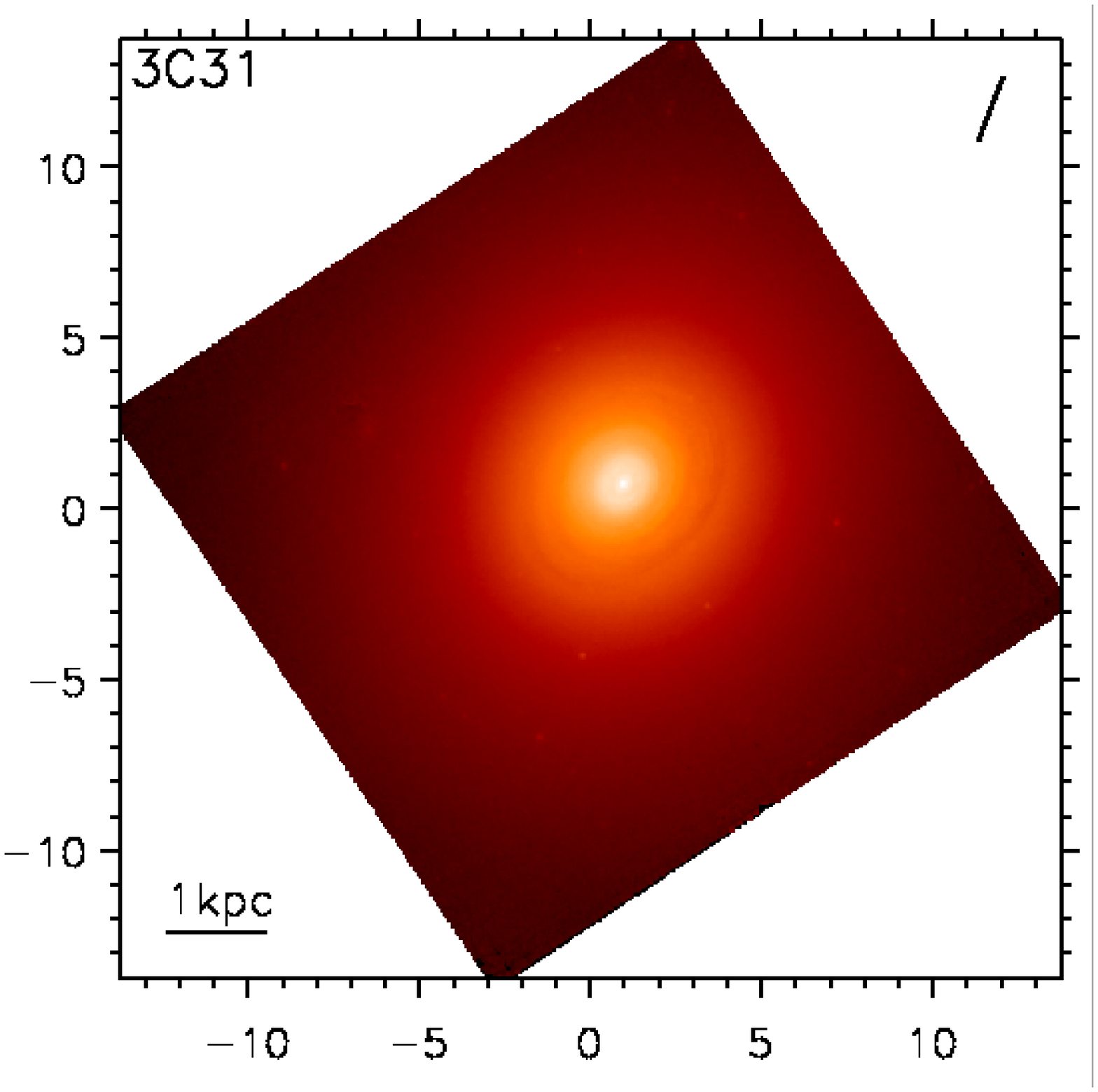}
\caption{HST/NICMOS F160W image of 3C31}
\end{figure}

\clearpage

\begin{figure}
\plotone{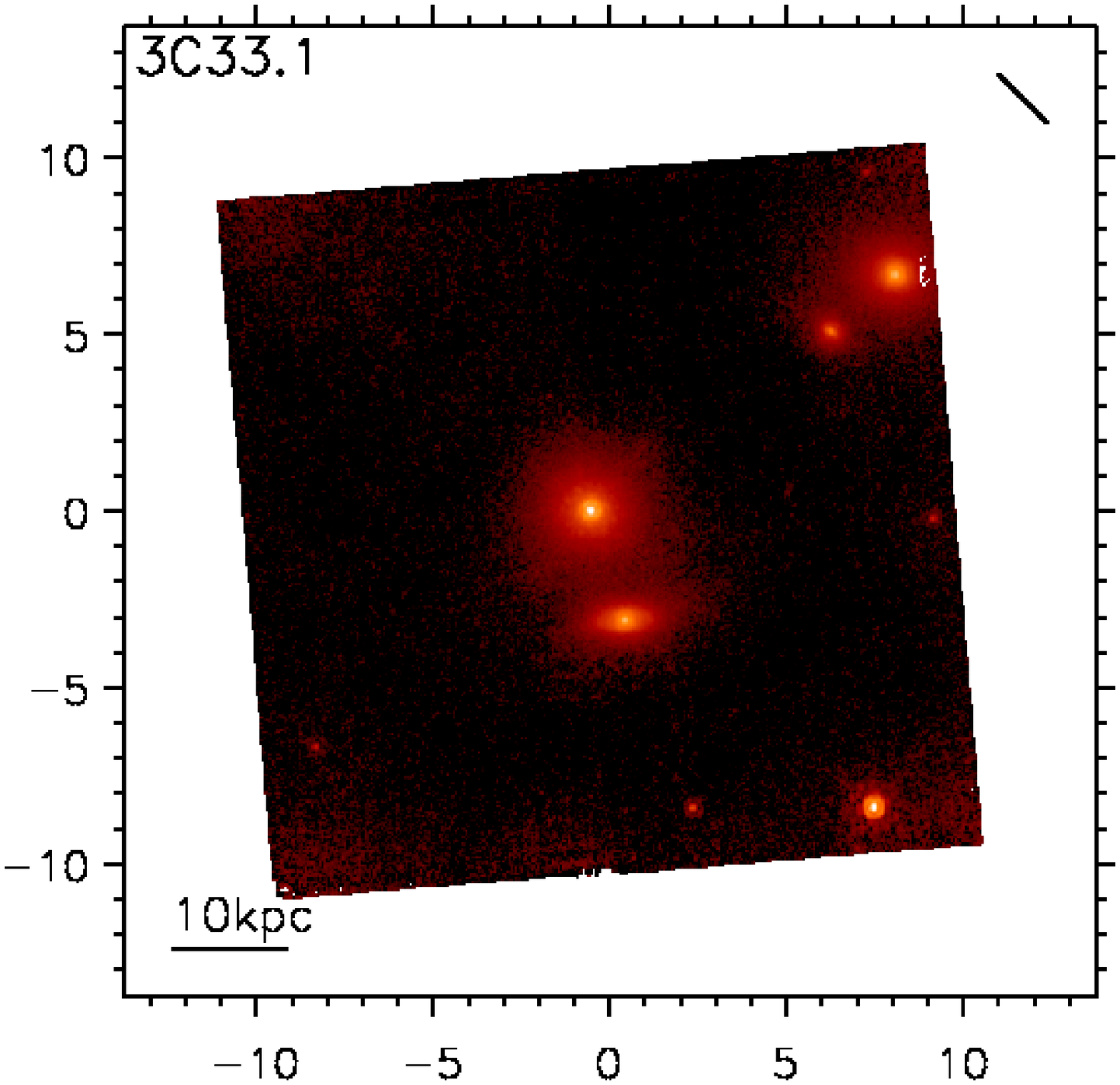}
\caption{HST/NICMOS F160W image of 3C33.1}
\end{figure}

\begin{figure}
\plotone{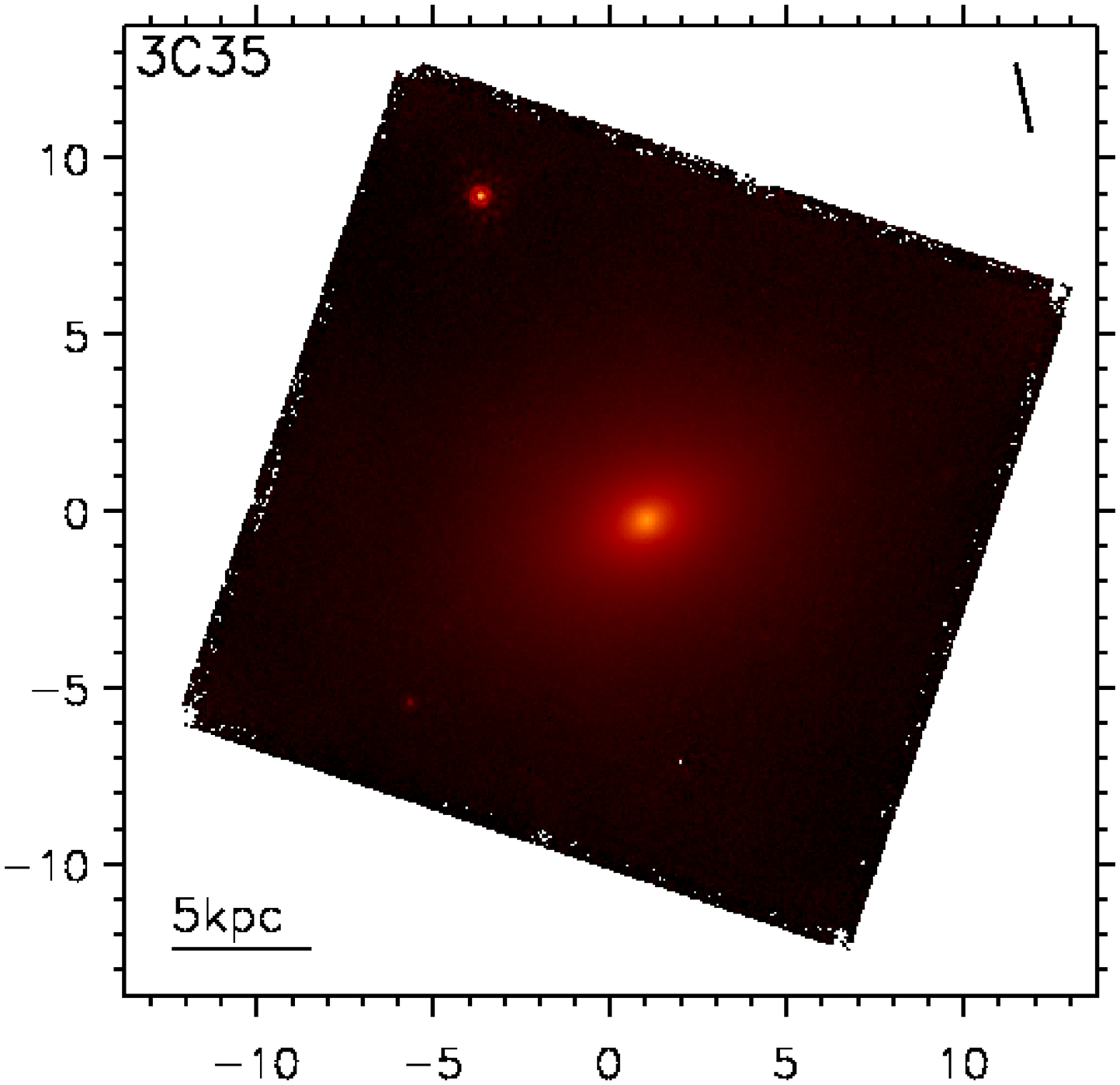}
\caption{HST/NICMOS F160W image of 3C35}
\end{figure}


\begin{figure}
\plotone{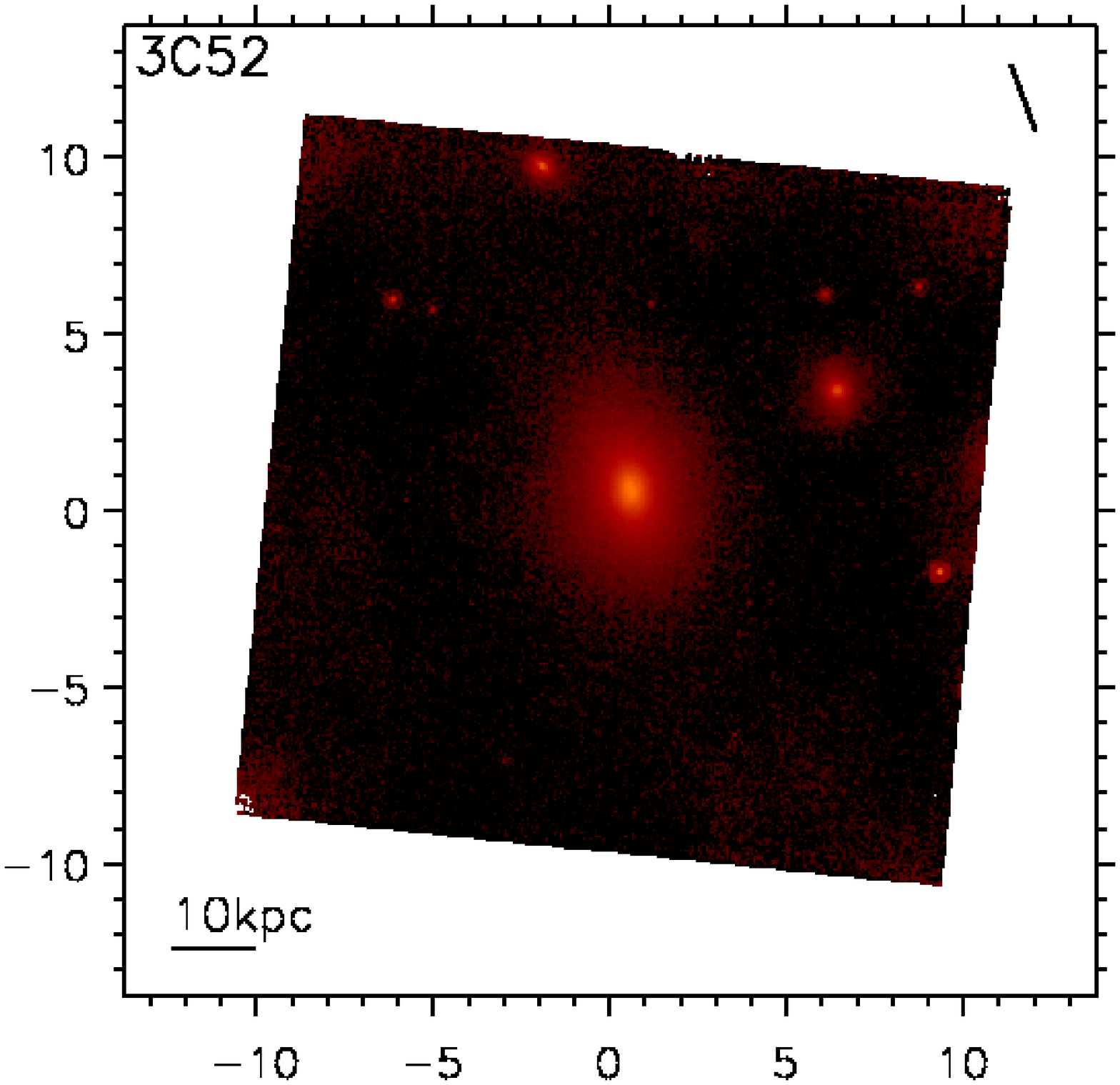}
\caption{HST/NICMOS F160W image of 3C52}
\end{figure}


\begin{figure}
\plotone{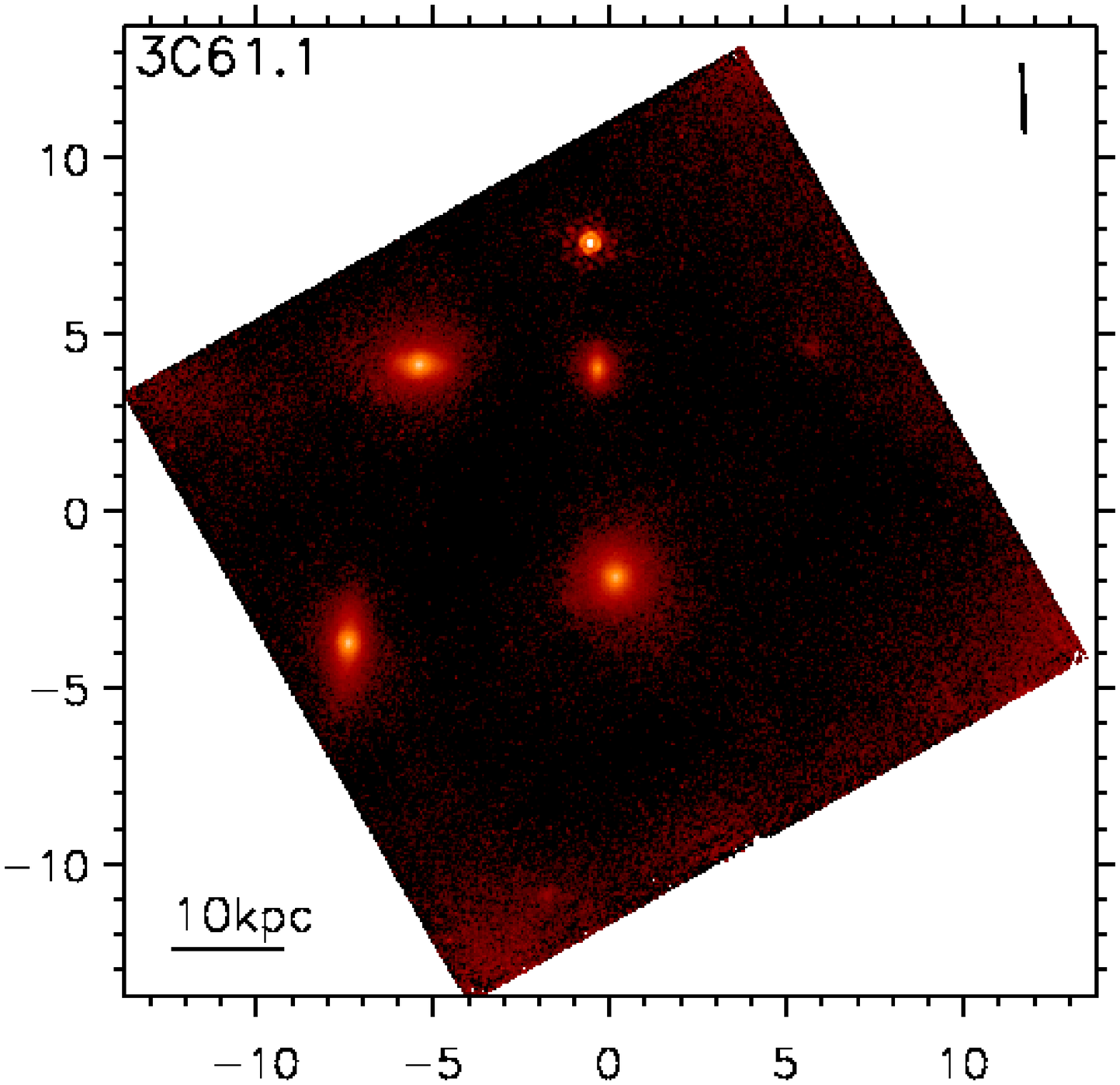}
\caption{HST/NICMOS F160W image of 3C61.1}
\end{figure}

\clearpage


\begin{figure}
\plotone{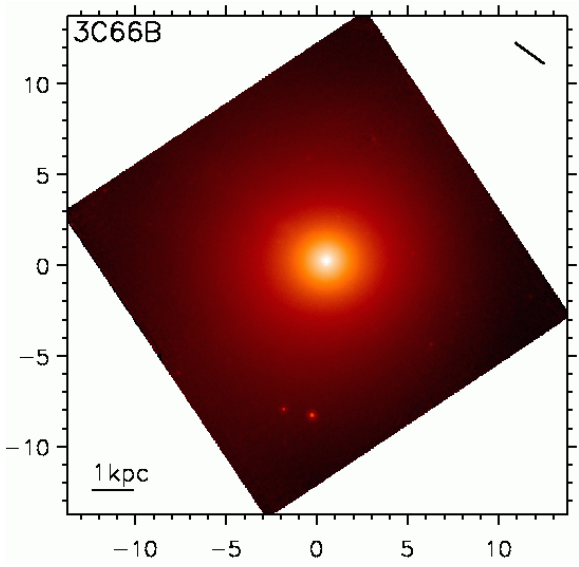}
\caption{HST/NICMOS F160W image of 3C66B}
\end{figure}

\begin{figure}
\plotone{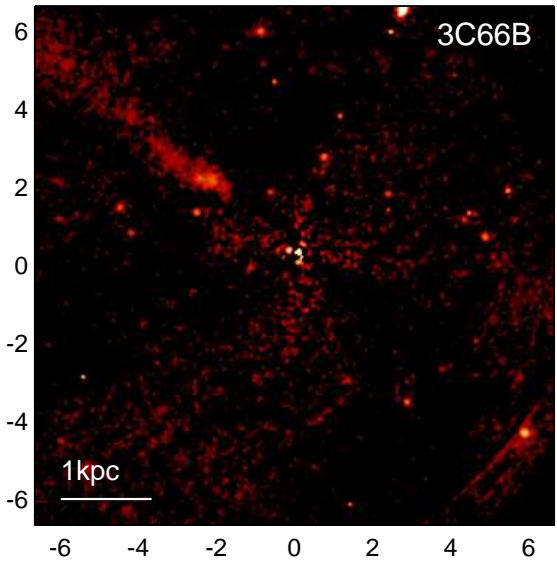}
\newline
\newline
\caption{Model-subtracted residual for 3C66B}
\end{figure}


\begin{figure}
\plotone{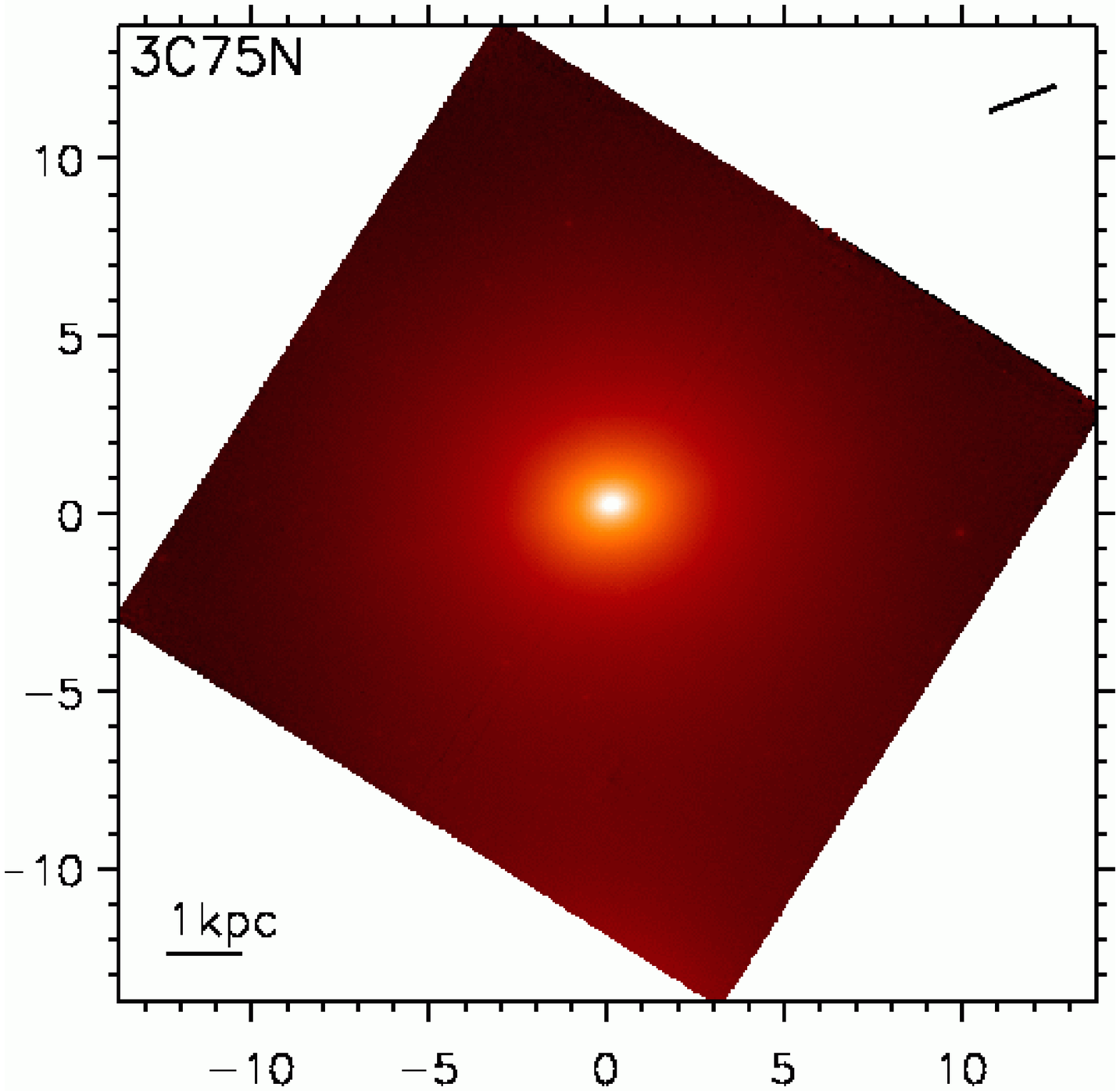}
\caption{HST/NICMOS F160W image of 3C75N}
\end{figure}


\begin{figure}
\plotone{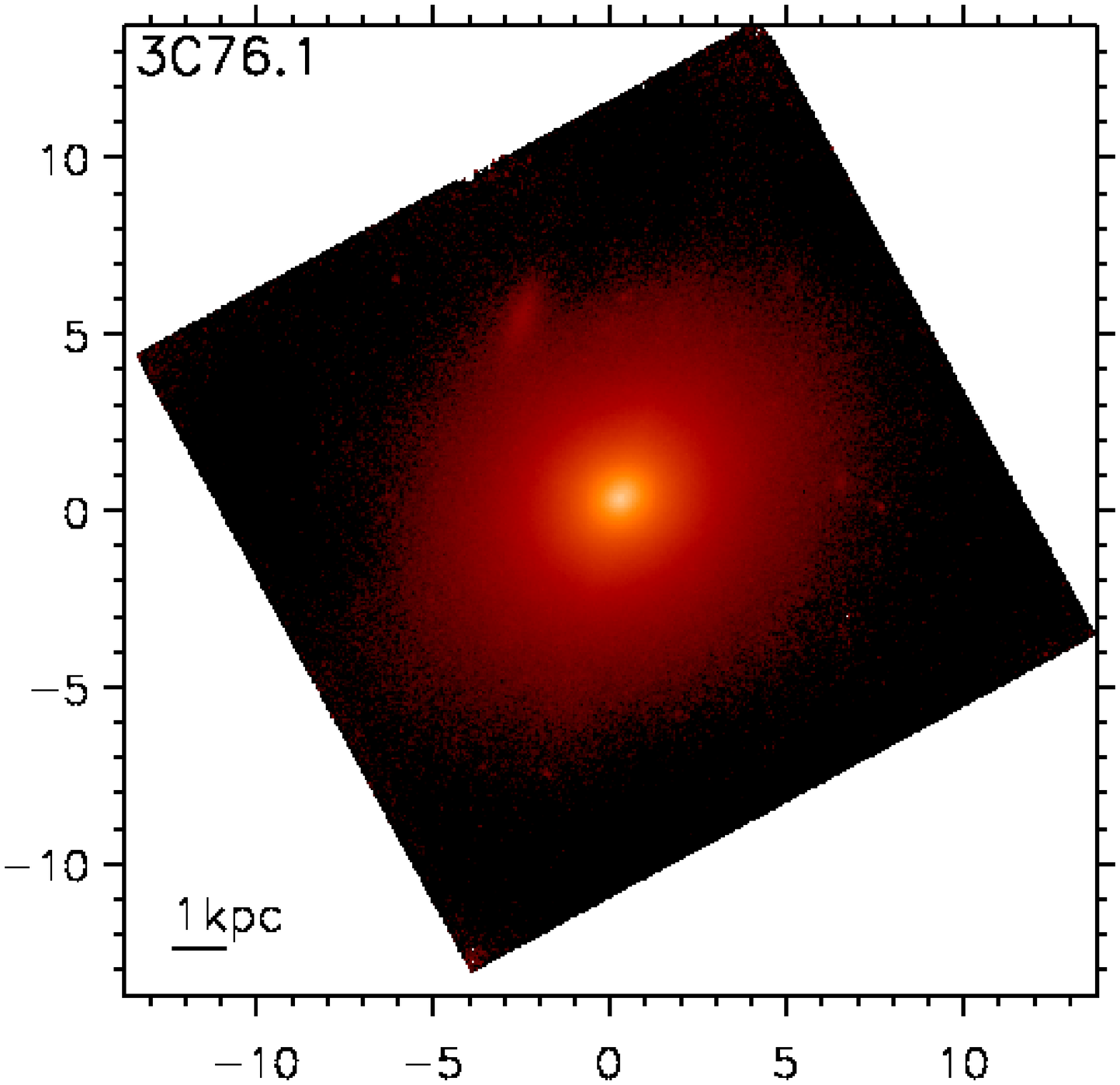}
\caption{HST/NICMOS F160W image of 3C76.1}
\end{figure}

\clearpage


\begin{figure}
\plotone{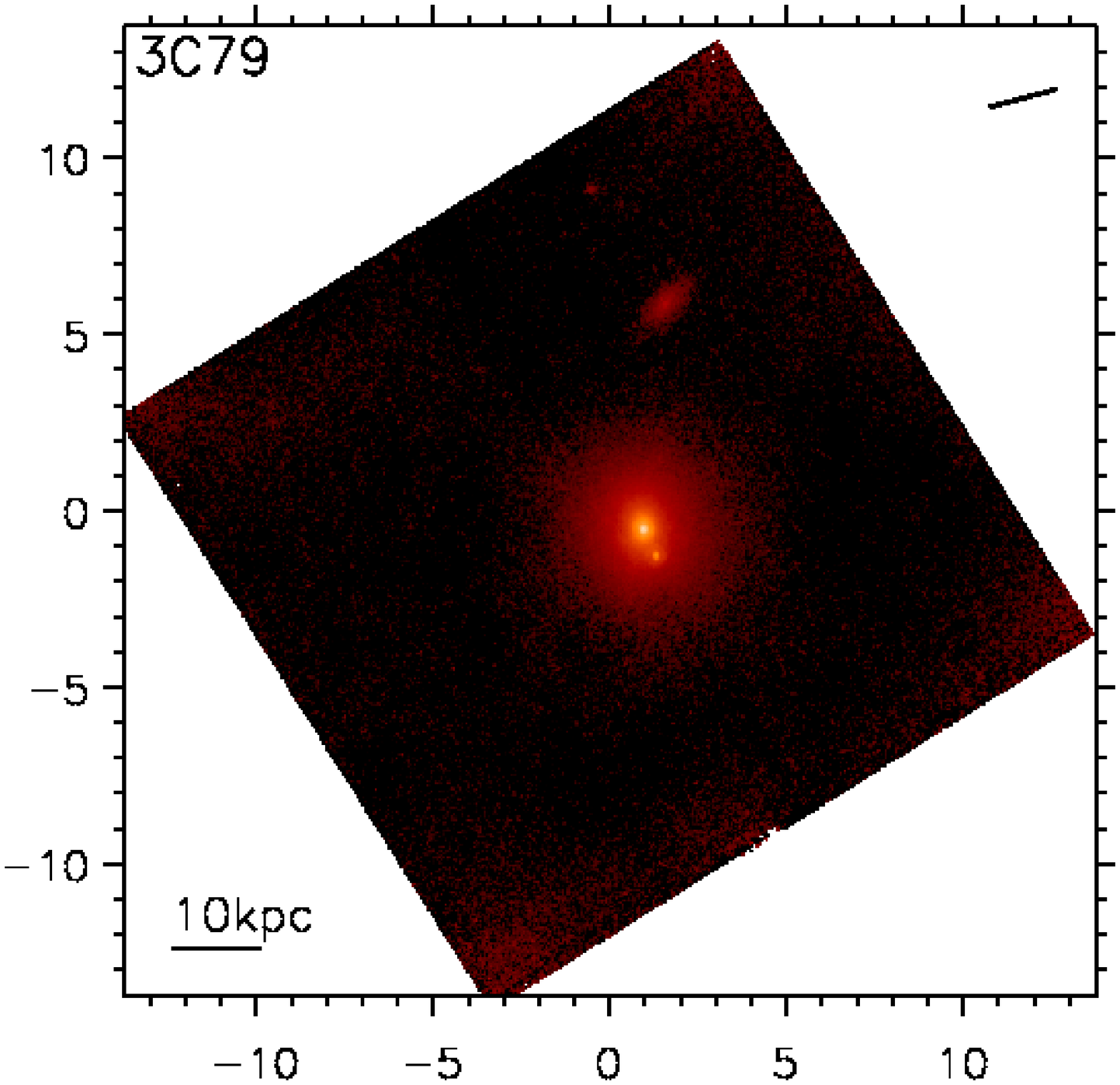}
\caption{HST/NICMOS F160W image of 3C79}
\end{figure}


\begin{figure}
\plotone{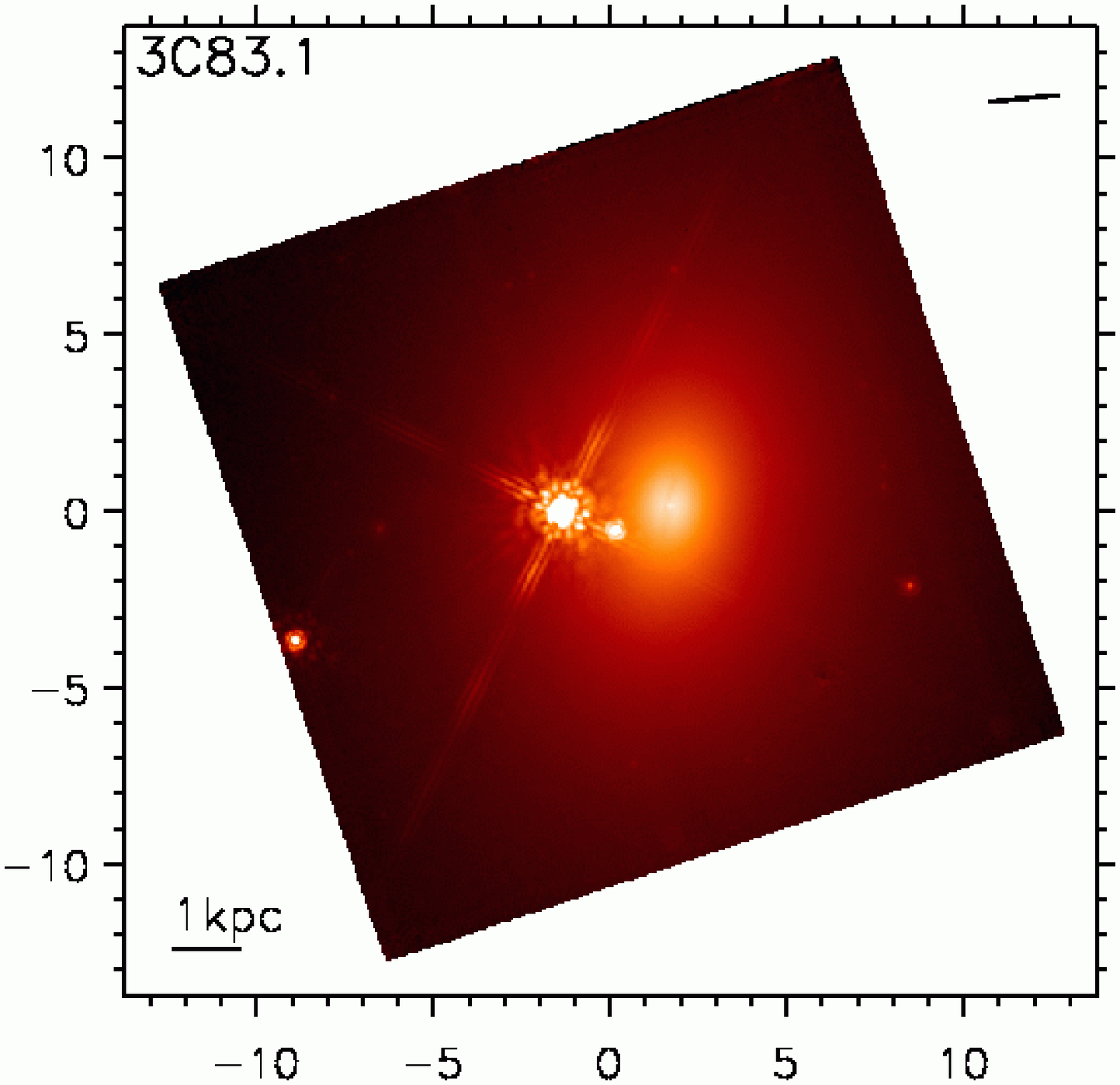}
\caption{HST/NICMOS F160W image of 3C83.1}
\end{figure}


\begin{figure}
\plotone{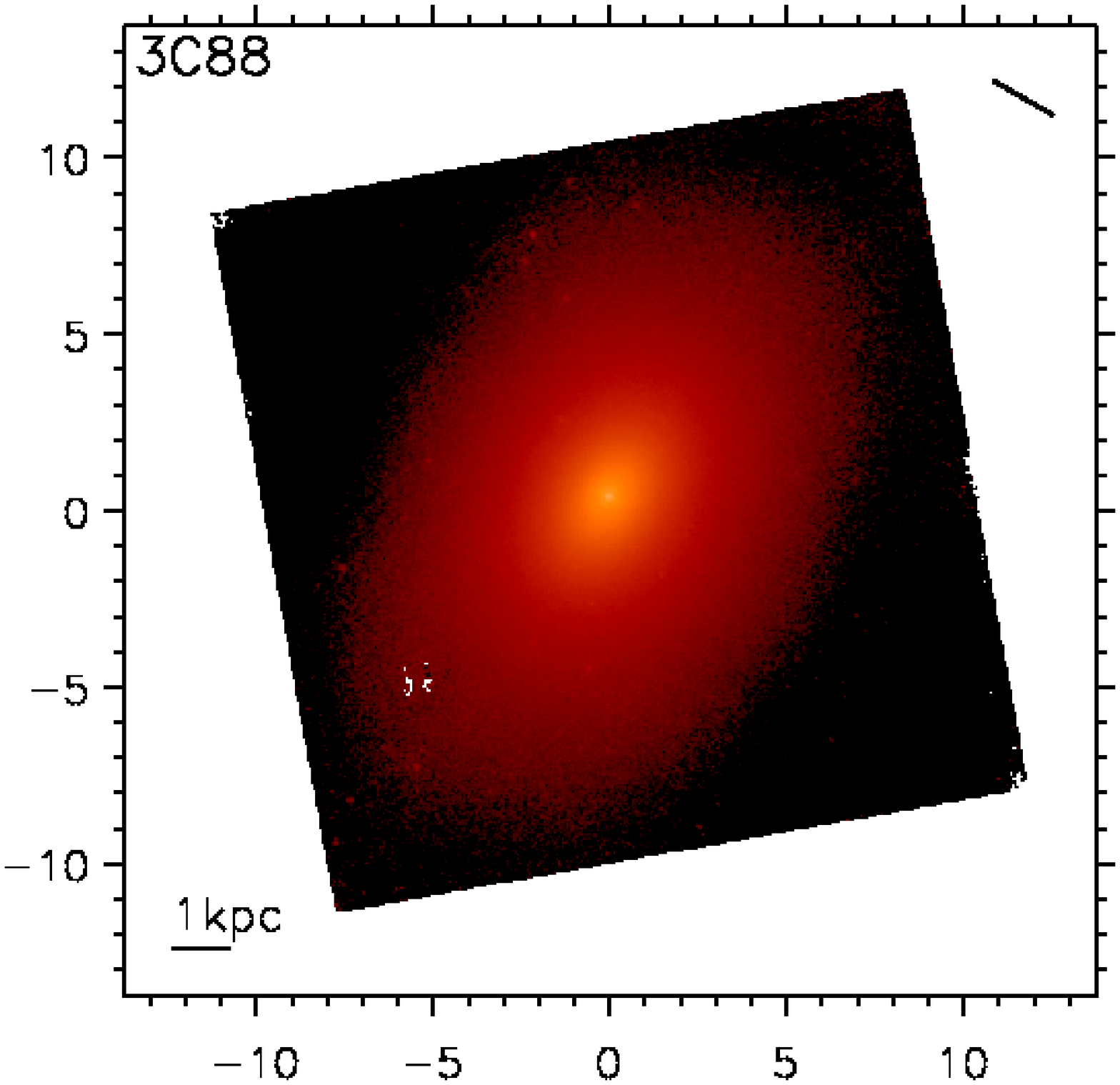}
\caption{HST/NICMOS F160W image of 3C88}
\end{figure}


\begin{figure}
\plotone{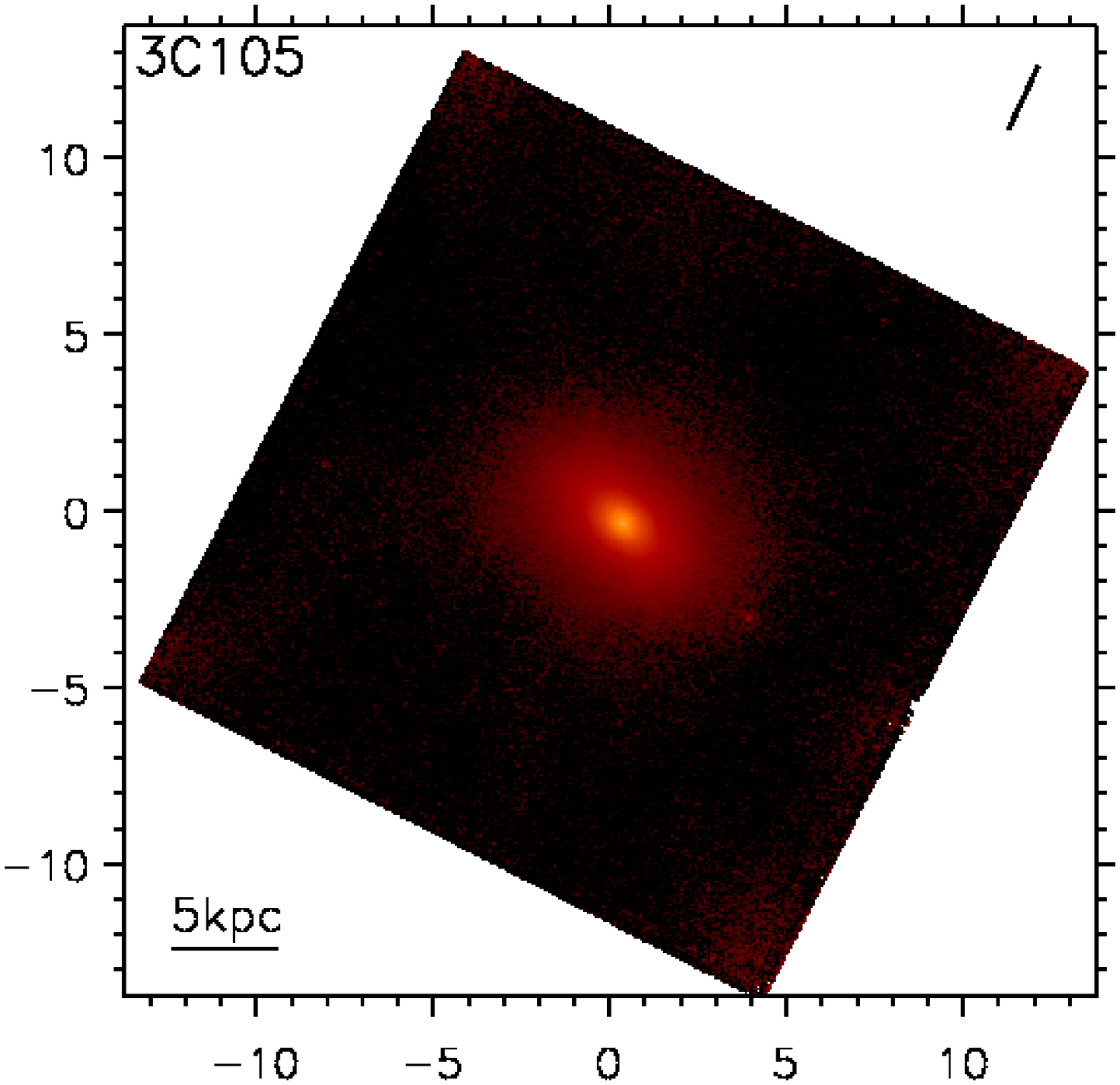}
\caption{HST/NICMOS F160W image of 3C105}
\end{figure}

\clearpage


\begin{figure}
\plotone{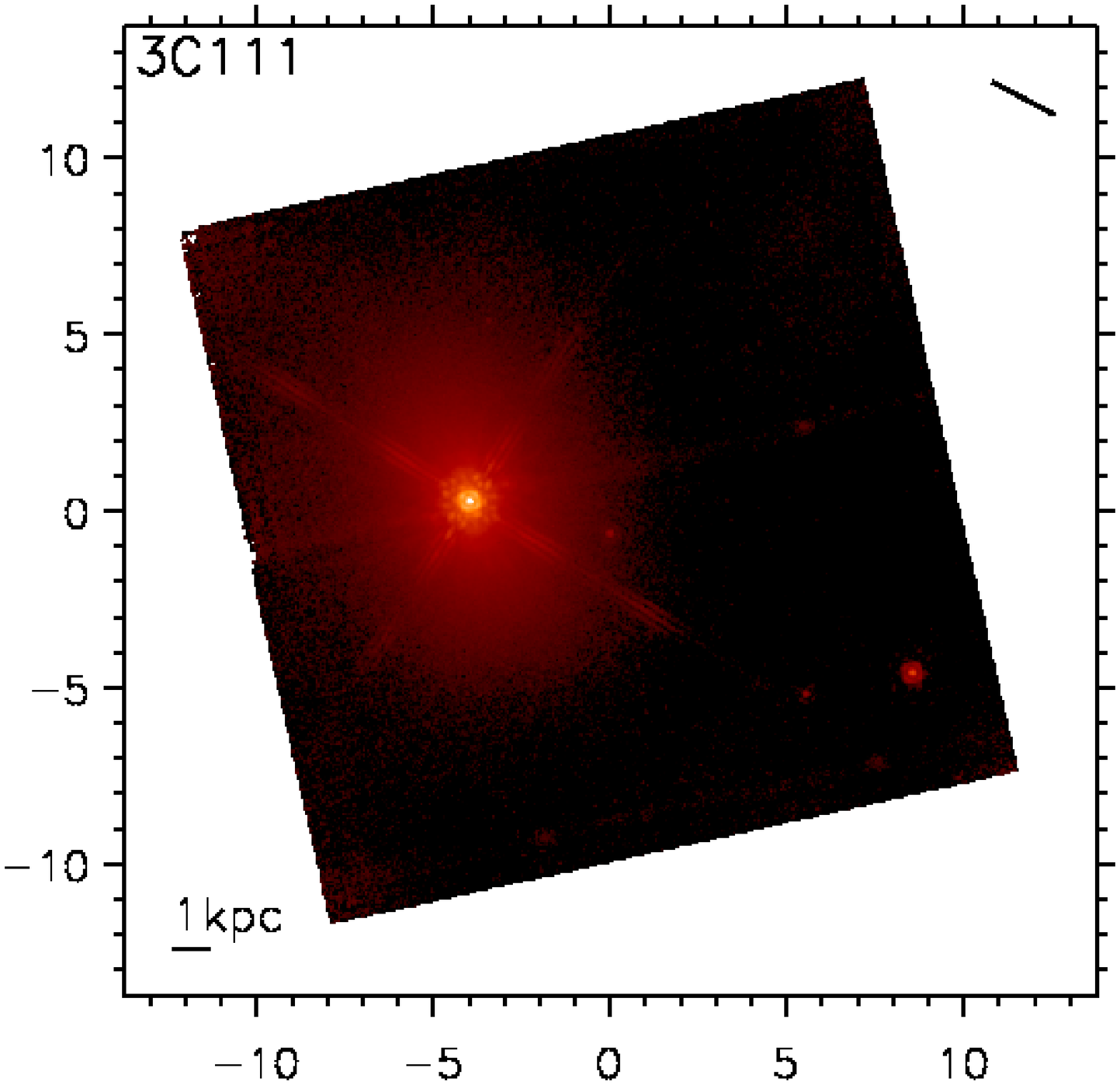}
\caption{HST/NICMOS F160W image of 3C111}
\end{figure}


\begin{figure}
\plotone{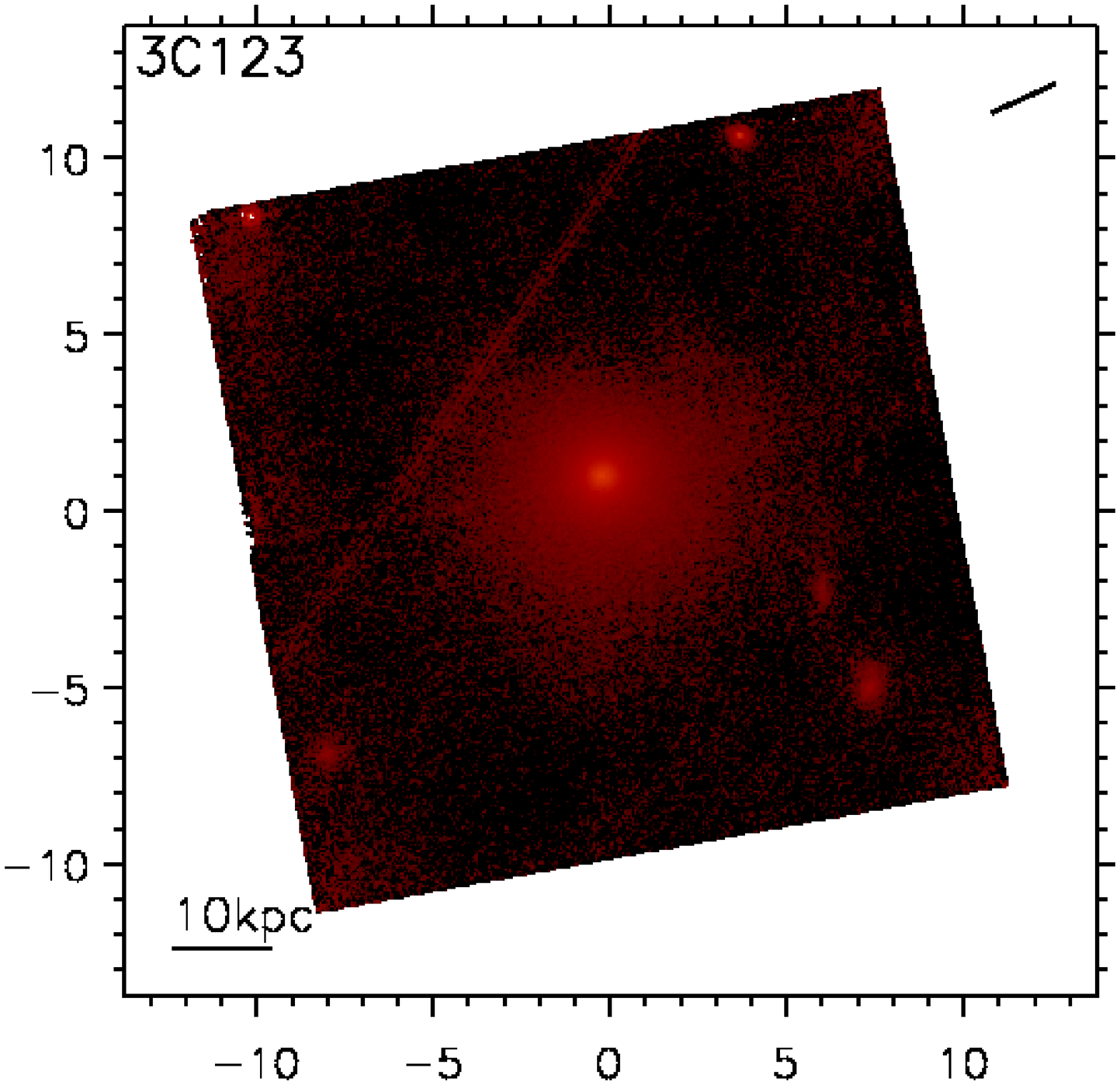}
\caption{HST/NICMOS F160W image of 3C123}
\end{figure}


\begin{figure}
\plotone{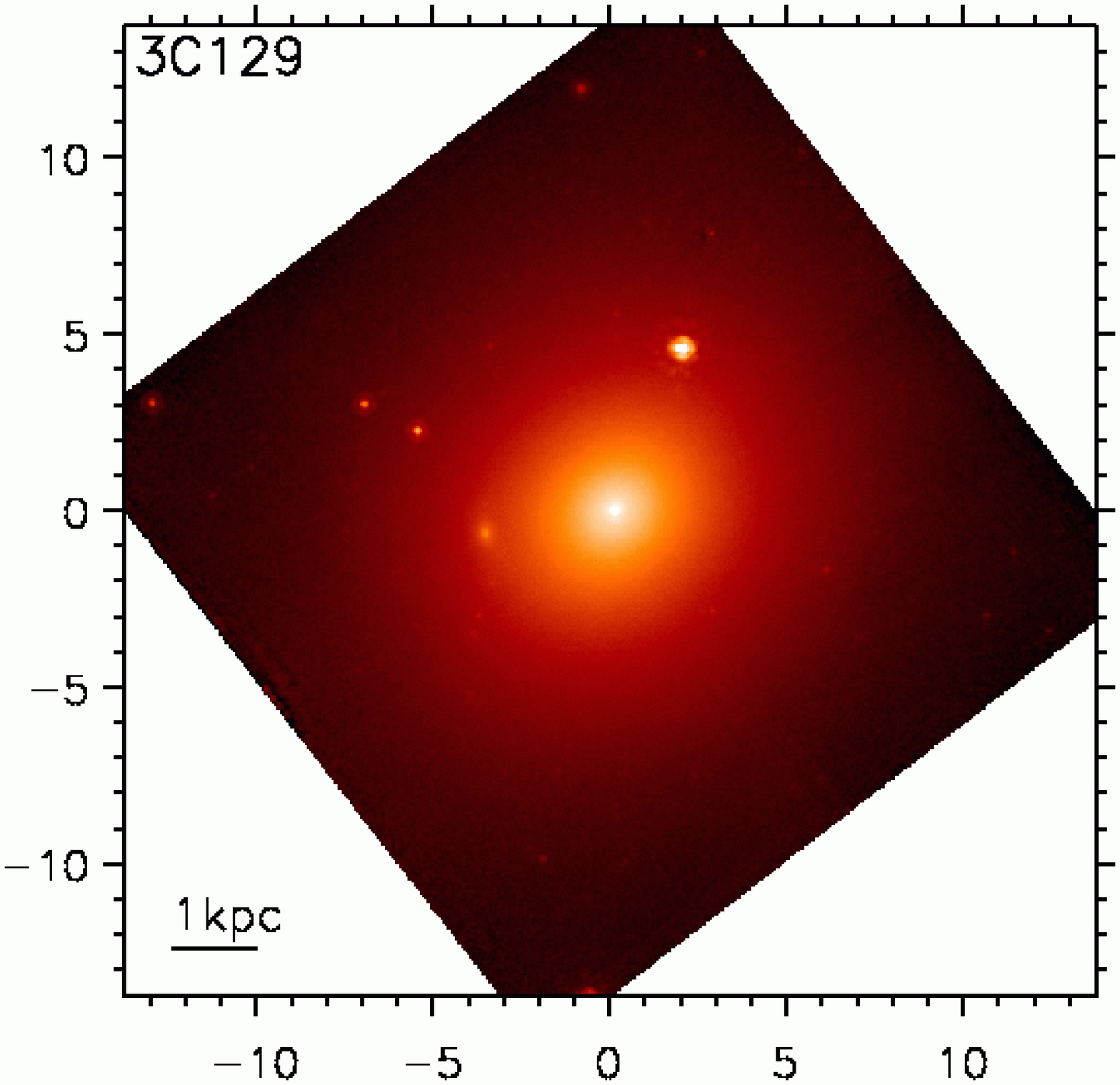}
\caption{HST/NICMOS F160W image of 3C129}
\end{figure}


\begin{figure}
\plotone{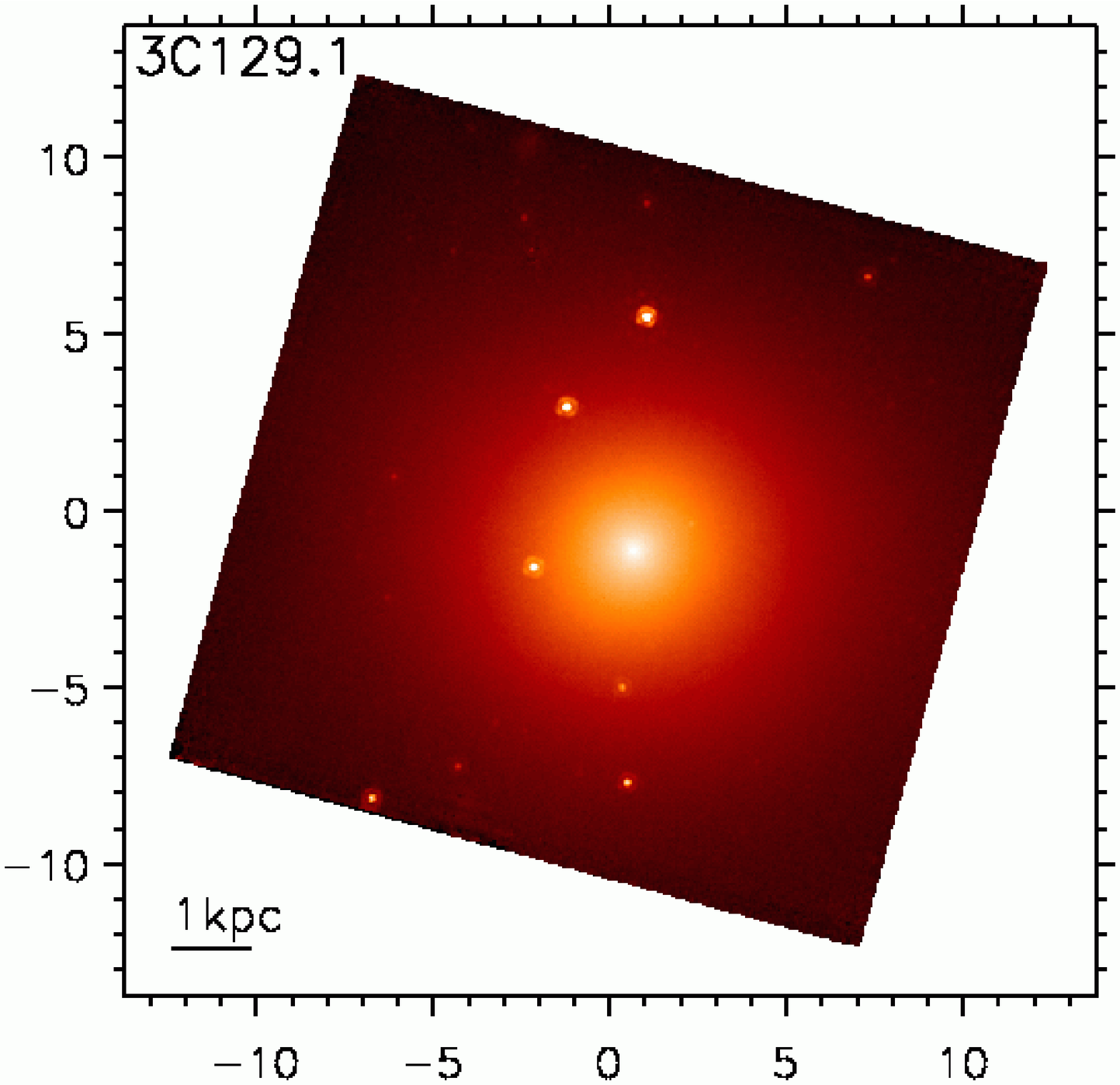}
\caption{HST/NICMOS F160W image of 3C129.1}
\end{figure}
 
\clearpage


\begin{figure}
\plotone{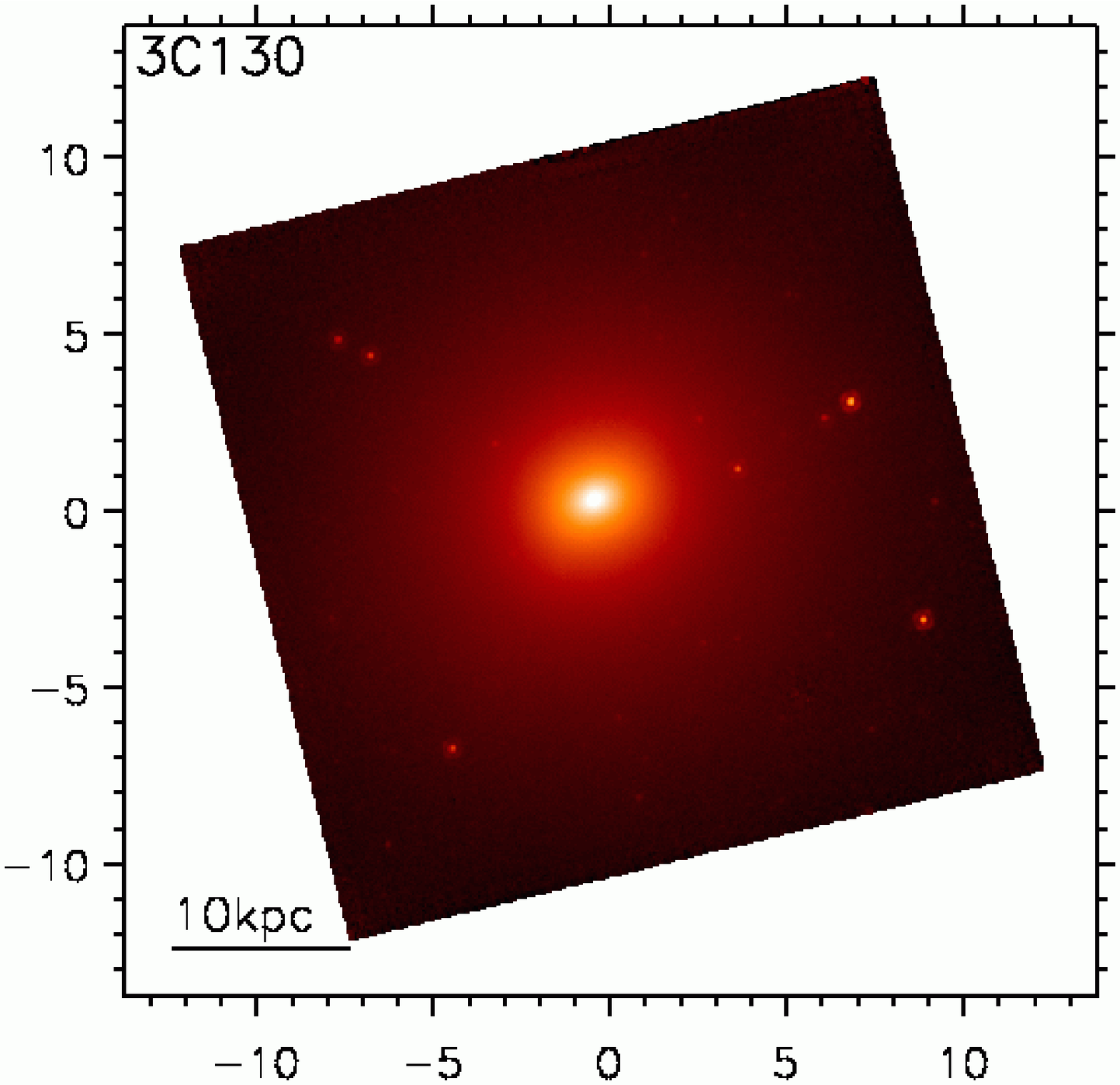}
\caption{HST/NICMOS F160W image of 3C130}
\end{figure}


\begin{figure}
\plotone{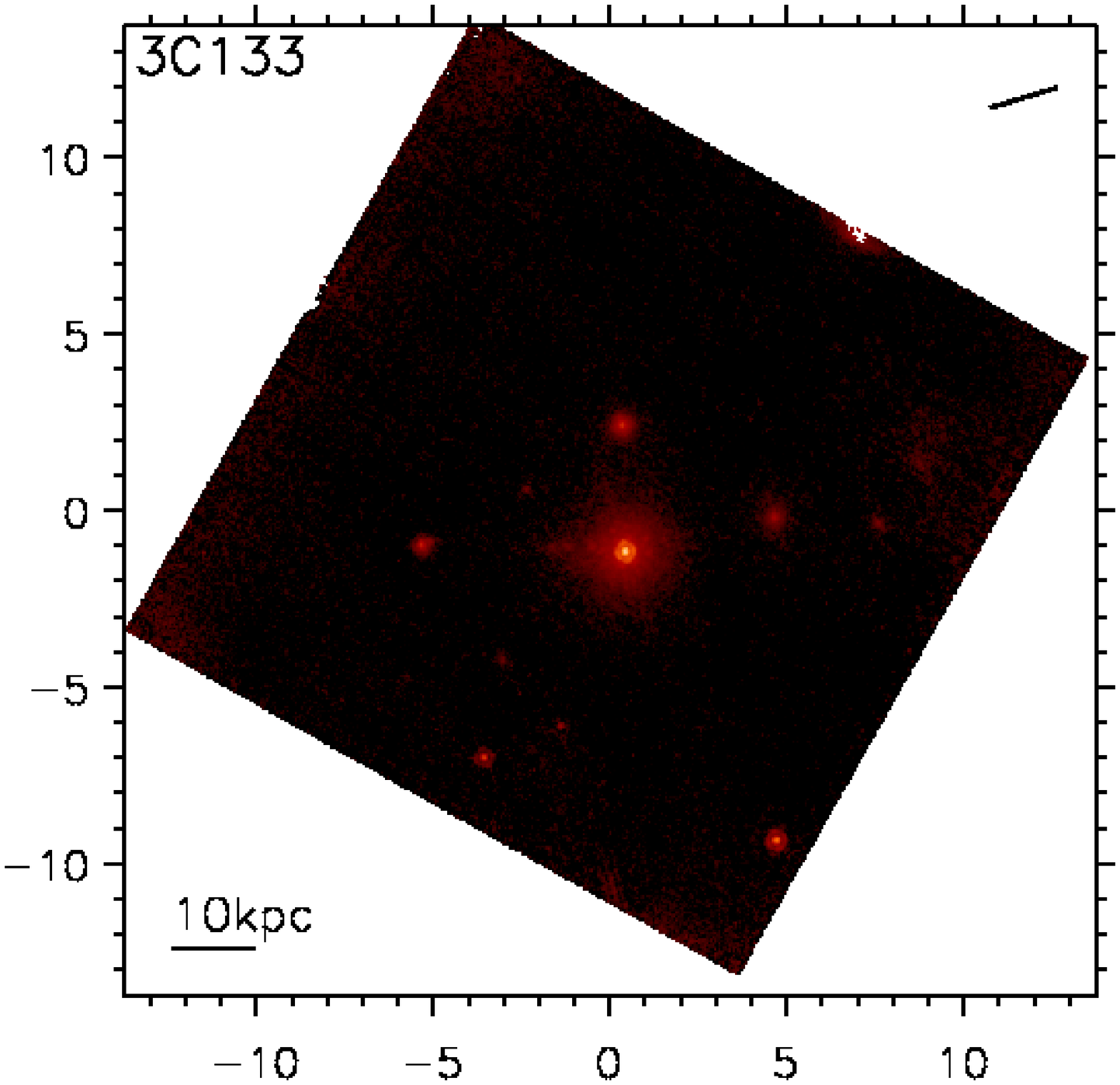}
\caption{HST/NICMOS F160W image of 3C133}
\end{figure}


\begin{figure}
\plotone{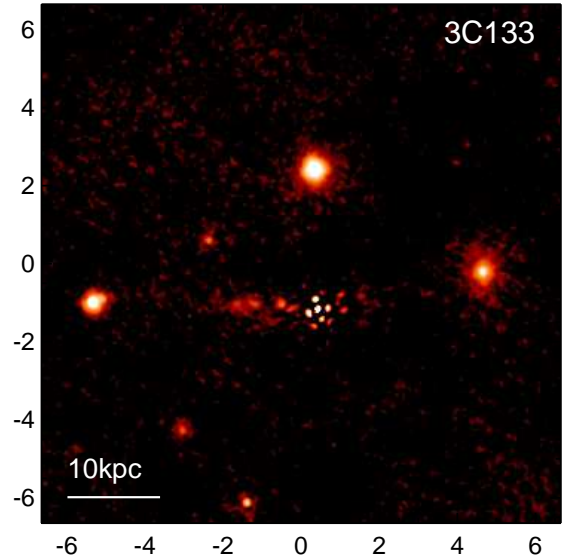}
\newline
\newline
\caption{Model-subtracted residual for 3C133}
\end{figure}

\begin{figure}
\plotone{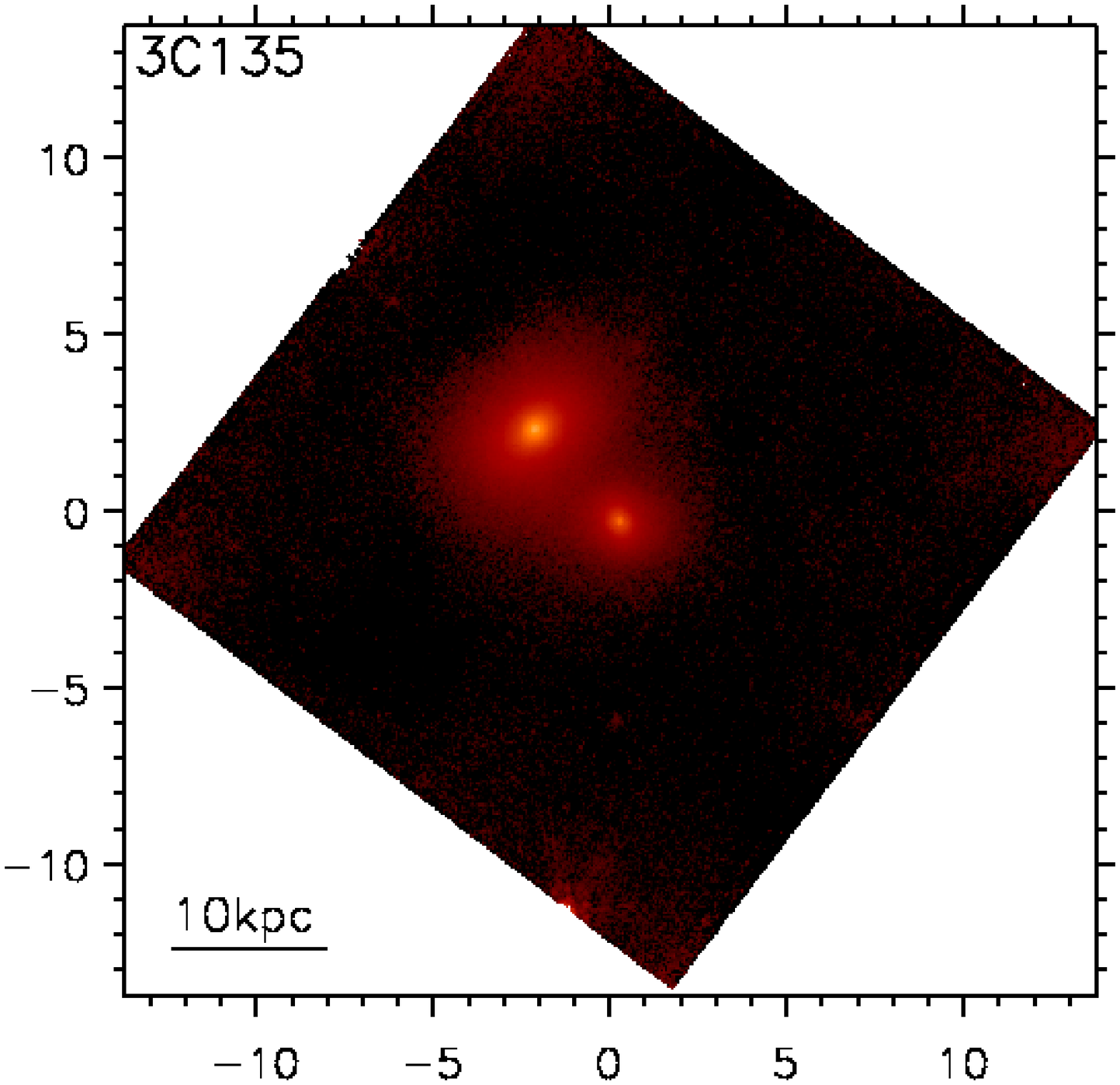}
\caption{HST/NICMOS F160W image of 3C135}
\end{figure}

\clearpage


\begin{figure}
\plotone{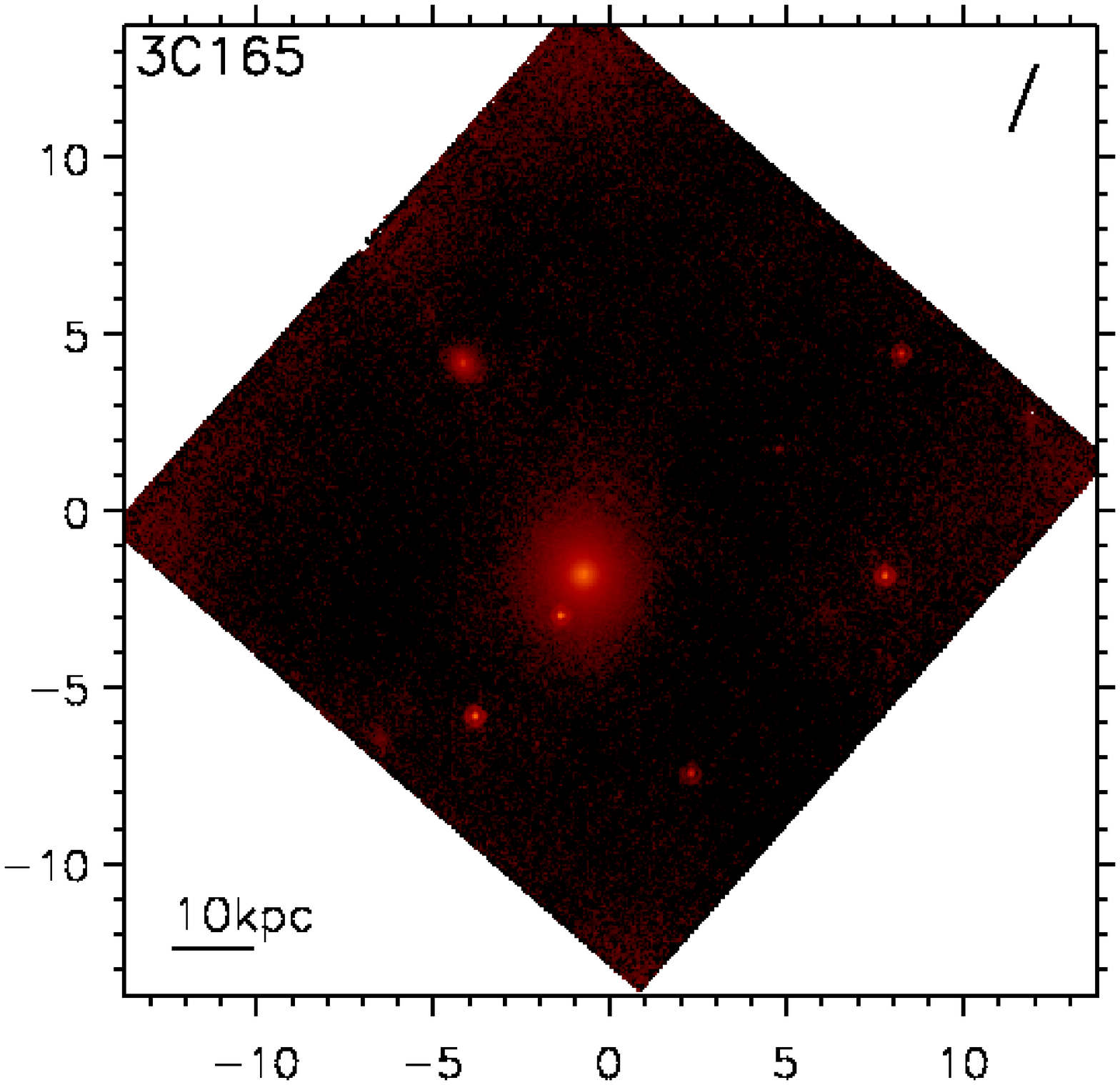}
\caption{HST/NICMOS F160W image of 3C165}
\end{figure}


\begin{figure}
\plotone{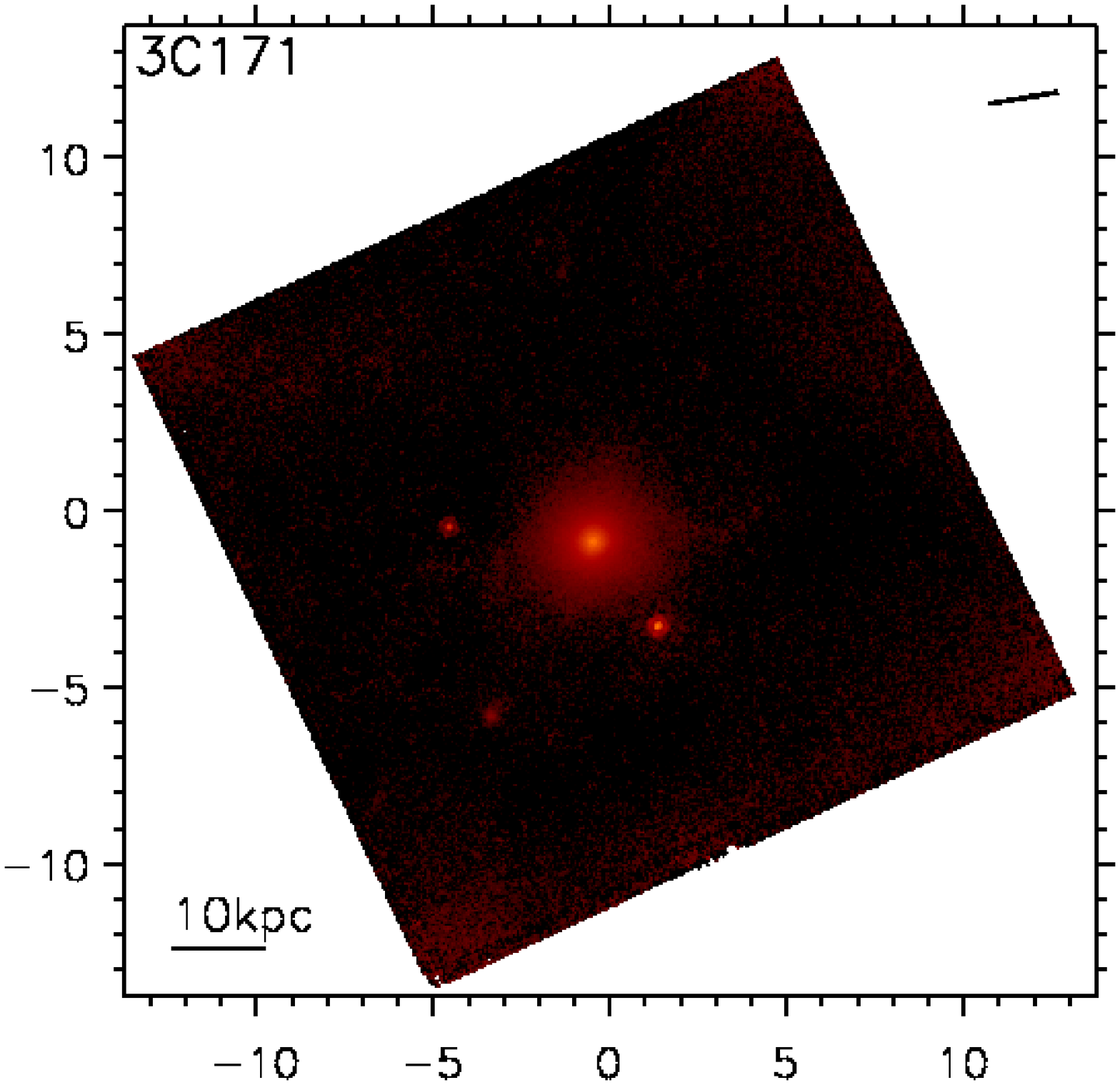}
\caption{HST/NICMOS F160W image of 3C171}
\end{figure}


\begin{figure}
\plotone{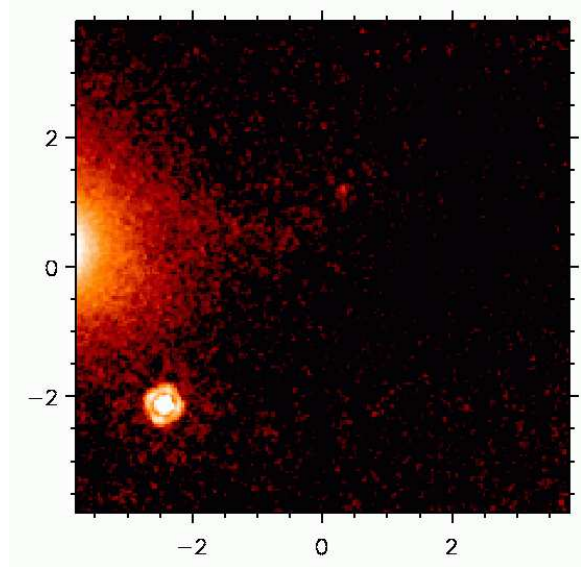}
\caption{Zoom of a  faint tail of emission to the  west of the nucleus
of 3C171}
\end{figure}


\begin{figure}
\plotone{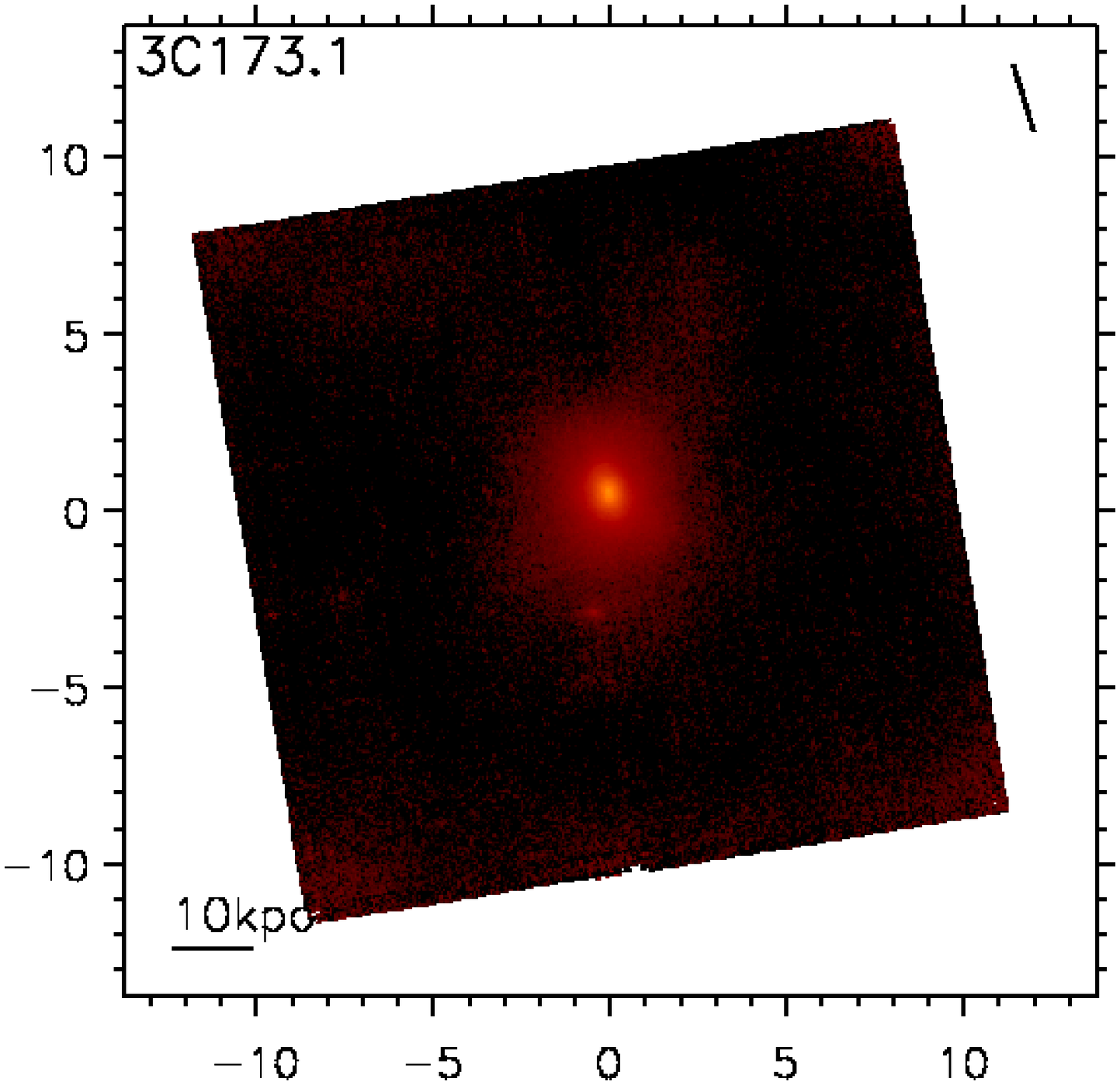}
\caption{HST/NICMOS F160W image of 3C173.1}
\end{figure}

\clearpage


\begin{figure}
\plotone{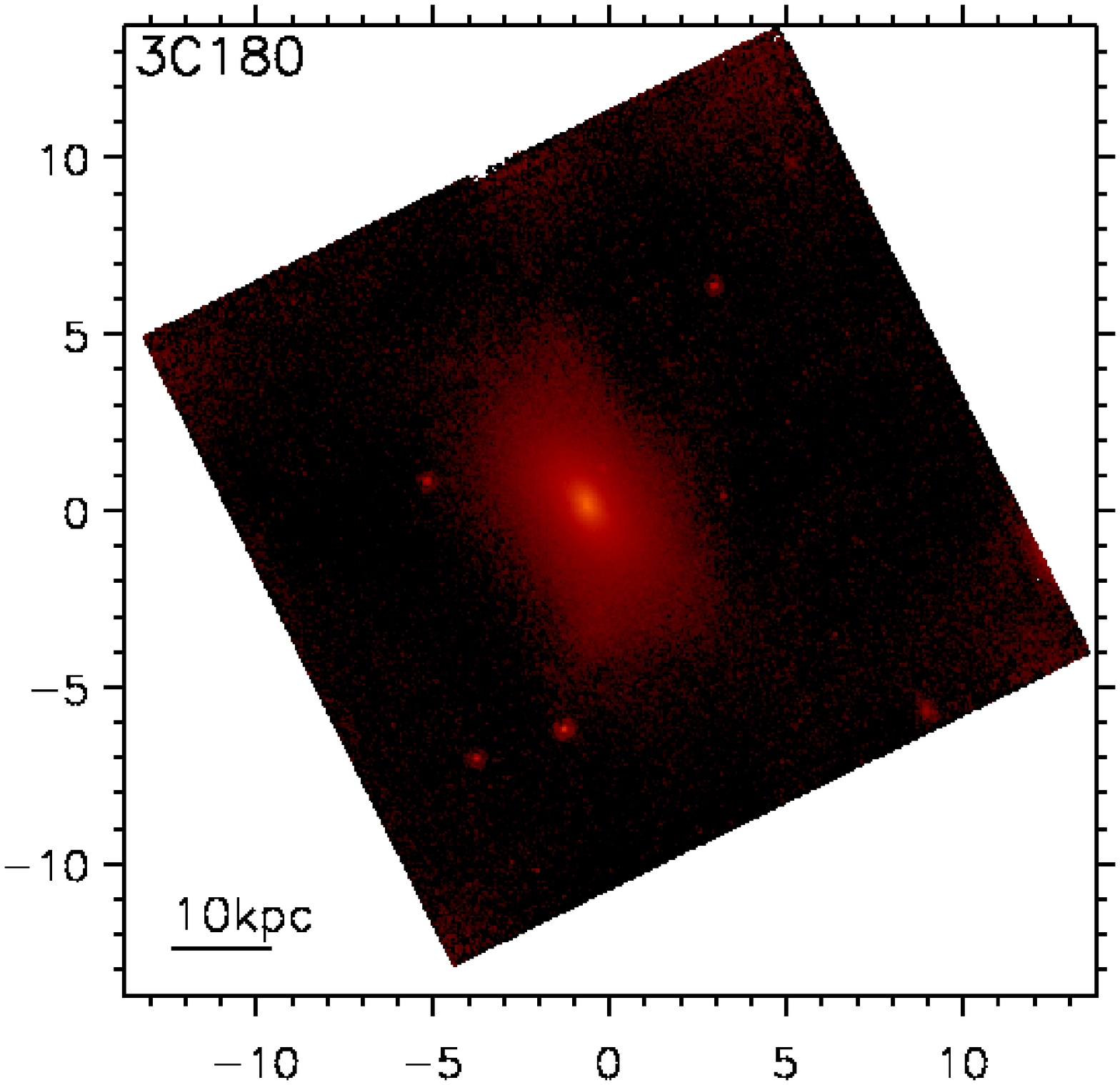}
\caption{HST/NICMOS F160W image of 3C180}
\end{figure}


\begin{figure}
\plotone{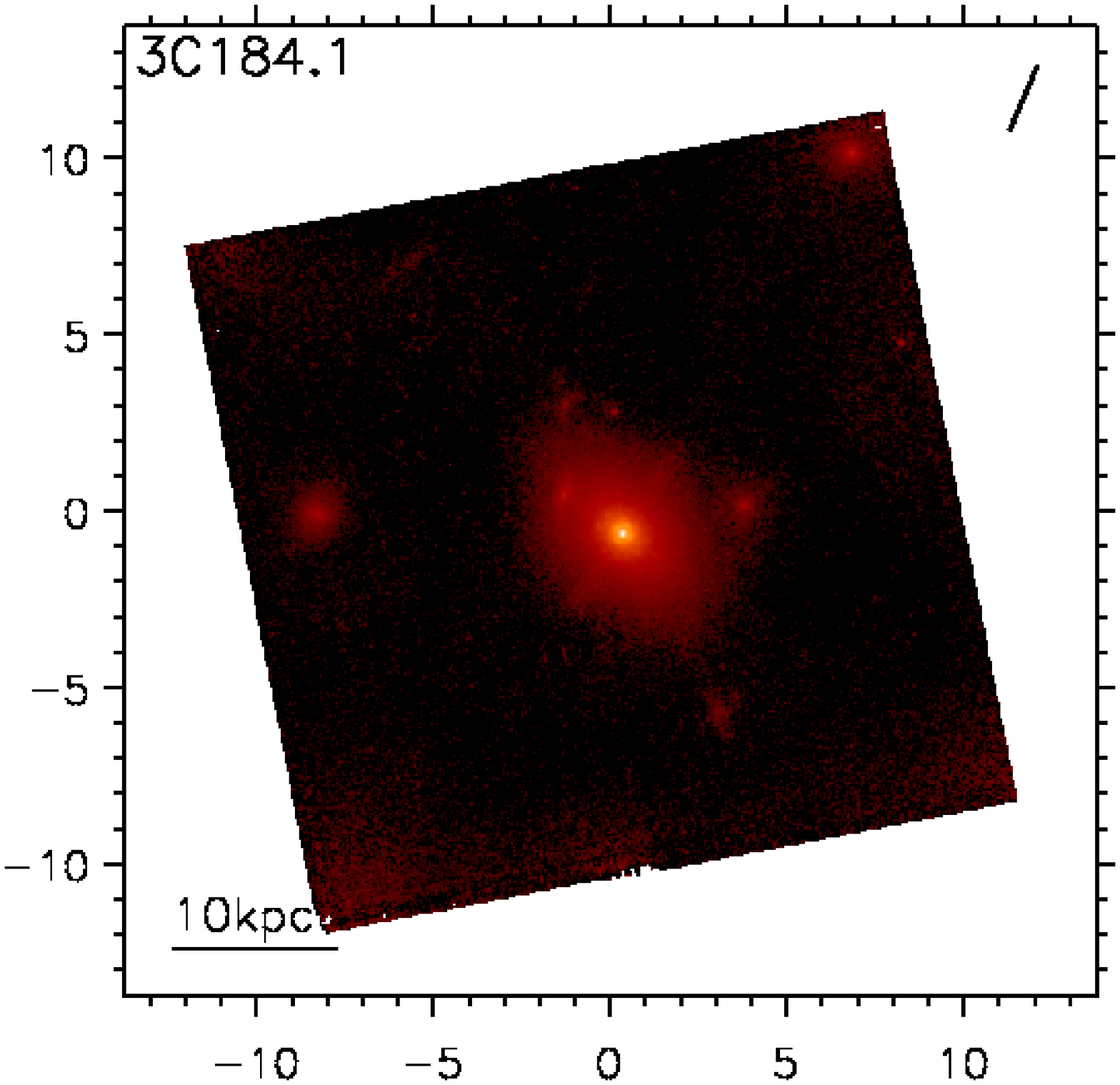}
\caption{HST/NICMOS F160W image of 3C184.1}
\end{figure}


\begin{figure}
\plotone{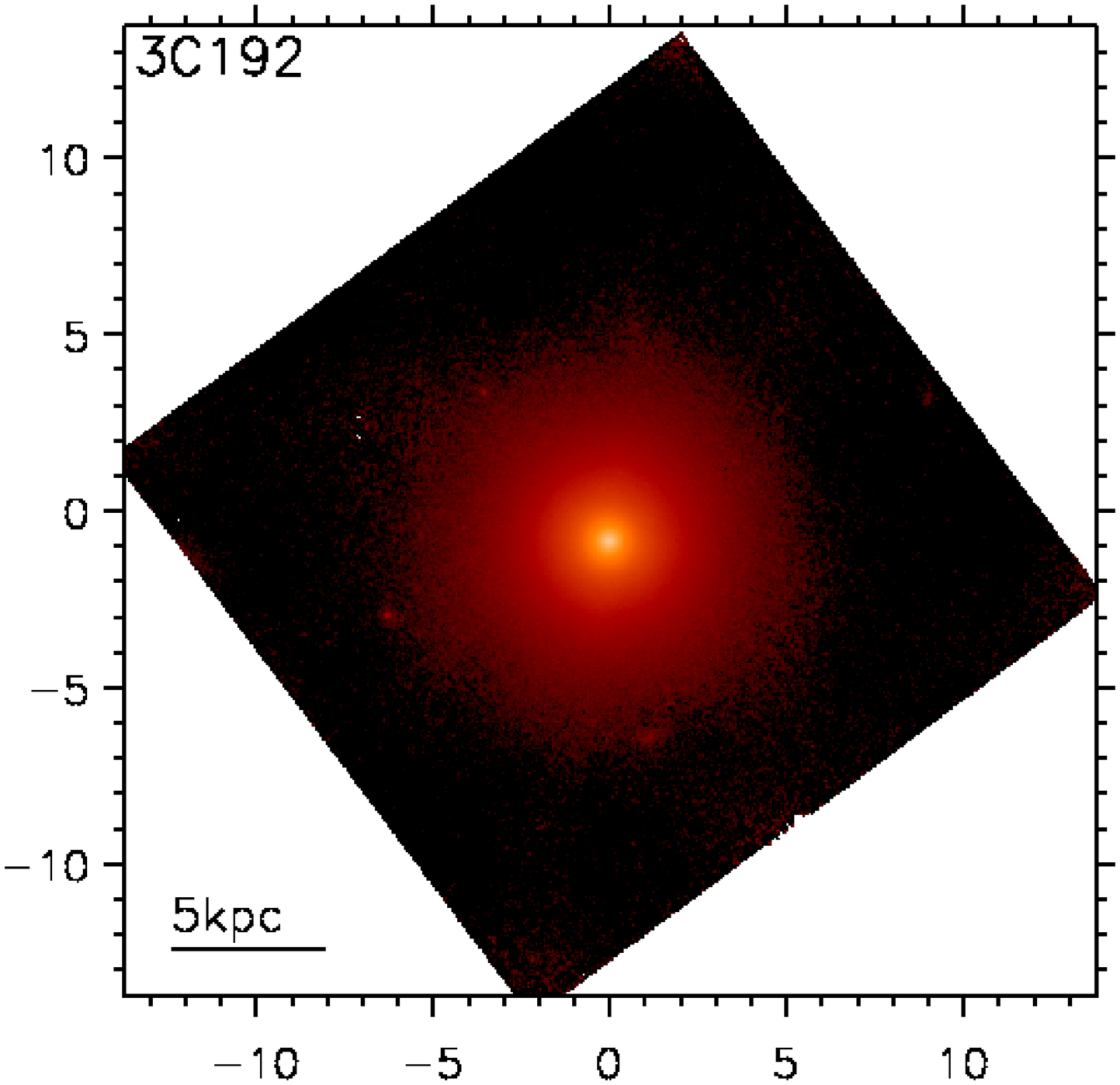}
\caption{HST/NICMOS F160W image of 3C192}
\end{figure}


\begin{figure}
\plotone{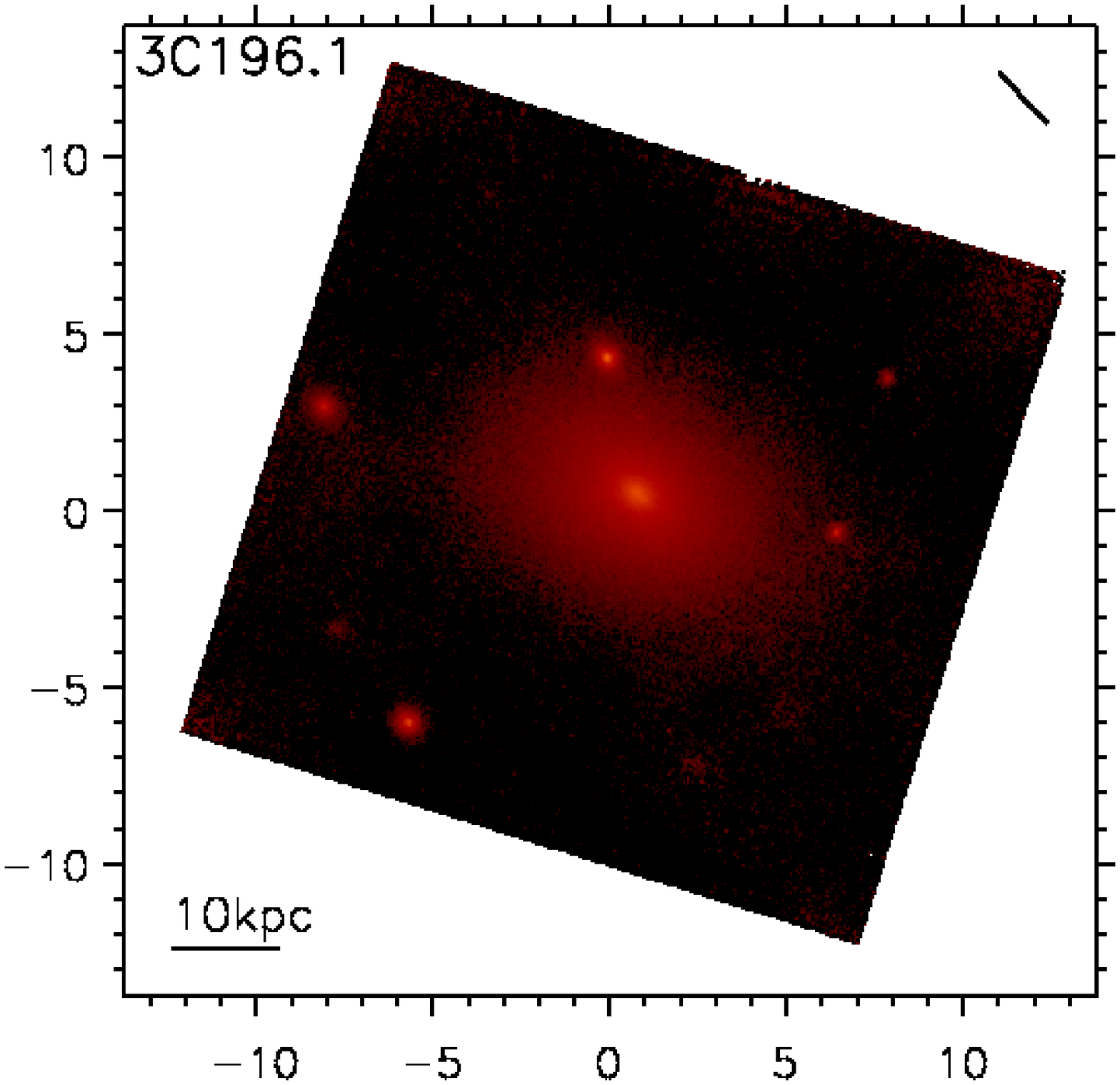}
\caption{HST/NICMOS F160W image of 3C196.1}
\end{figure}

\clearpage


\begin{figure}
\plotone{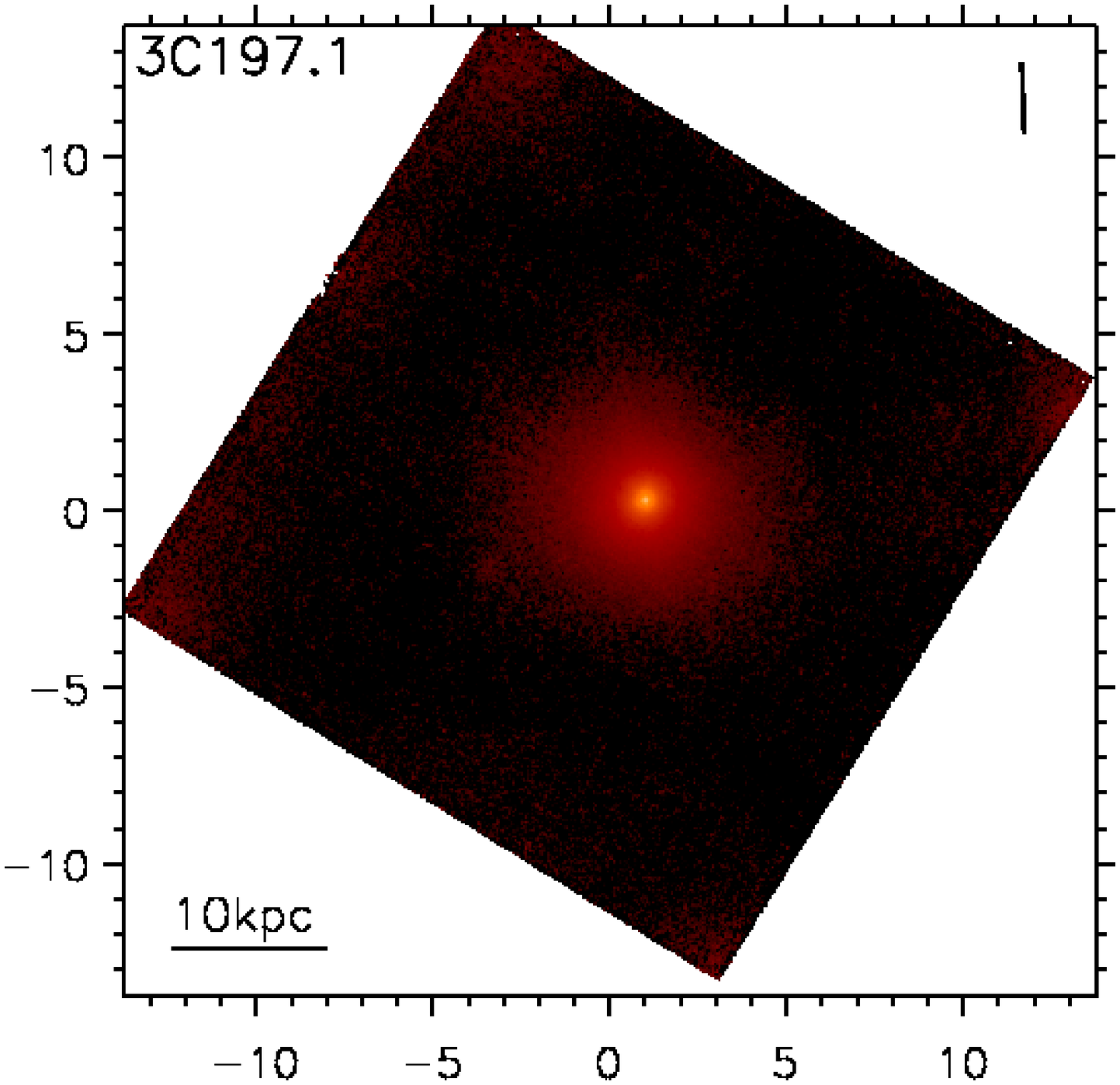}
\caption{HST/NICMOS F160W image of 3C197.1}
\end{figure}


\begin{figure}
\plotone{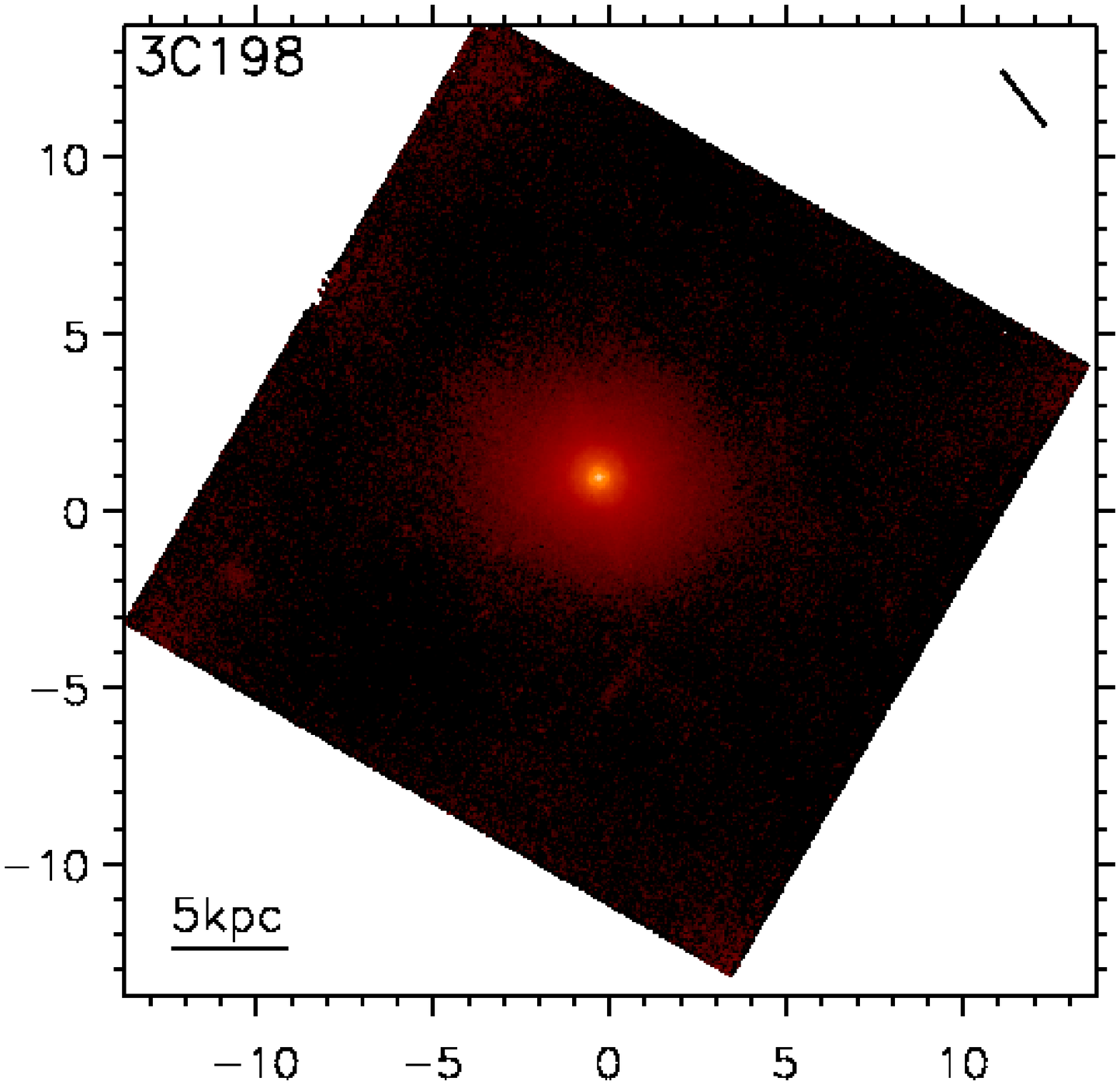}
\caption{HST/NICMOS F160W image of 3C198}
\end{figure}

\begin{figure}
\plotone{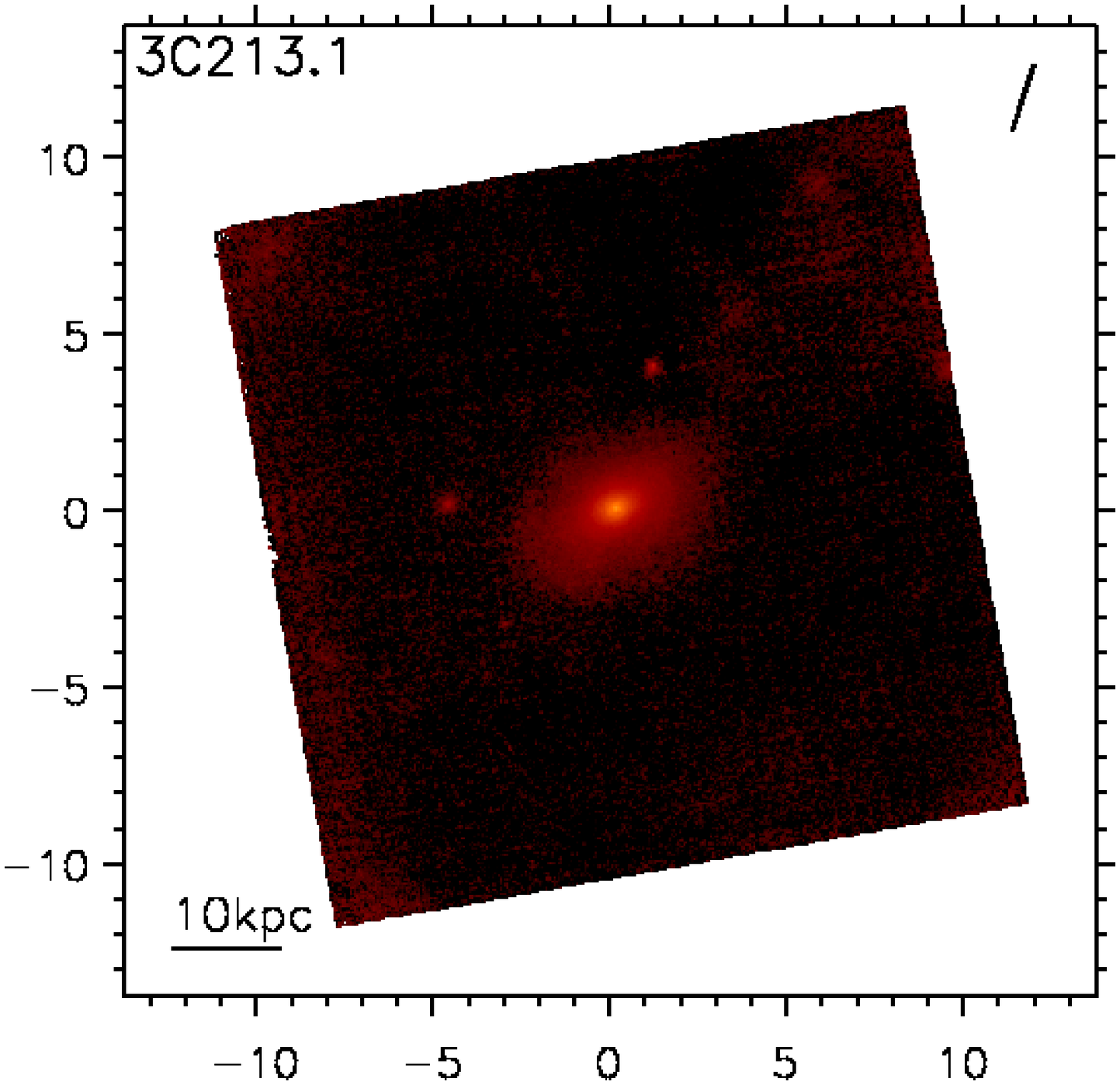}
\caption{HST/NICMOS F160W image of 3C213.1}
\end{figure}


\begin{figure}
\plotone{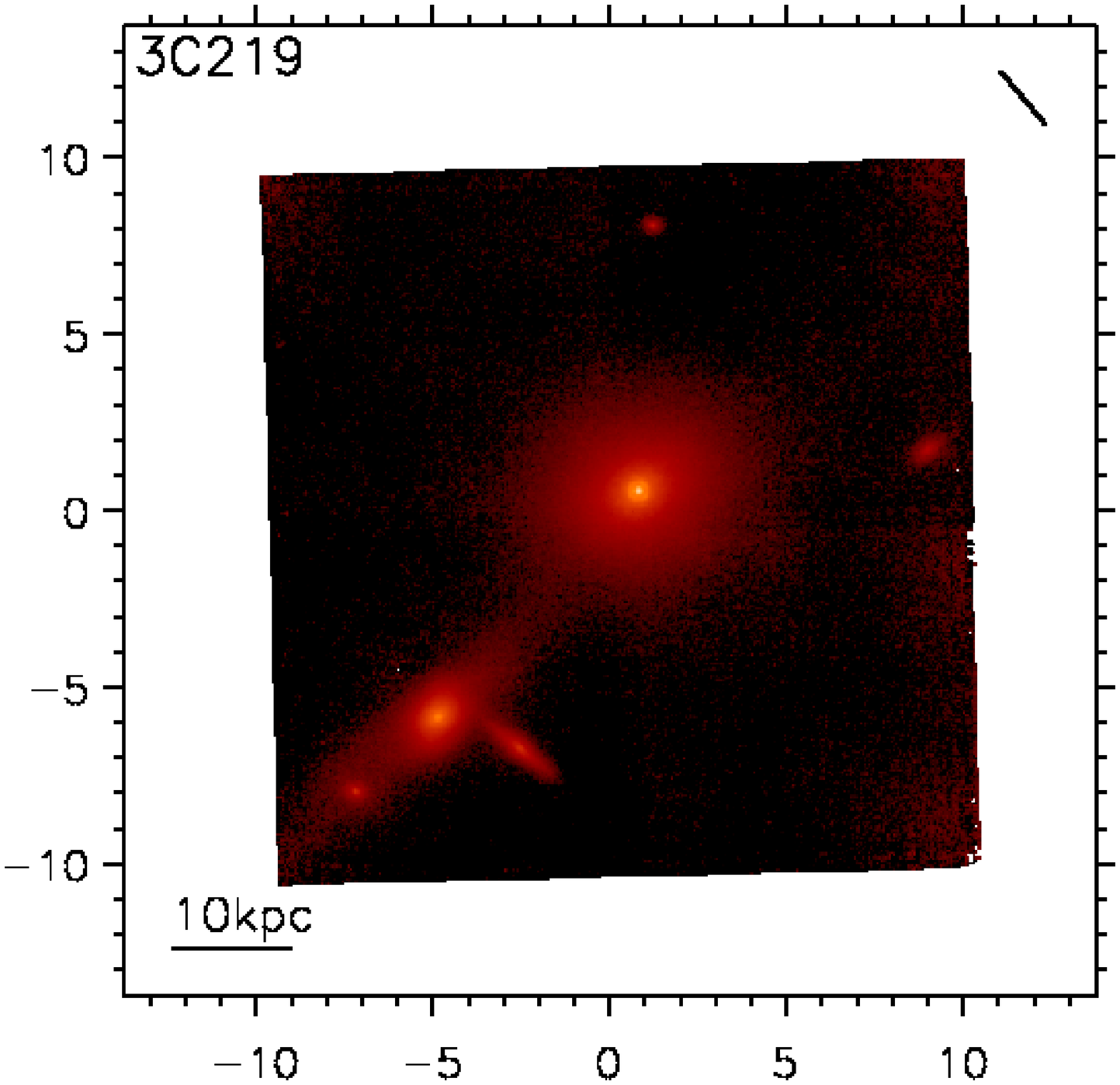}
\caption{HST/NICMOS F160W image of 3C219}
\end{figure}

\clearpage


\begin{figure}
\plotone{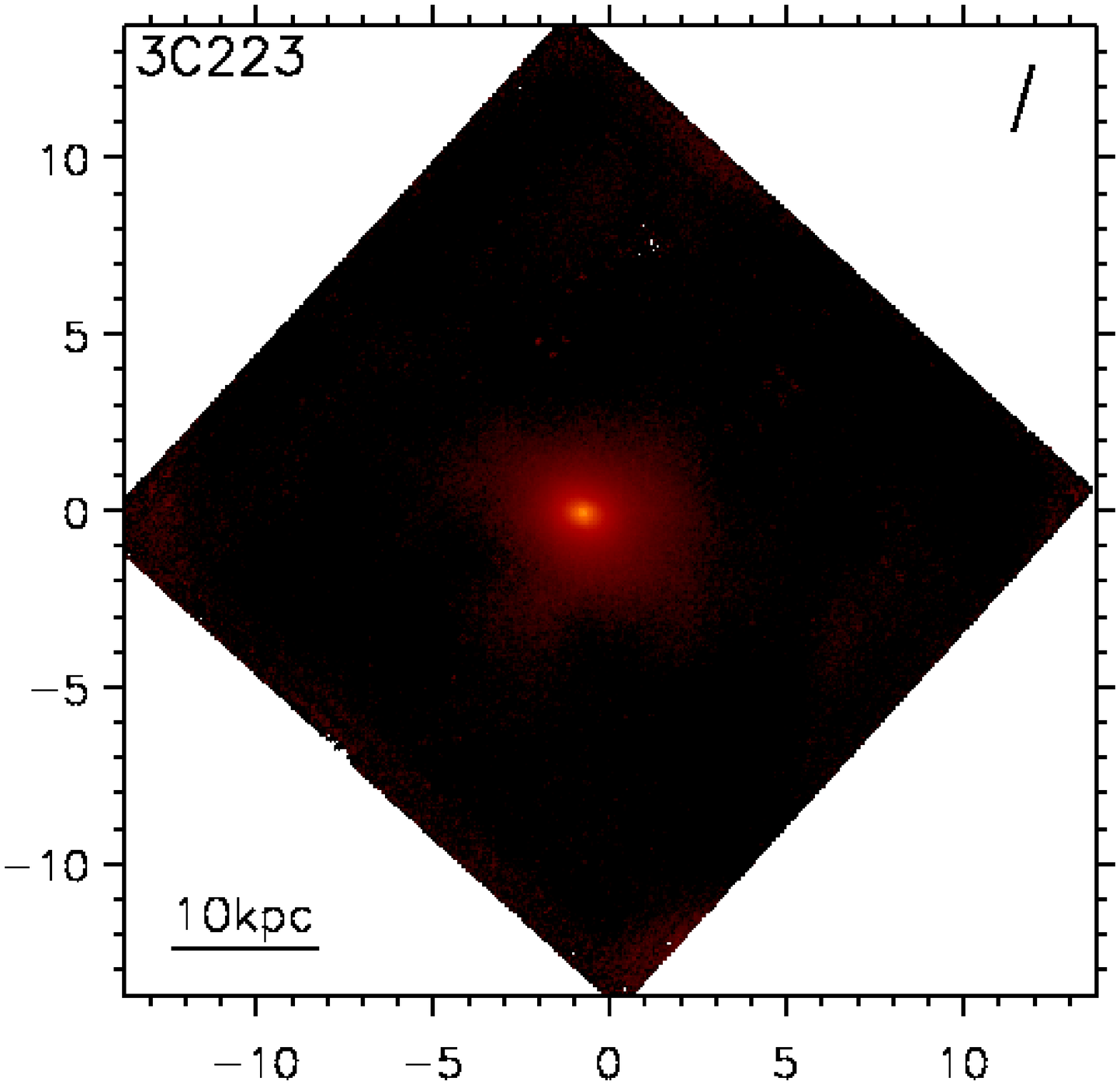}
\caption{HST/NICMOS F160W image of 3C223}
\end{figure}


\begin{figure}
\plotone{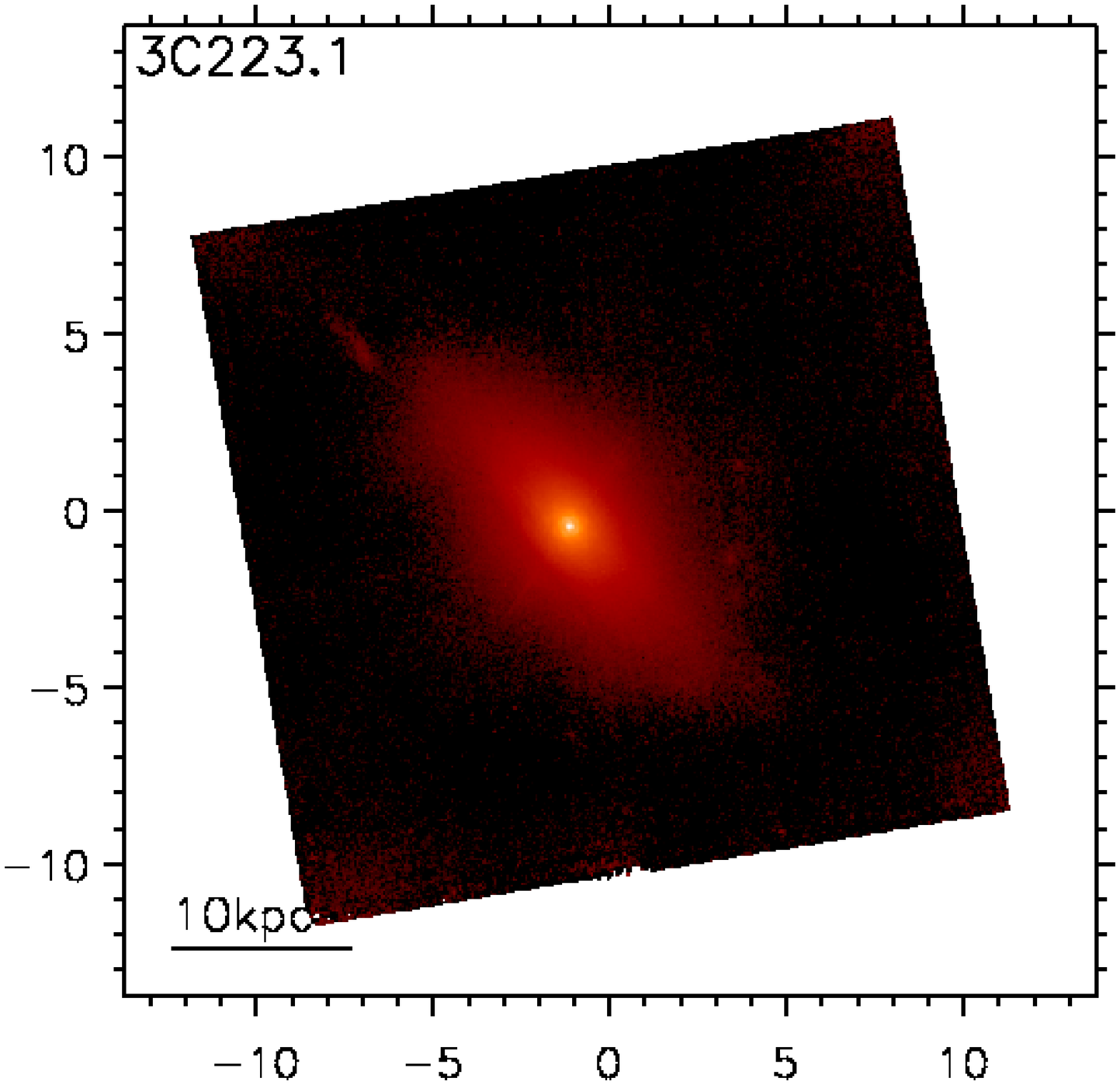}
\caption{HST/NICMOS F160W image of 3C223.1}
\end{figure}


\begin{figure}
\plotone{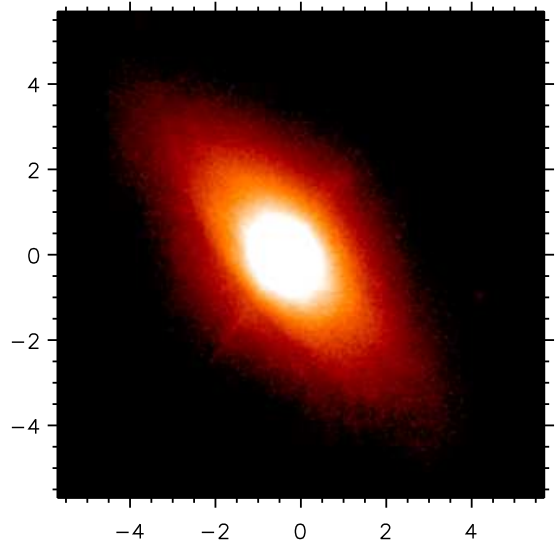}
\caption{Thin dust disk to the southeast of 3C223.1}
\end{figure}


\begin{figure}
\plotone{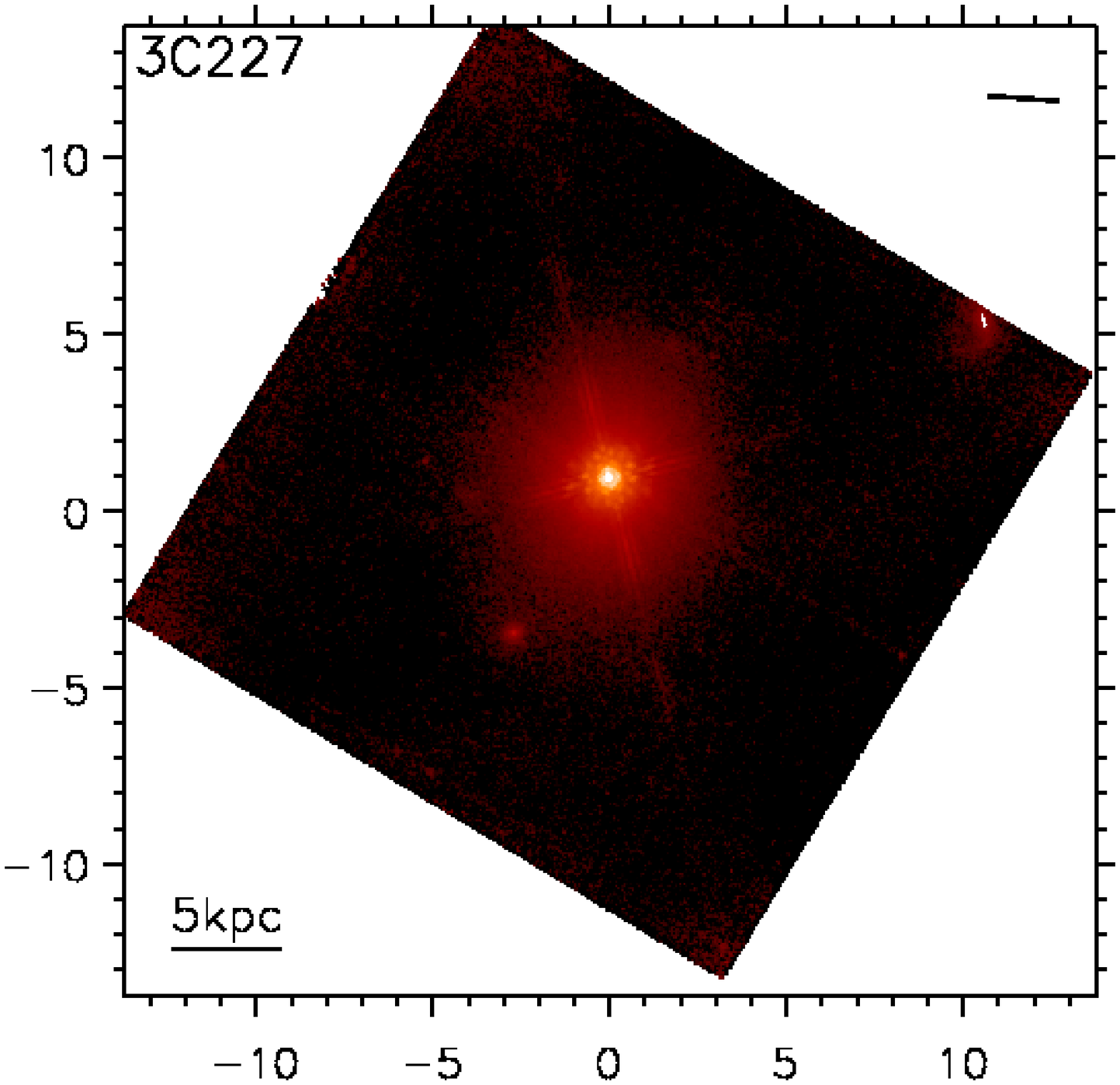}
\caption{HST/NICMOS F160W image of 3C227}
\end{figure}

\clearpage


\begin{figure}
\plotone{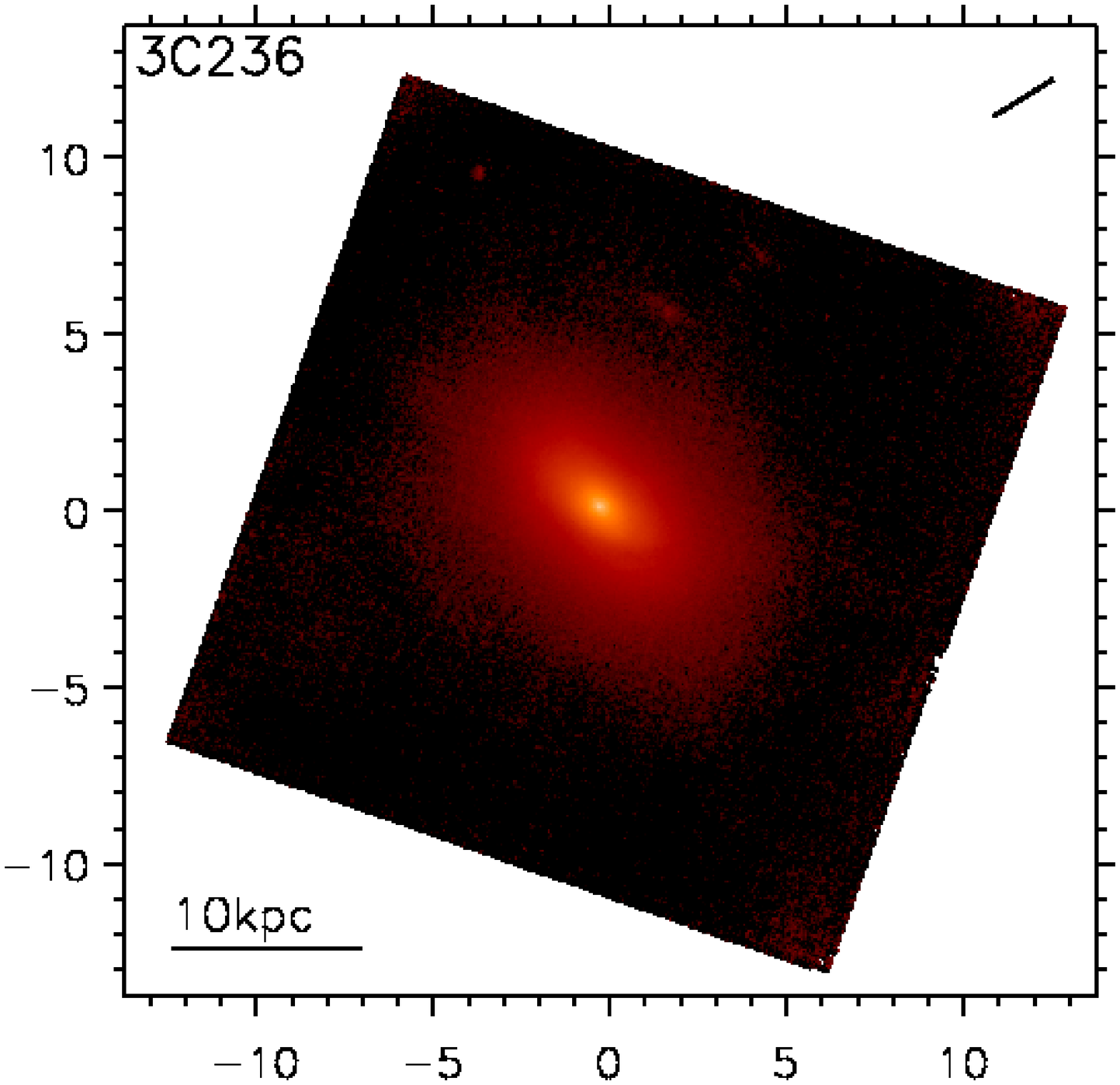}
\caption{HST/NICMOS F160W image of 3C236}
\end{figure}


\begin{figure}
\plotone{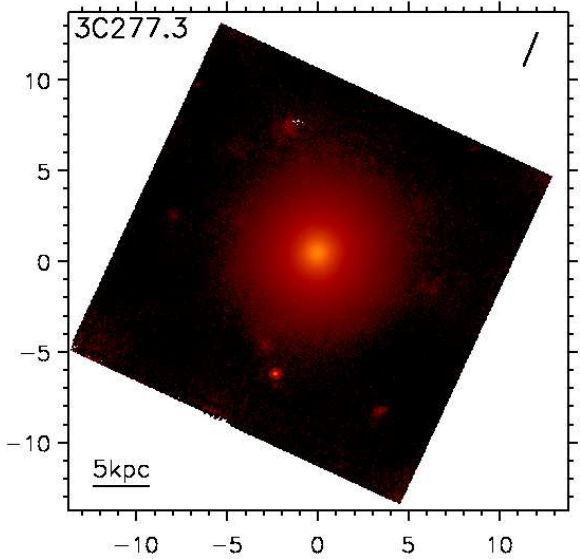}
\caption{HST/NICMOS F160W image of 3C277.3}
\end{figure}


\begin{figure}
\plotone{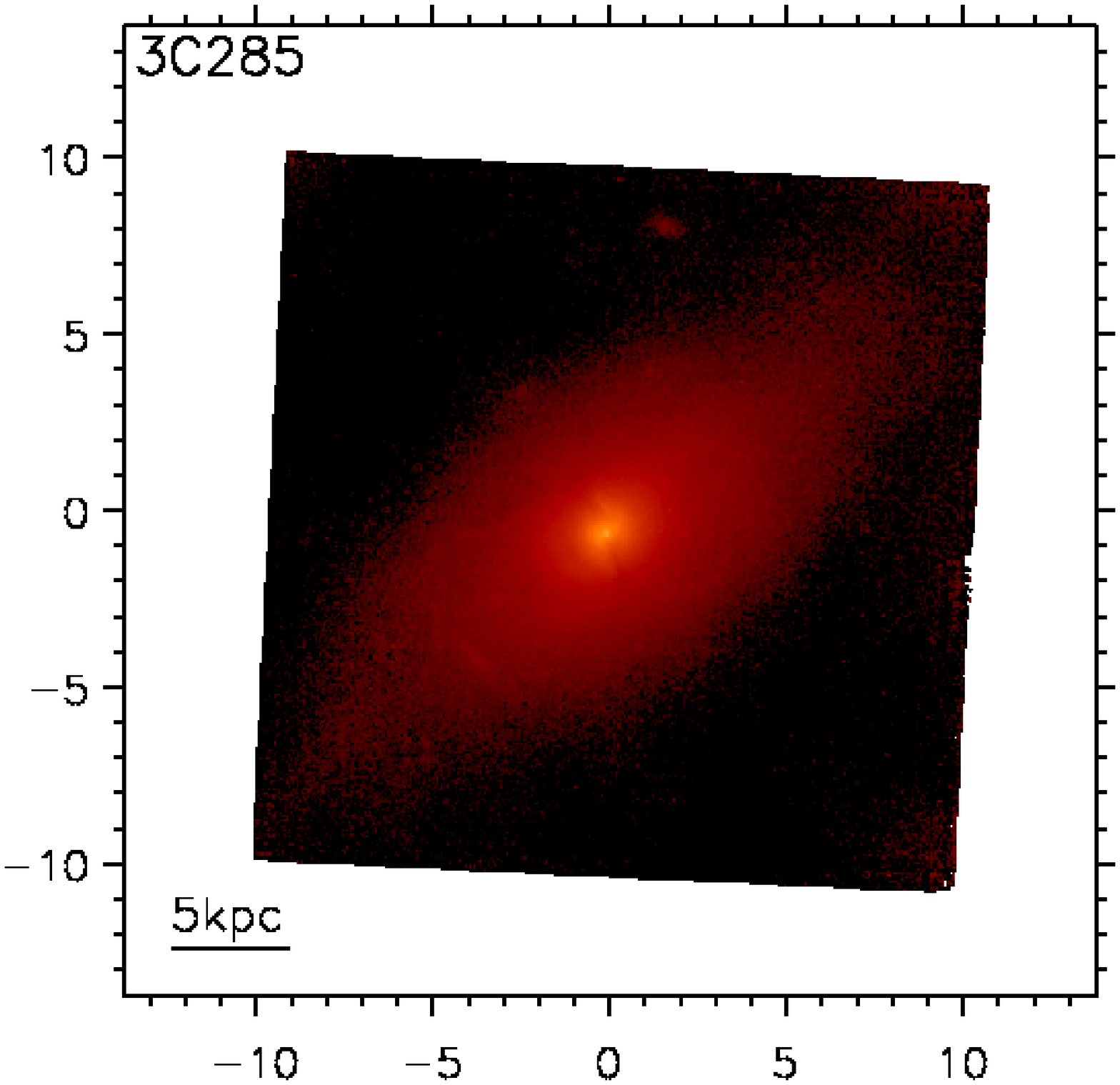}
\caption{HST/NICMOS F160W image of 3C285}
\end{figure}


\begin{figure}
\plotone{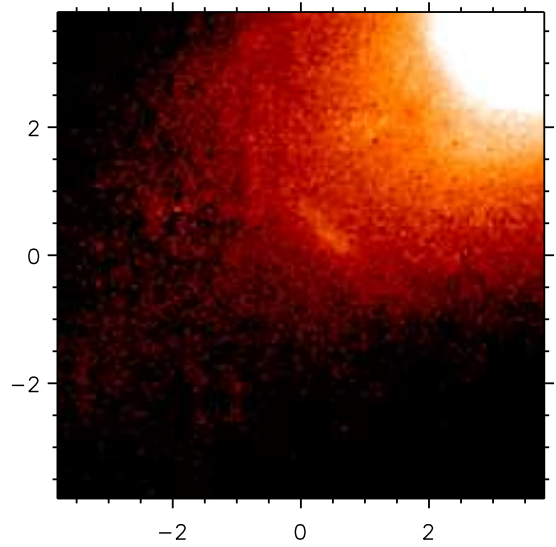}
\caption{Zoom of faint arc-like source southeast of 3C285}
\end{figure}

\clearpage


\begin{figure}
\plotone{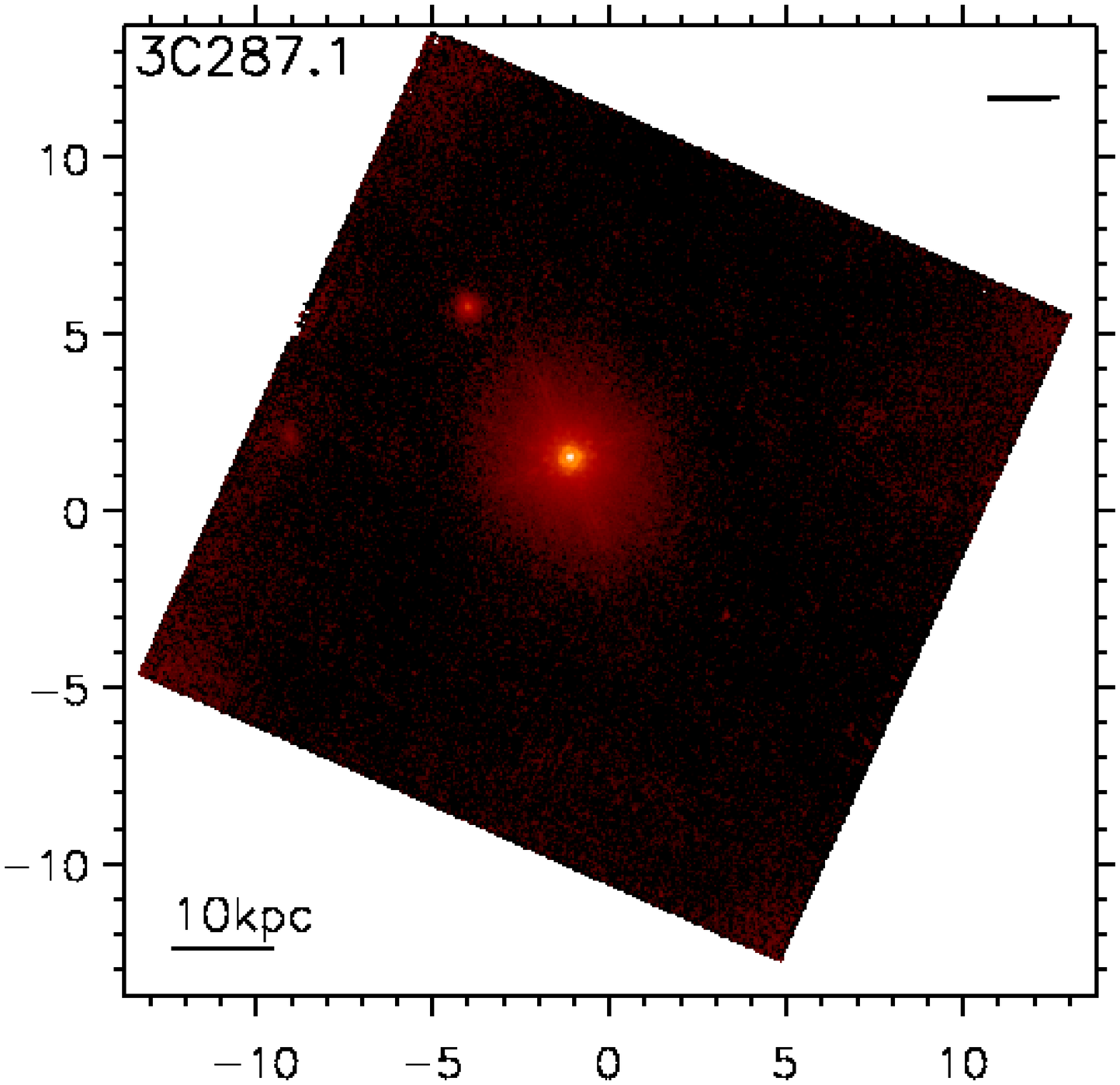}
\caption{HST/NICMOS F160W image of 3C287.1}
\end{figure}


\begin{figure}
\plotone{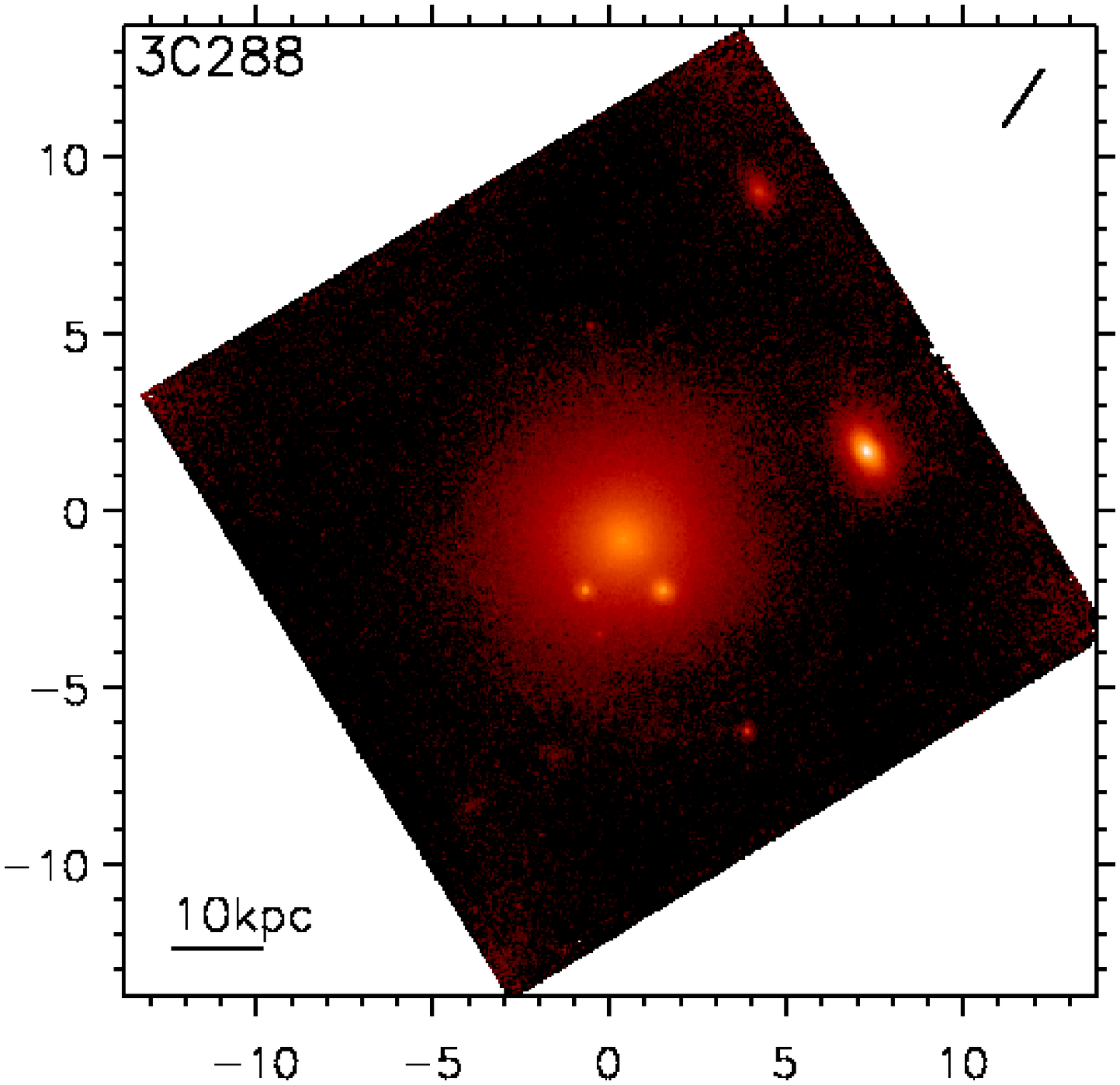}
\caption{HST/NICMOS F160W image of 3C288}
\end{figure}


\begin{figure}
\plotone{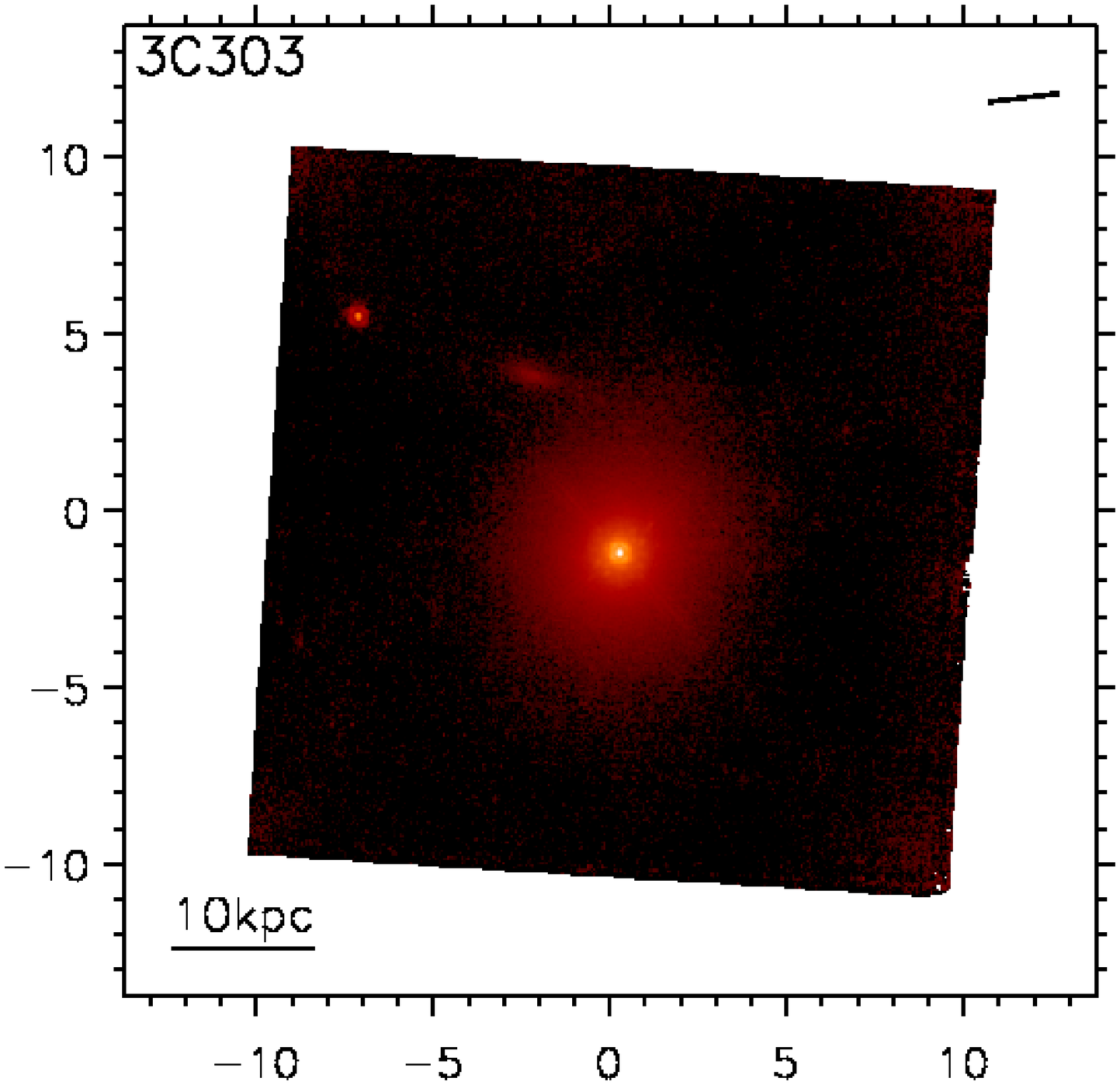}
\caption{HST/NICMOS F160W image of 3C303}
\end{figure}


\begin{figure}
\plotone{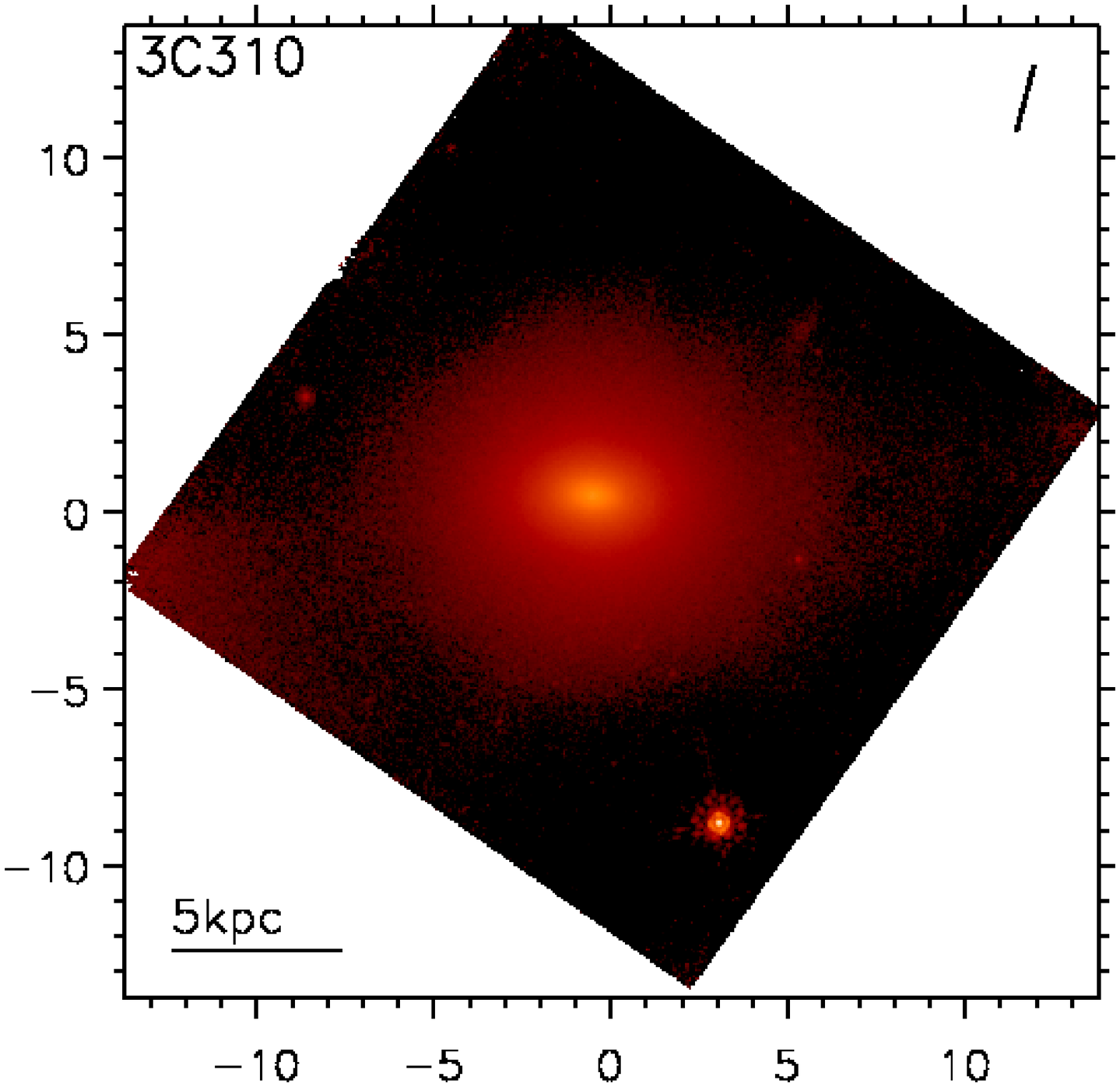}
\caption{HST/NICMOS F160W image of 3C310}
\end{figure}

\clearpage


\begin{figure}
\plotone{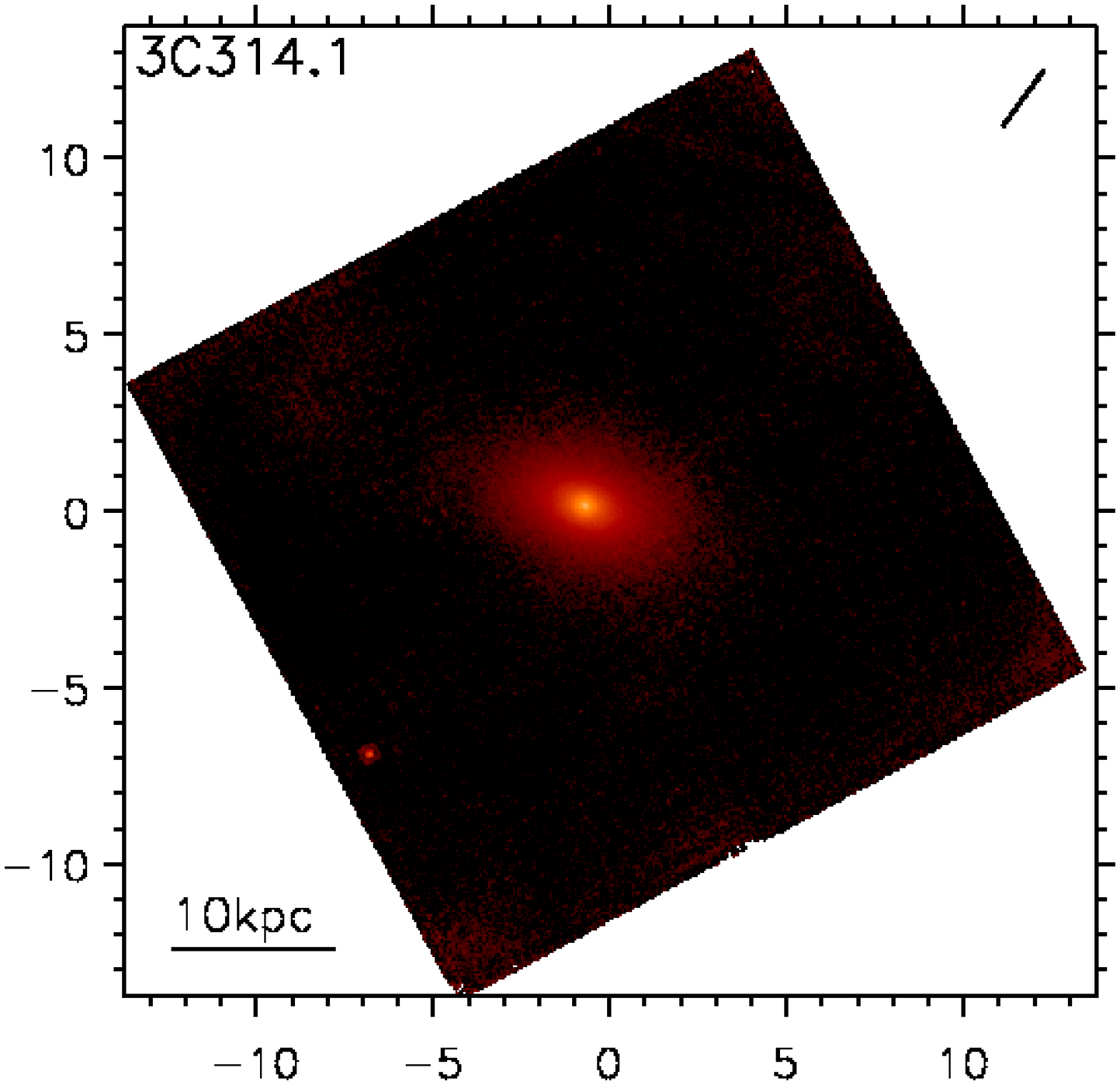}
\caption{HST/NICMOS F160W image of 3C314.1}
\end{figure}


\begin{figure}
\plotone{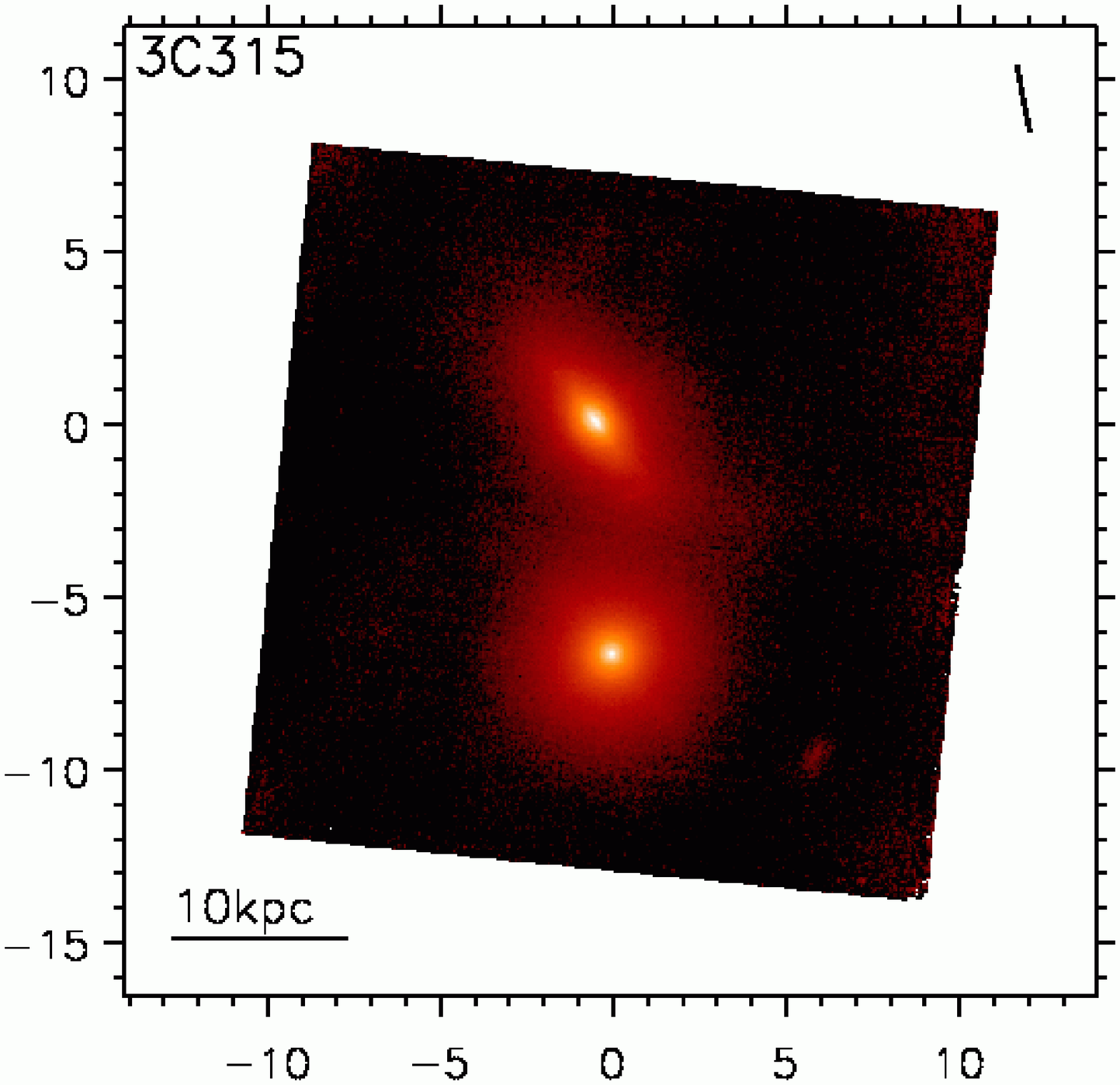}
\caption{HST/NICMOS F160W image of 3C315}
\end{figure}


\begin{figure}
\plotone{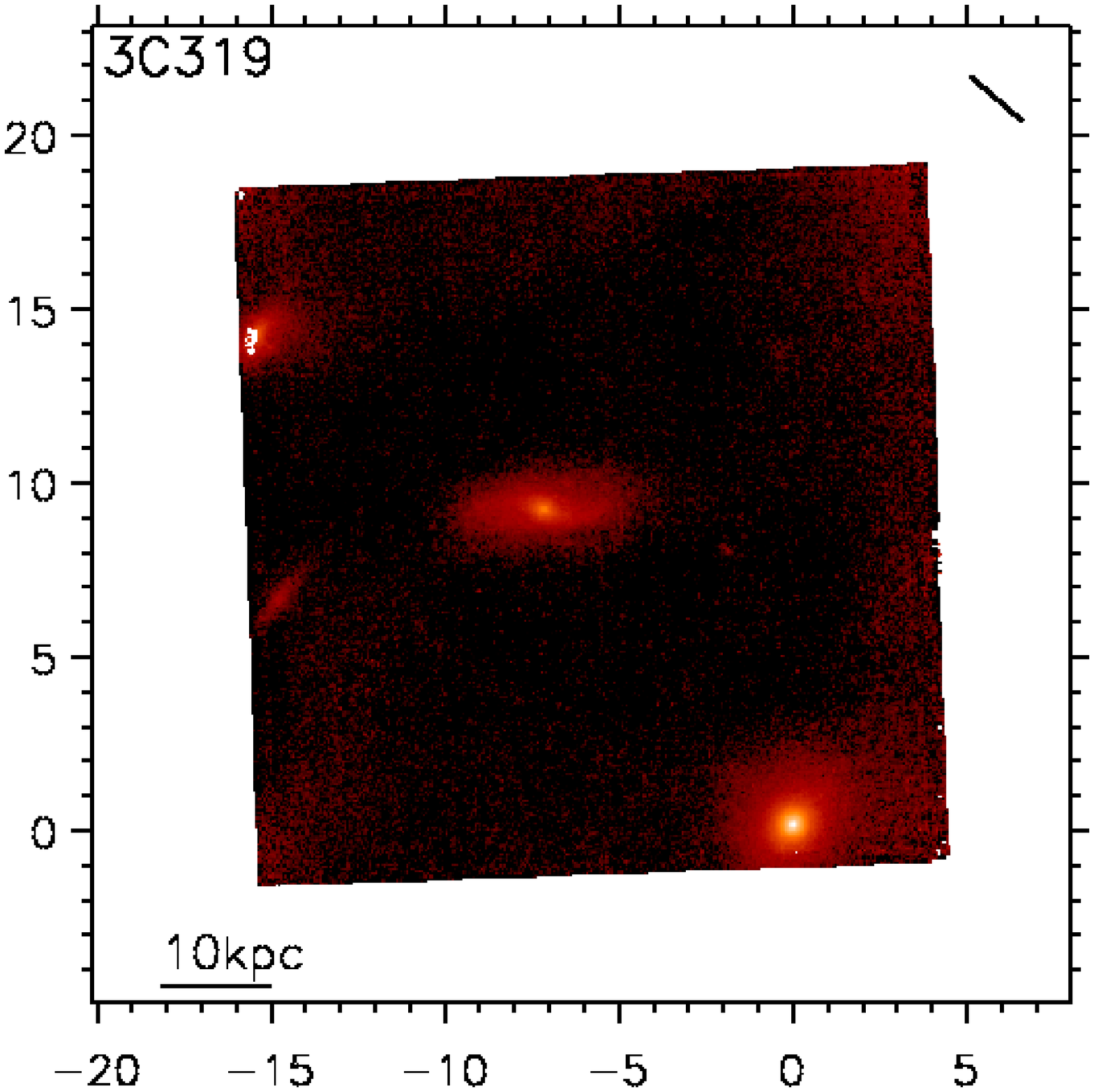}
\caption{HST/NICMOS F160W image of 3C319}
\end{figure}


\begin{figure}
\plotone{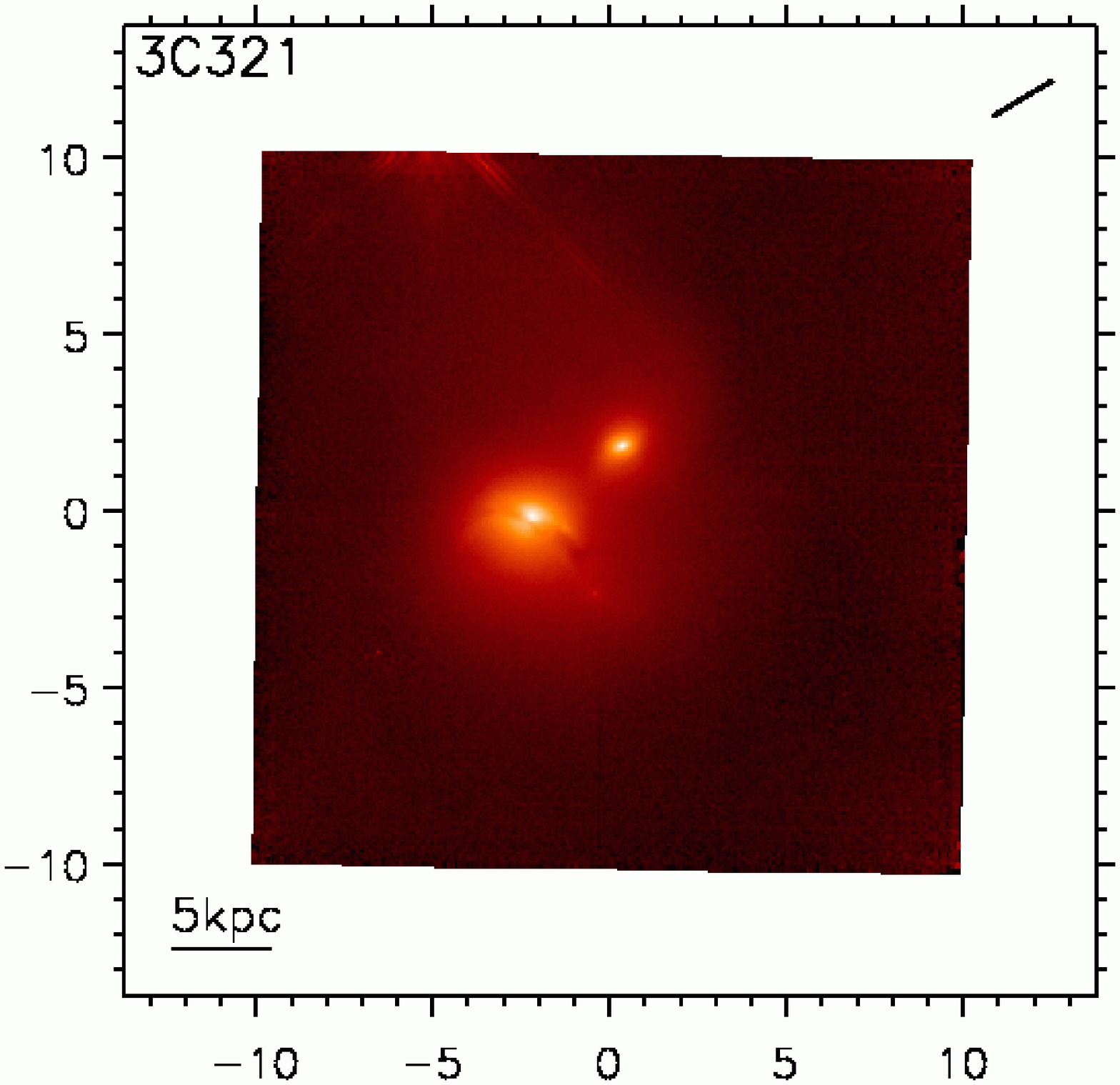}
\caption{HST/NICMOS F160W image of 3C321}
\end{figure}

\clearpage


\begin{figure}
\plotone{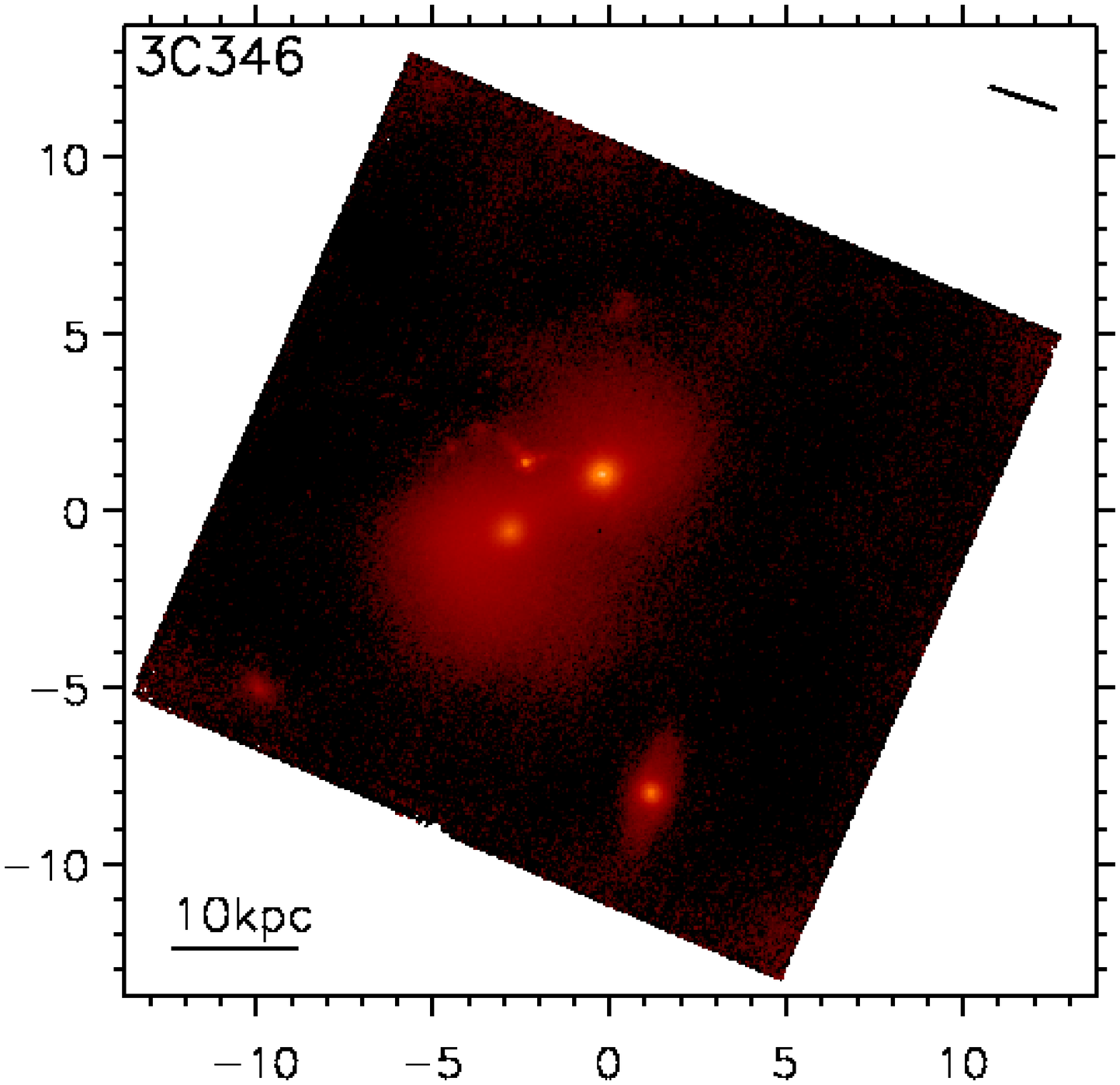}
\caption{HST/NICMOS F160W image of 3C346}
\end{figure}


\begin{figure}
\plotone{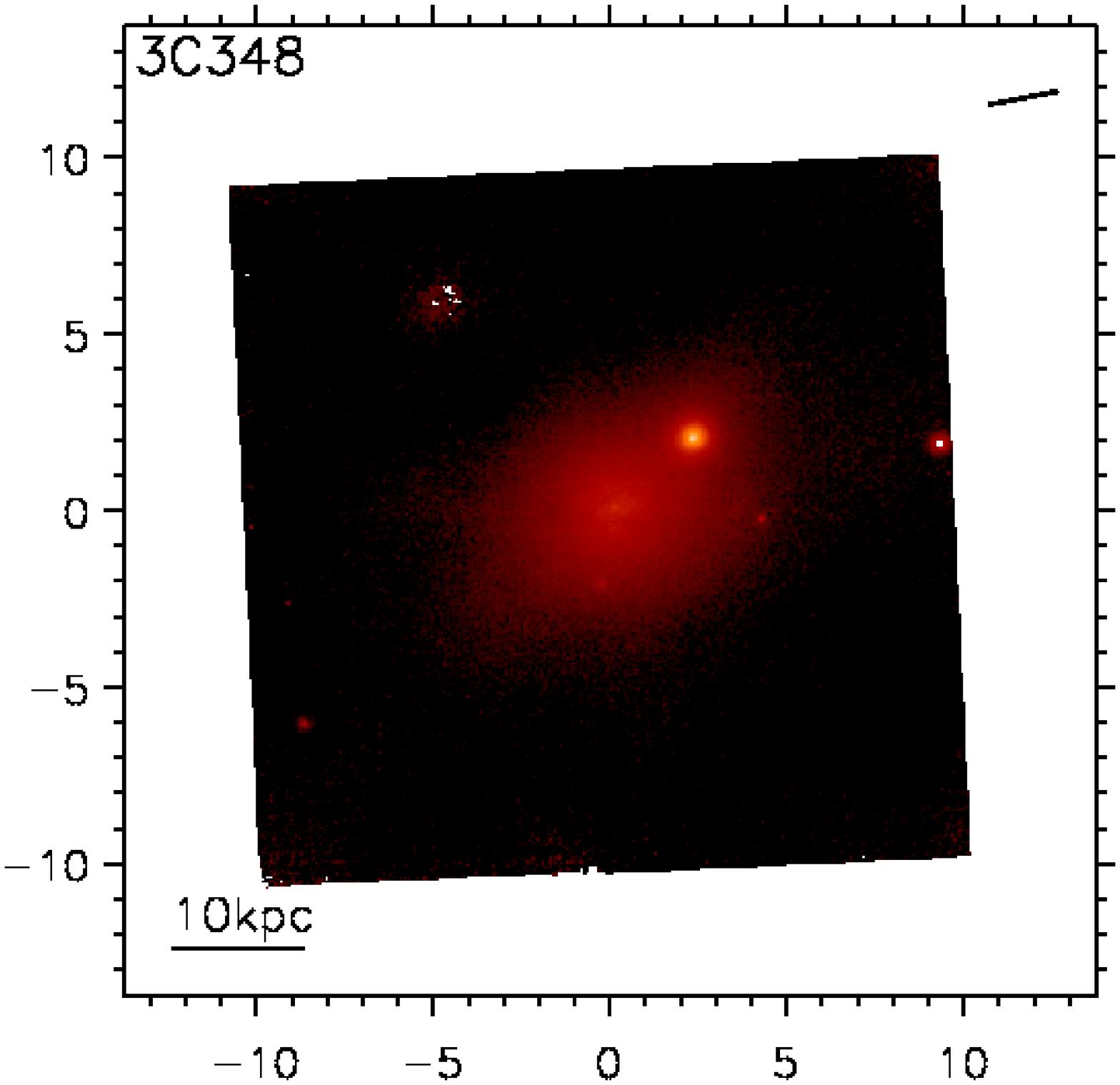}
\caption{HST/NICMOS F160W image of 3C348}
\end{figure}
 

\begin{figure}
\plotone{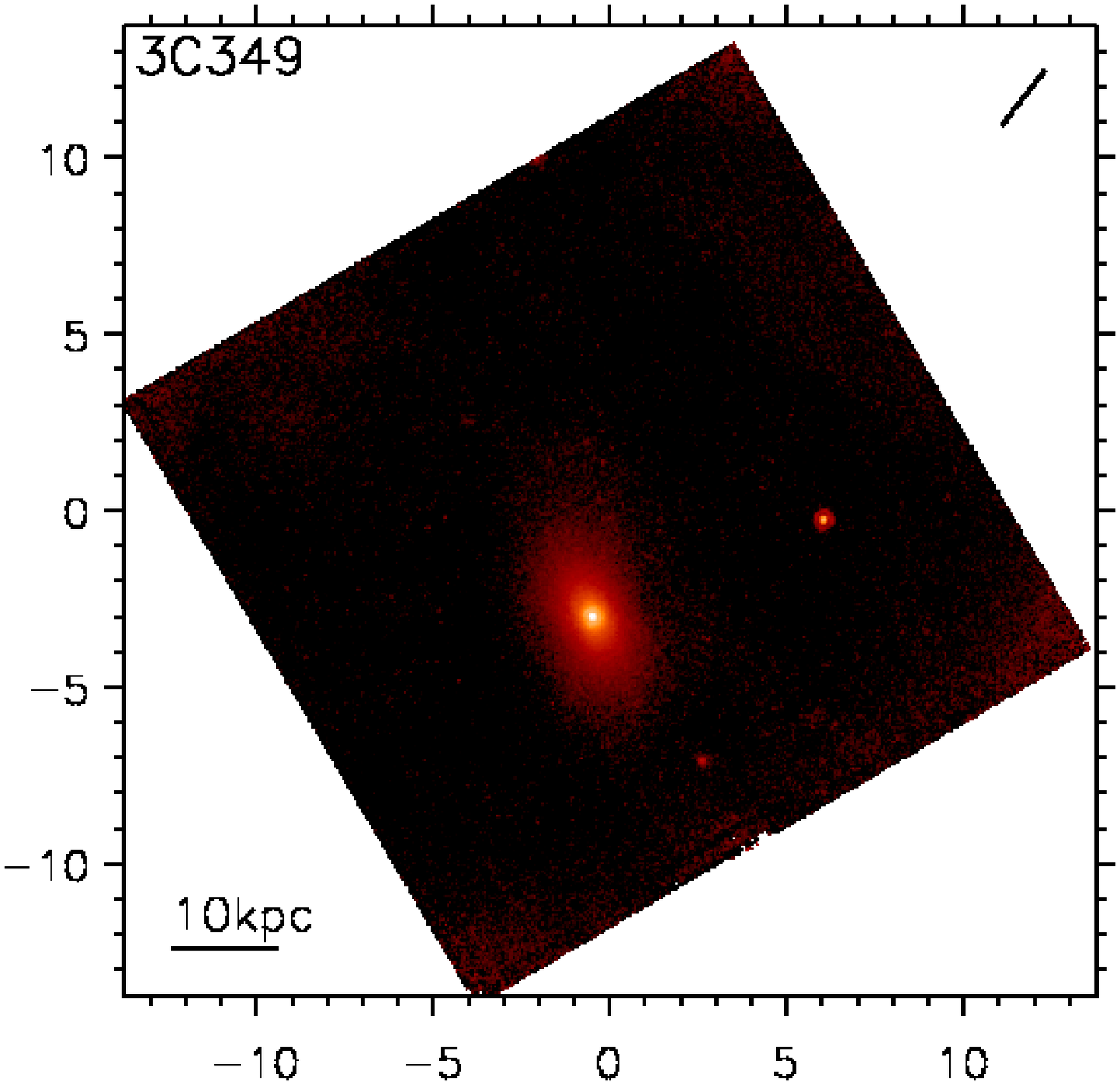}
\caption{HST/NICMOS F160W image of 3C349}
\end{figure}


\begin{figure}
\plotone{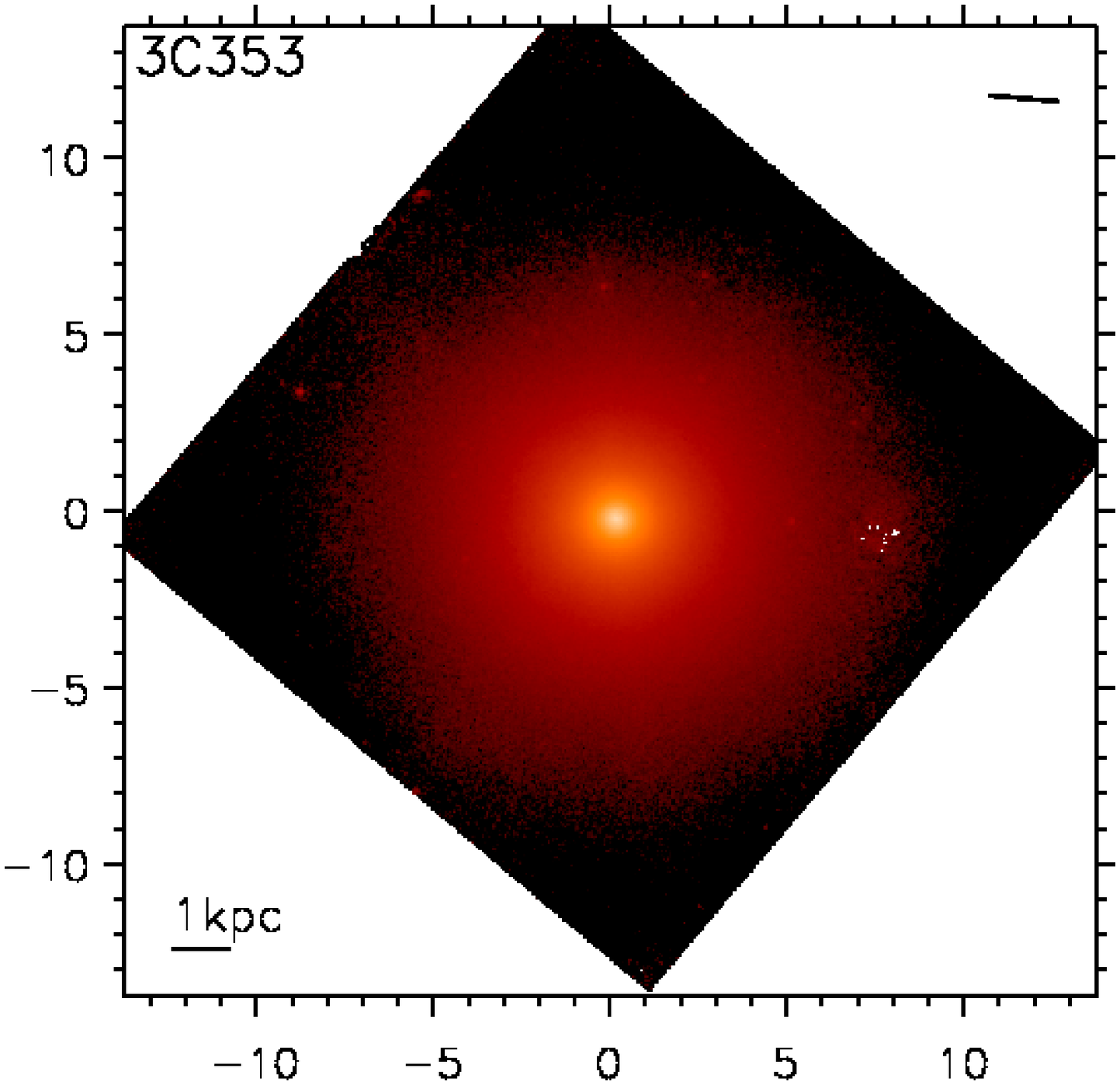}
\caption{HST/NICMOS F160W image of 3C353}
\end{figure}

\clearpage

\begin{figure}
\plotone{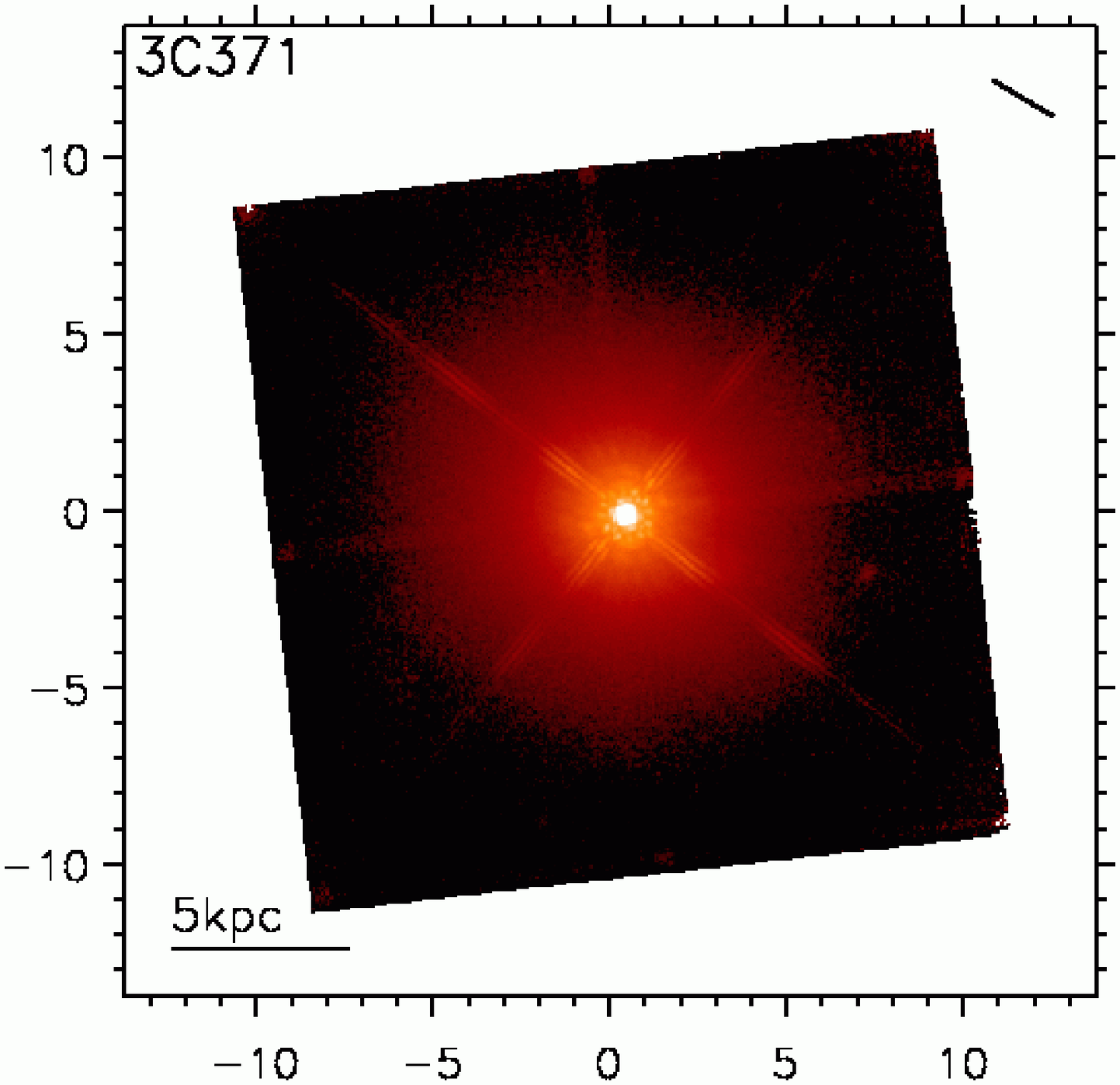}
\caption{HST/NICMOS F160W image of 3C371}
\end{figure}


\begin{figure}
\plotone{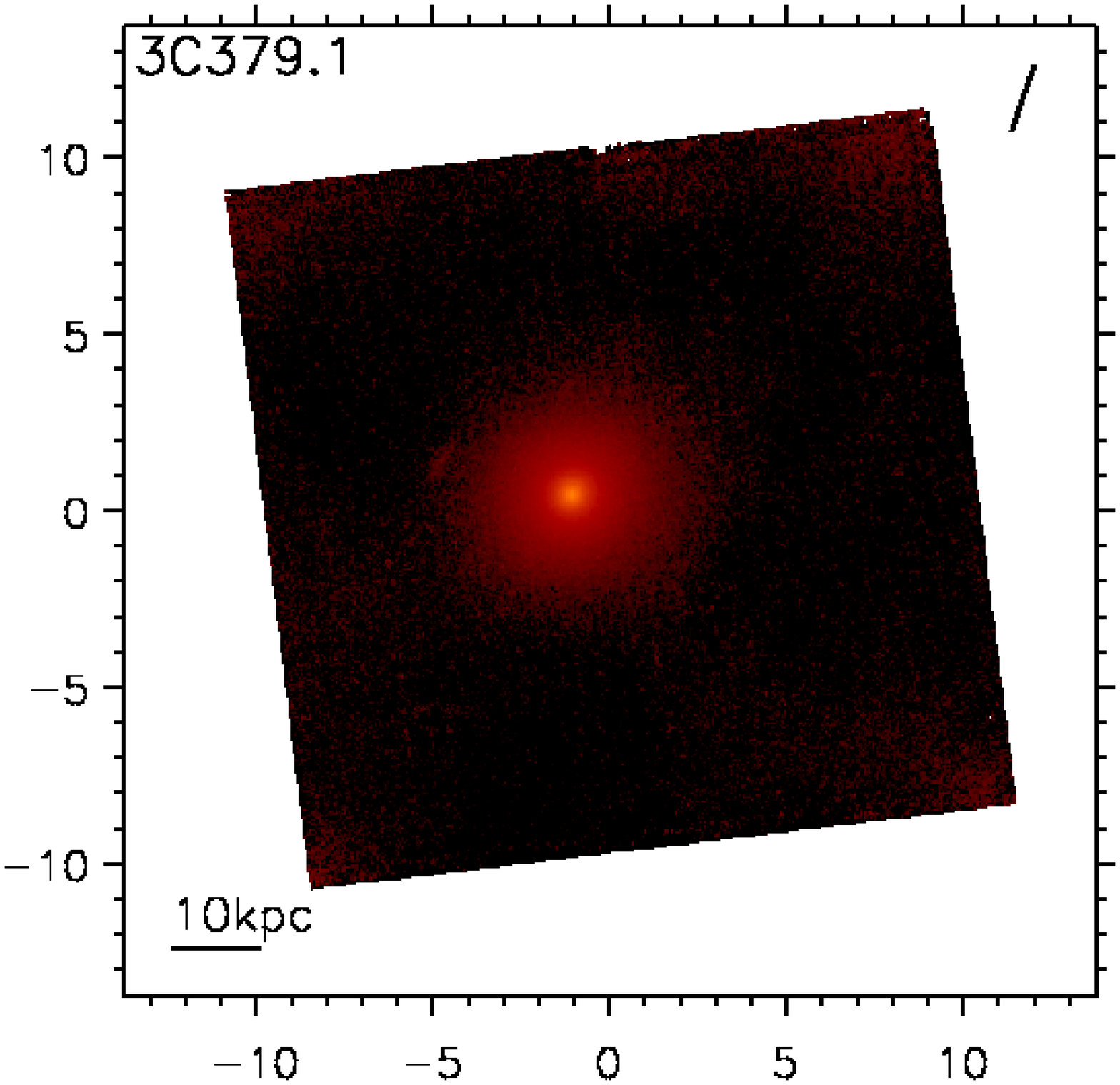}
\caption{HST/NICMOS F160W image of 3C379.1}
\end{figure}


\begin{figure}
\plotone{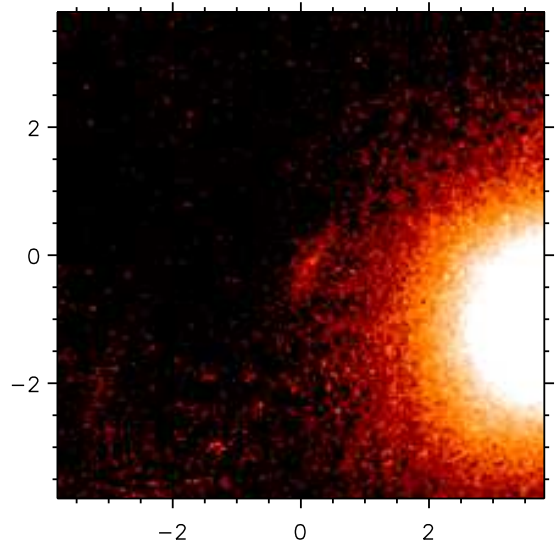}
\caption{Zoom of arc-like feature to the west of 3C379.1}
\end{figure}


\begin{figure}
\plotone{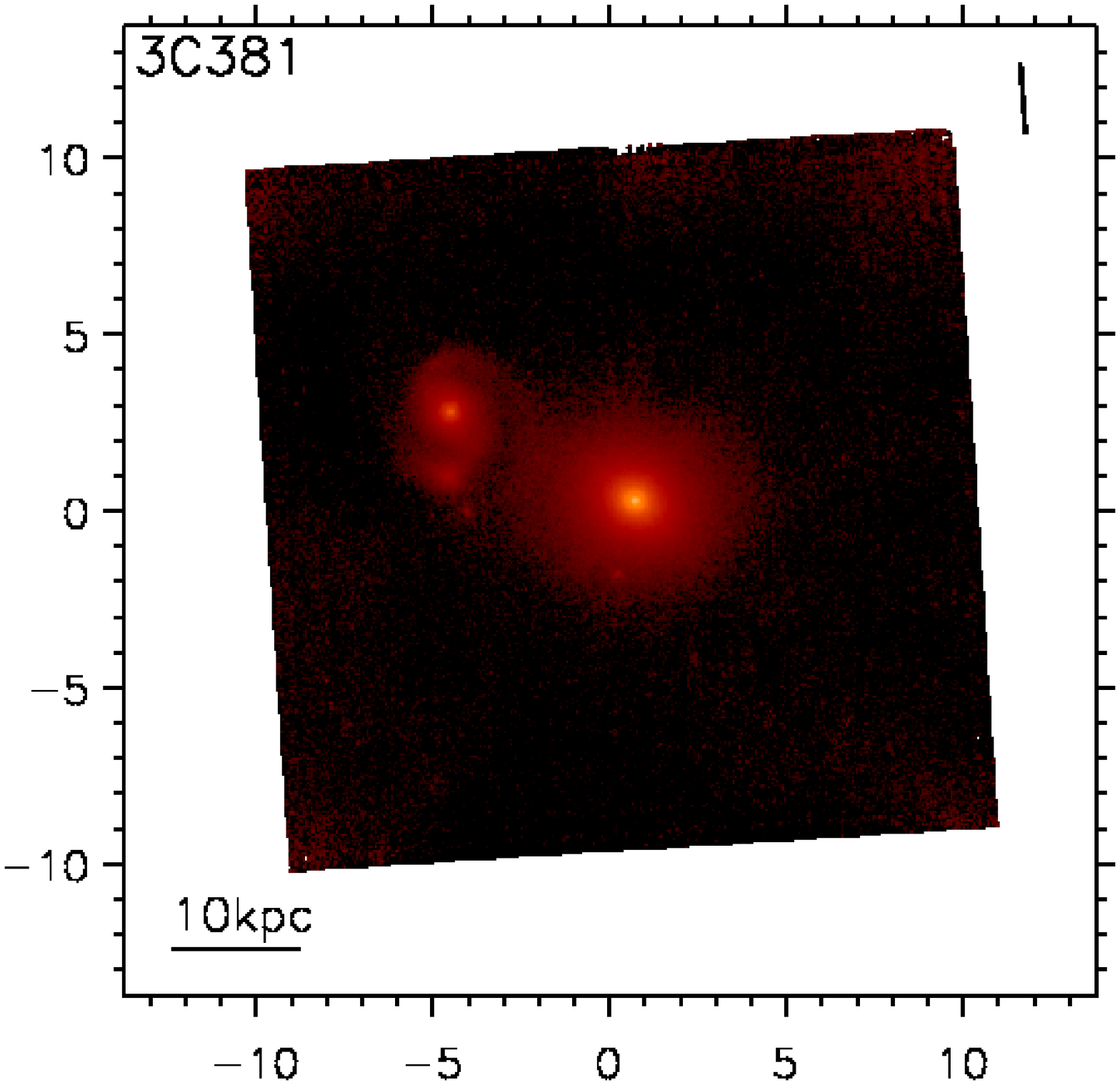}
\caption{HST/NICMOS F160W image of 3C381}
\end{figure}

\clearpage


\begin{figure}
\plotone{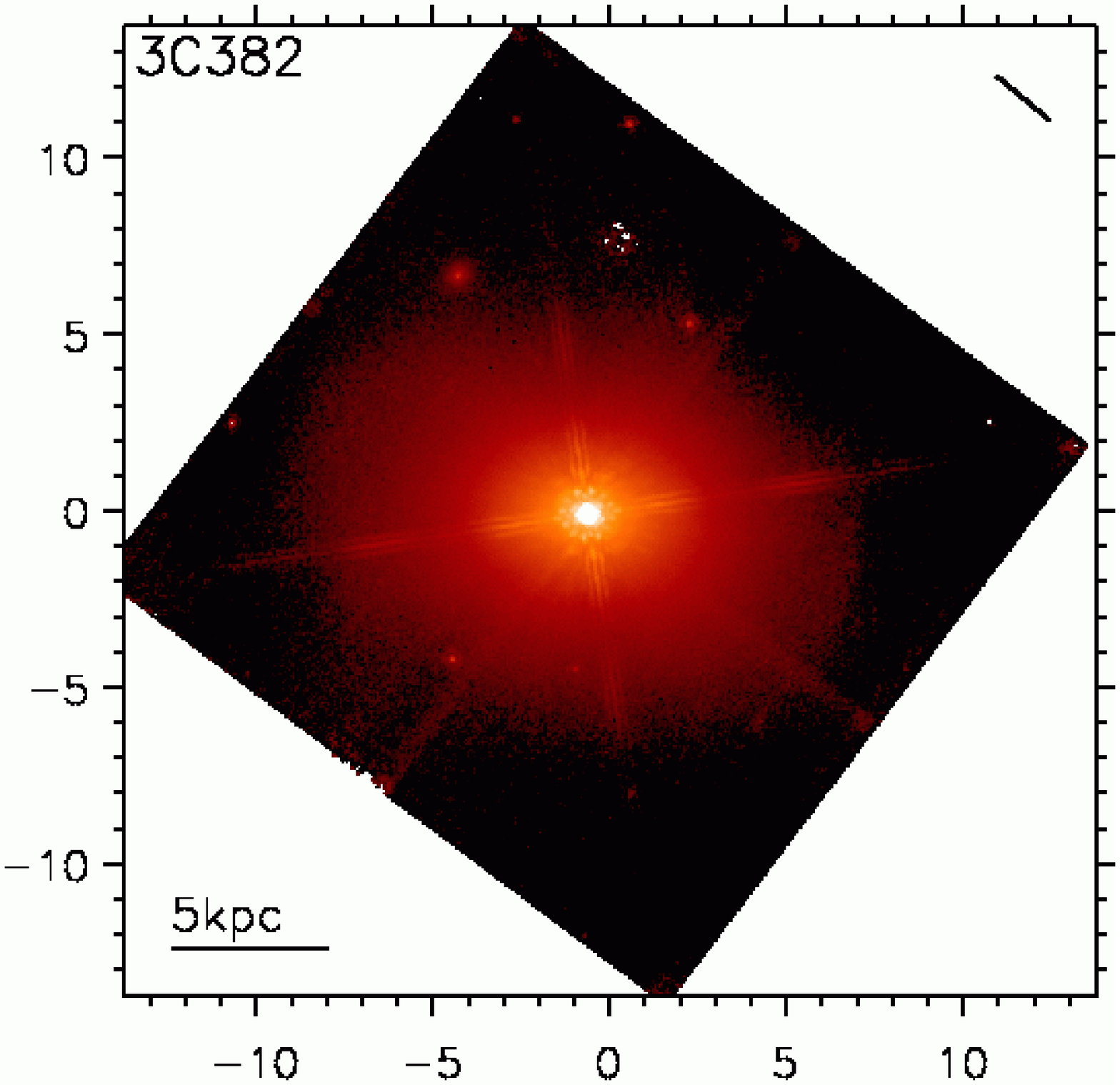}
\caption{HST/NICMOS F160W image of 3C382}
\end{figure}


\begin{figure}
\plotone{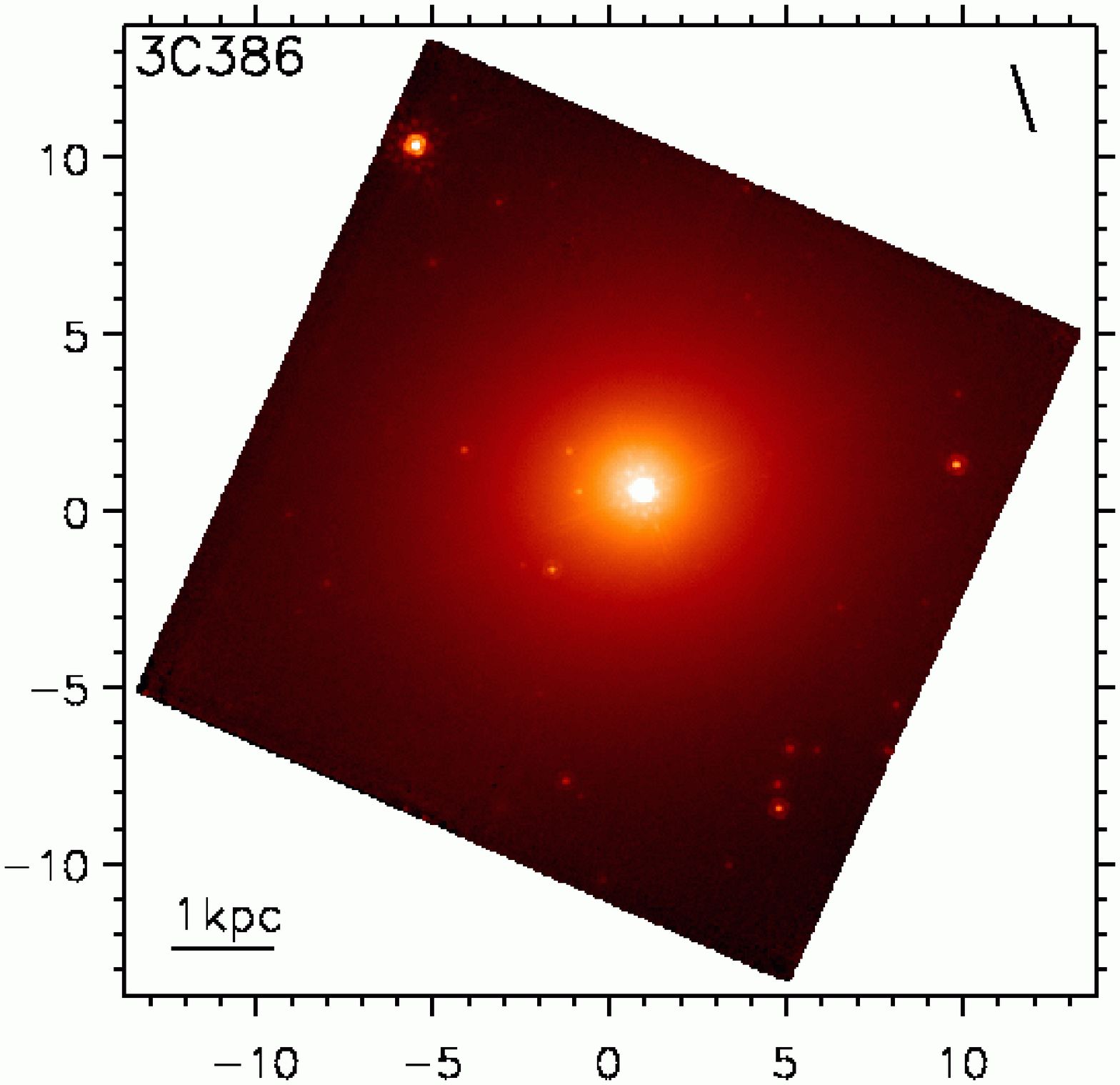}
\caption{HST/NICMOS F160W image of 3C386}
\end{figure}


\begin{figure}
\plotone{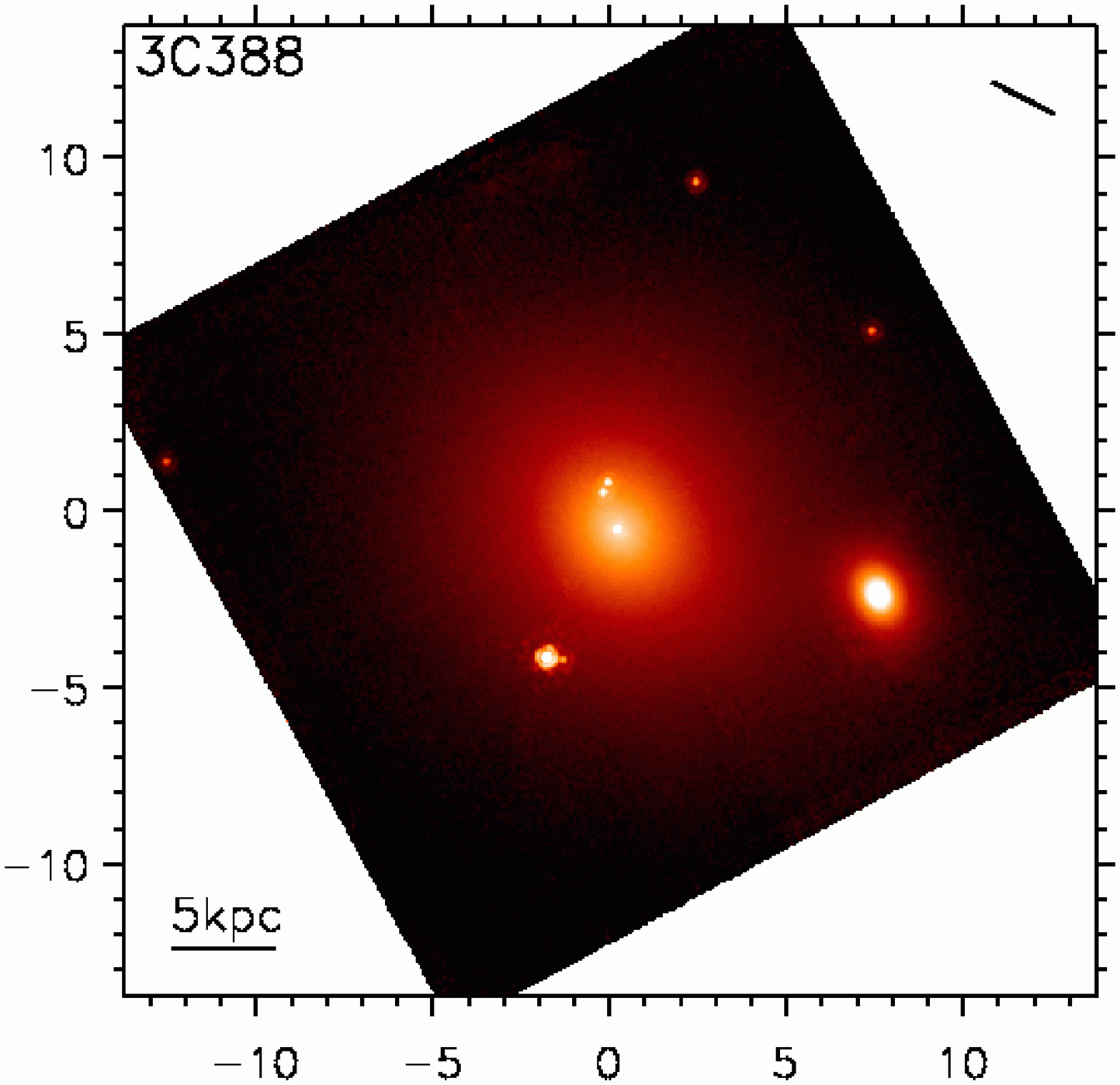}
\caption{HST/NICMOS F160W image of 3C388}
\end{figure}


\begin{figure}
\plotone{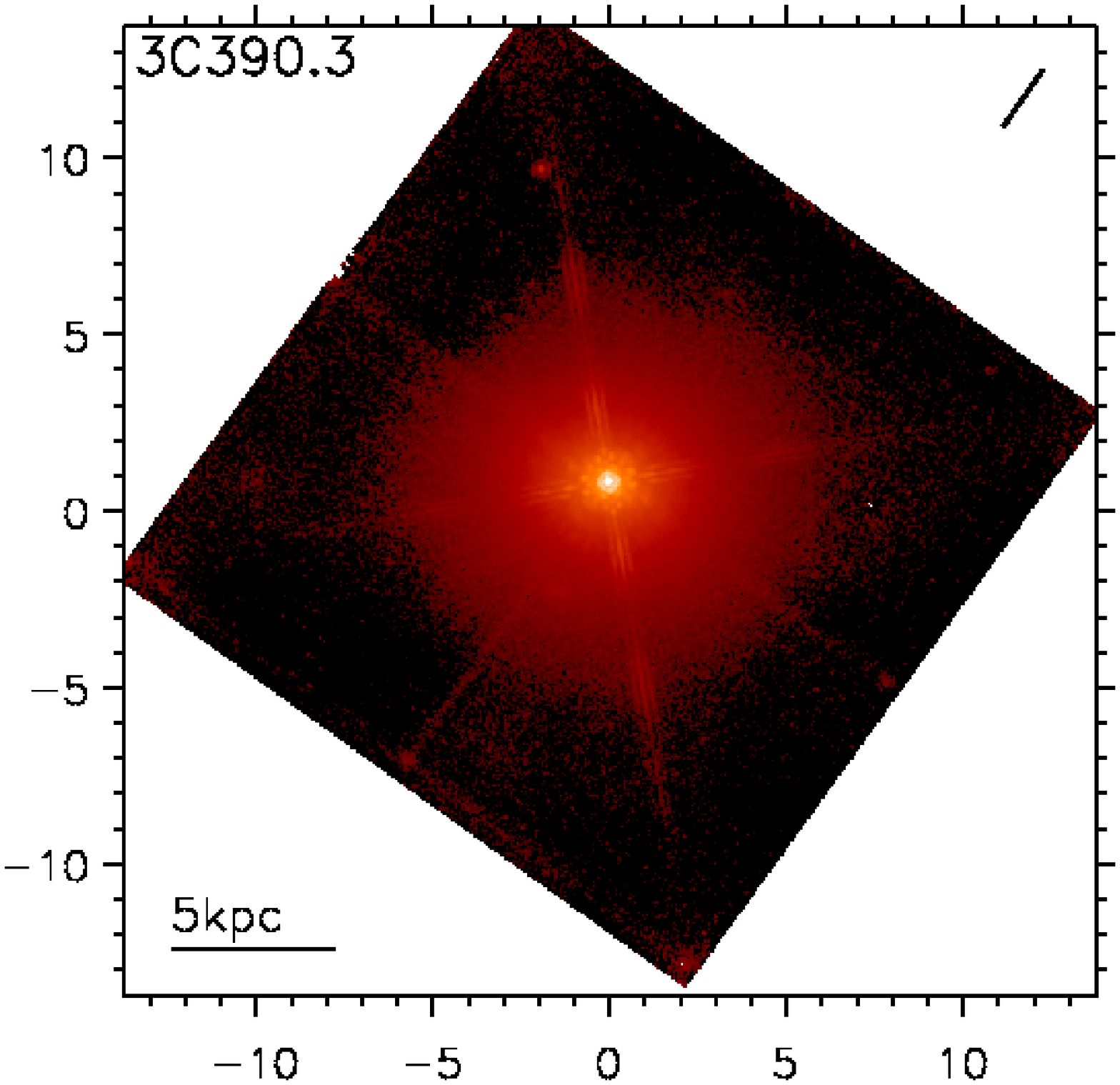}
\caption{HST/NICMOS F160W image of 3C390.3}
\end{figure}

\clearpage


\begin{figure}
\plotone{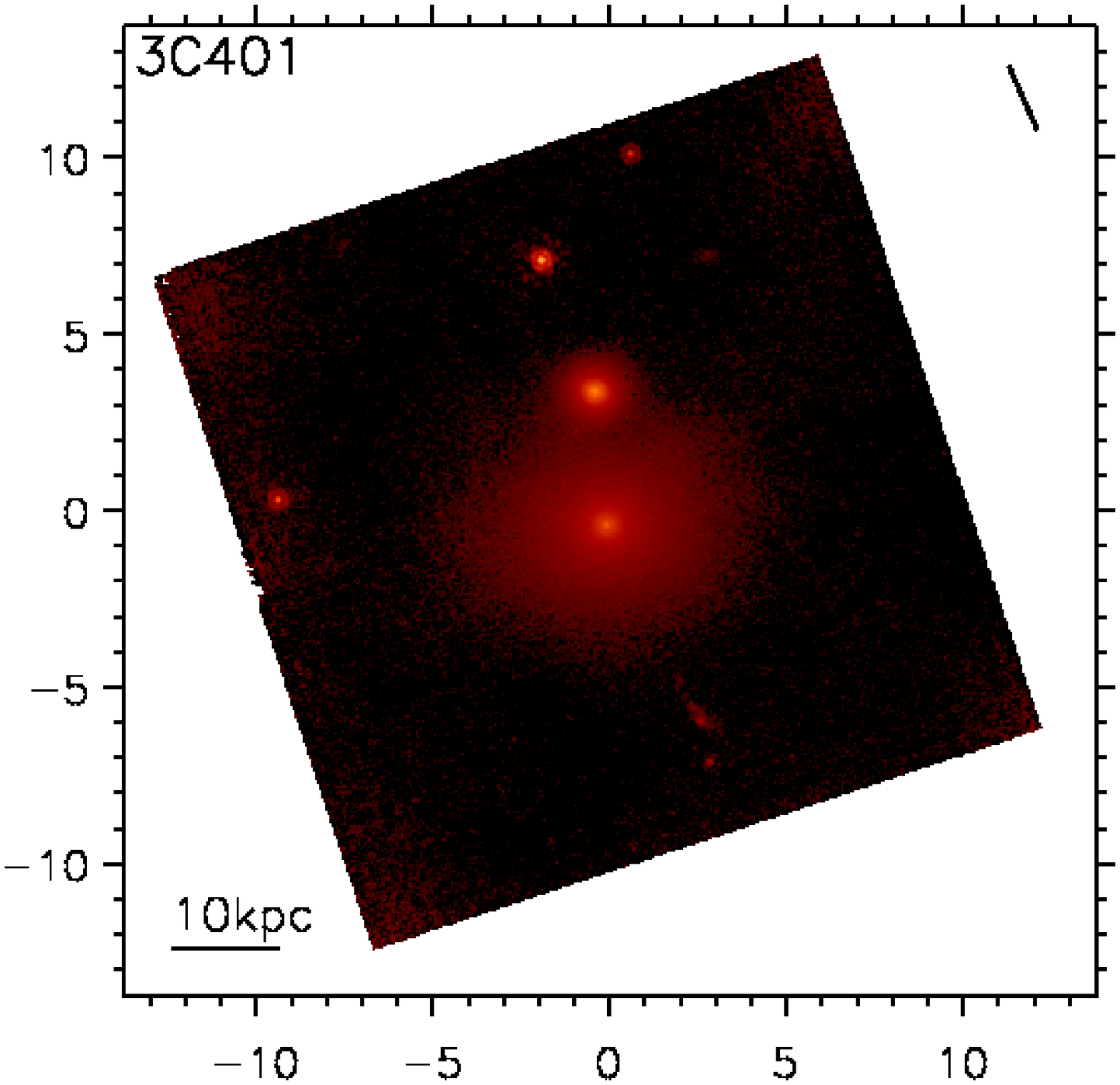}
\caption{HST/NICMOS F160W image of 3C401}
\end{figure} 

\begin{figure}
\plotone{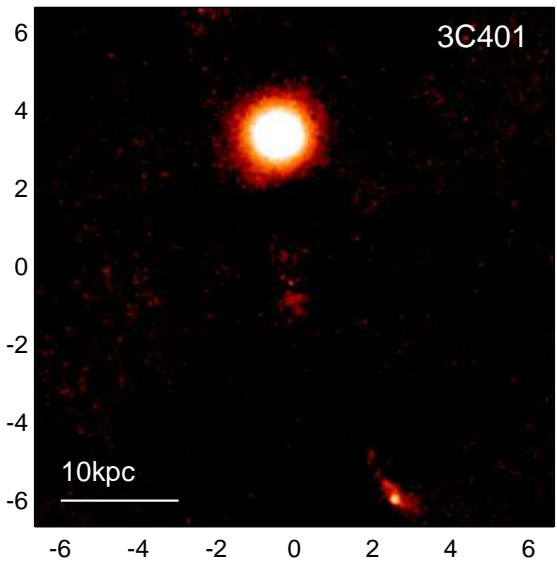}
\newline
\newline
\caption{Model-subtracted residual for 3C401}
\end{figure}


\begin{figure}
\plotone{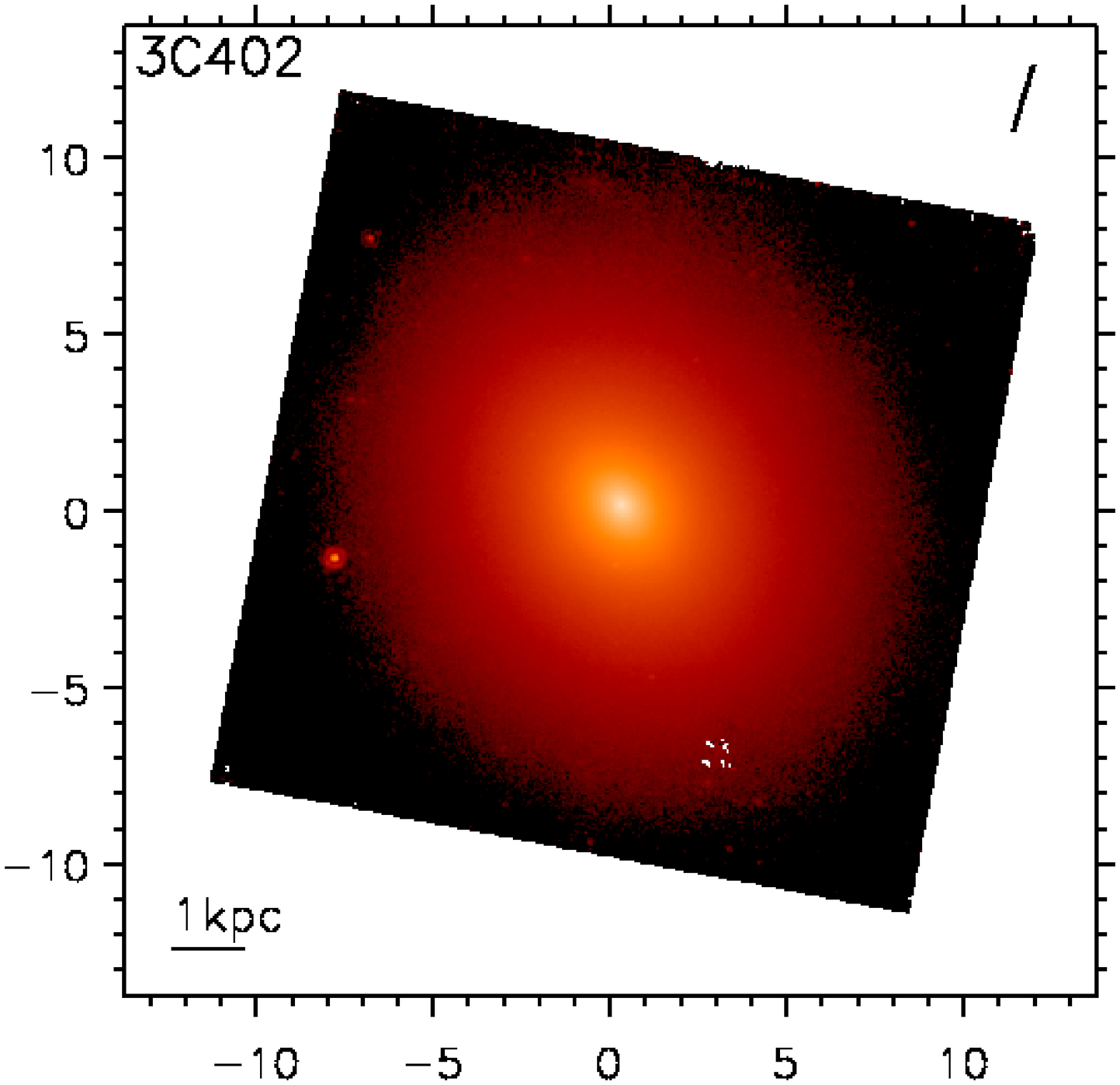}
\caption{HST/NICMOS F160W image of 3C402}
\end{figure}


\begin{figure}
\plotone{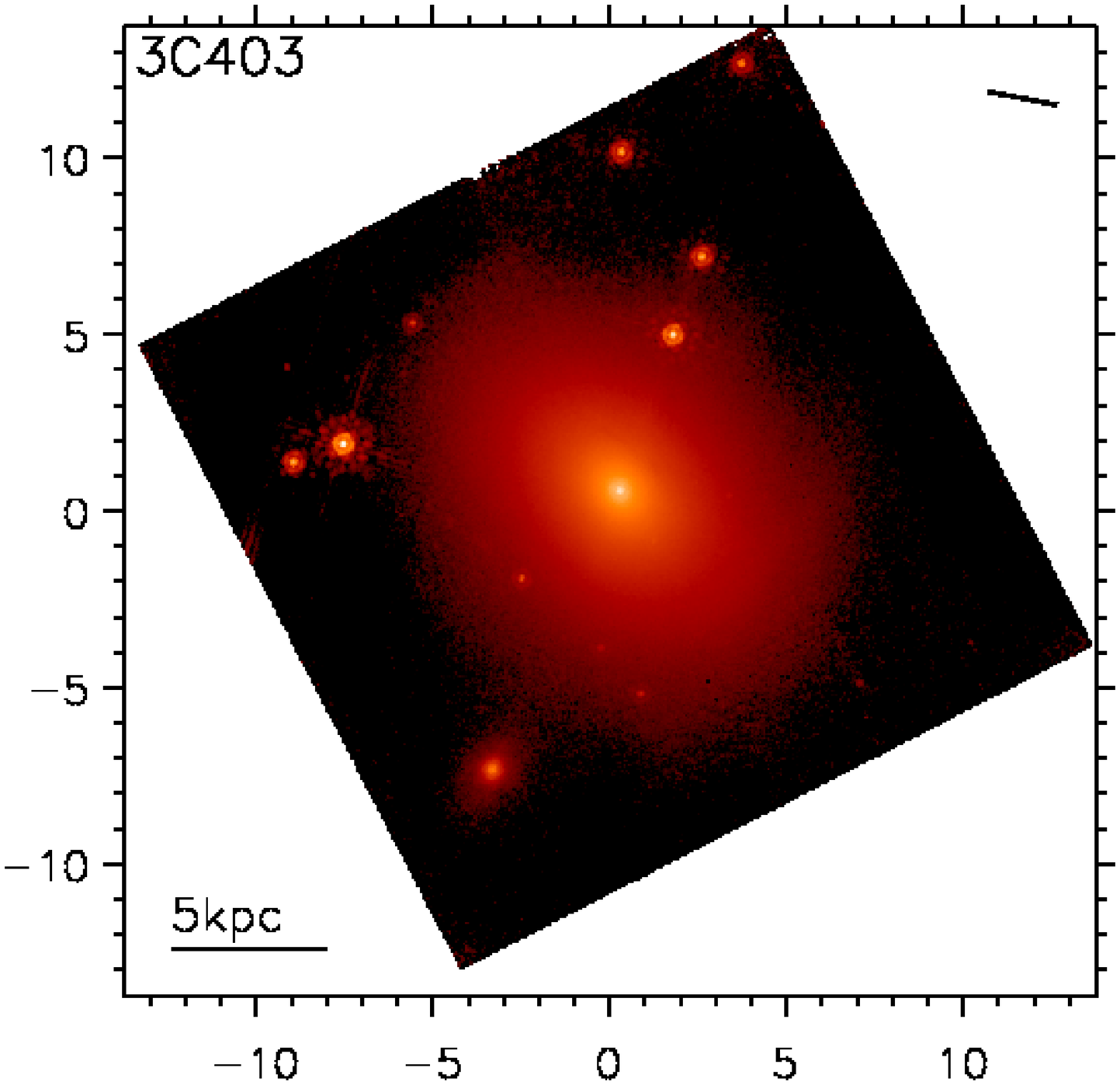}
\caption{HST/NICMOS F160W image of 3C403}
\end{figure}

\clearpage


\begin{figure}
\plotone{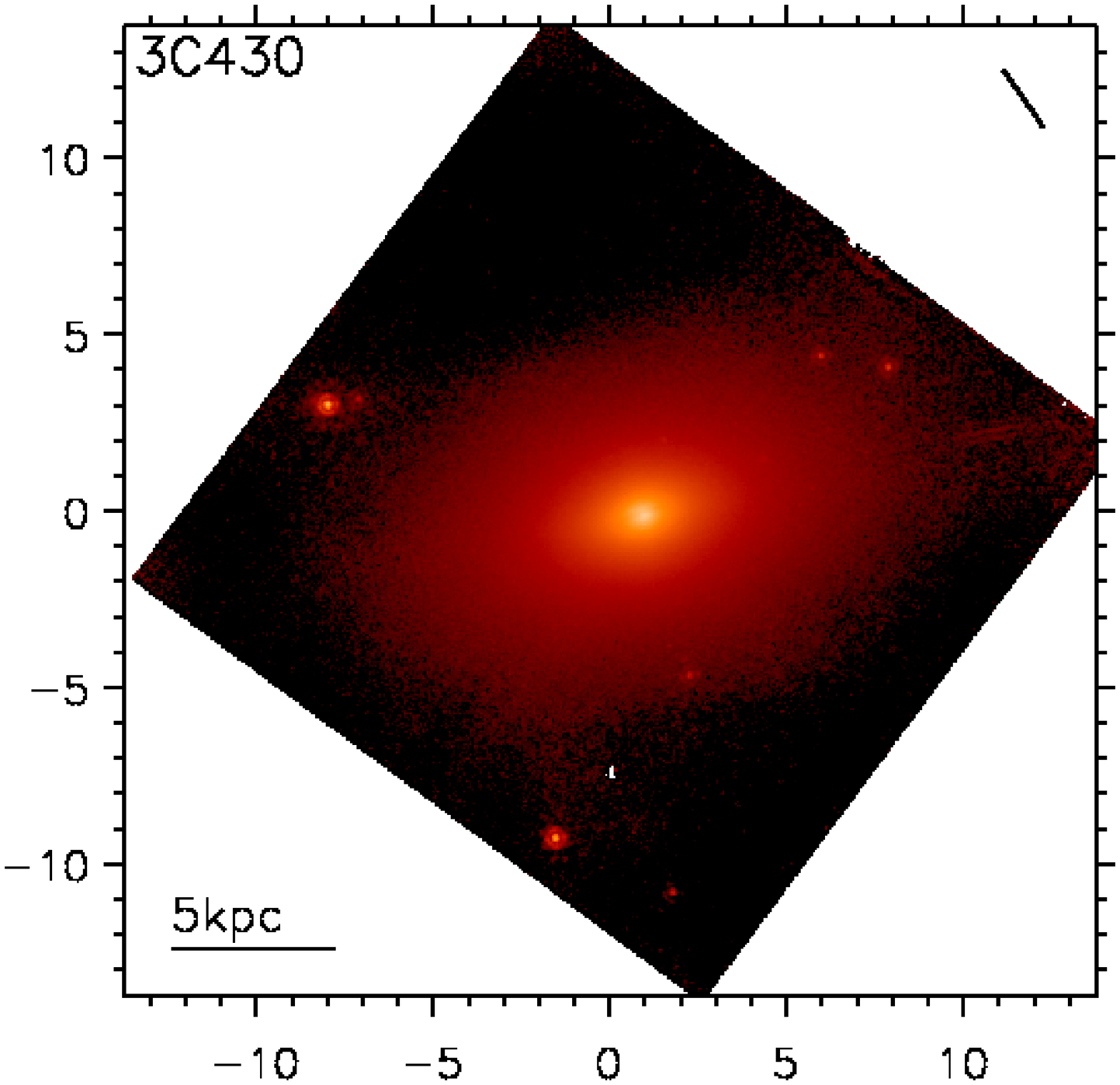}
\caption{HST/NICMOS F160W image of 3C430}
\end{figure}


\begin{figure}
\plotone{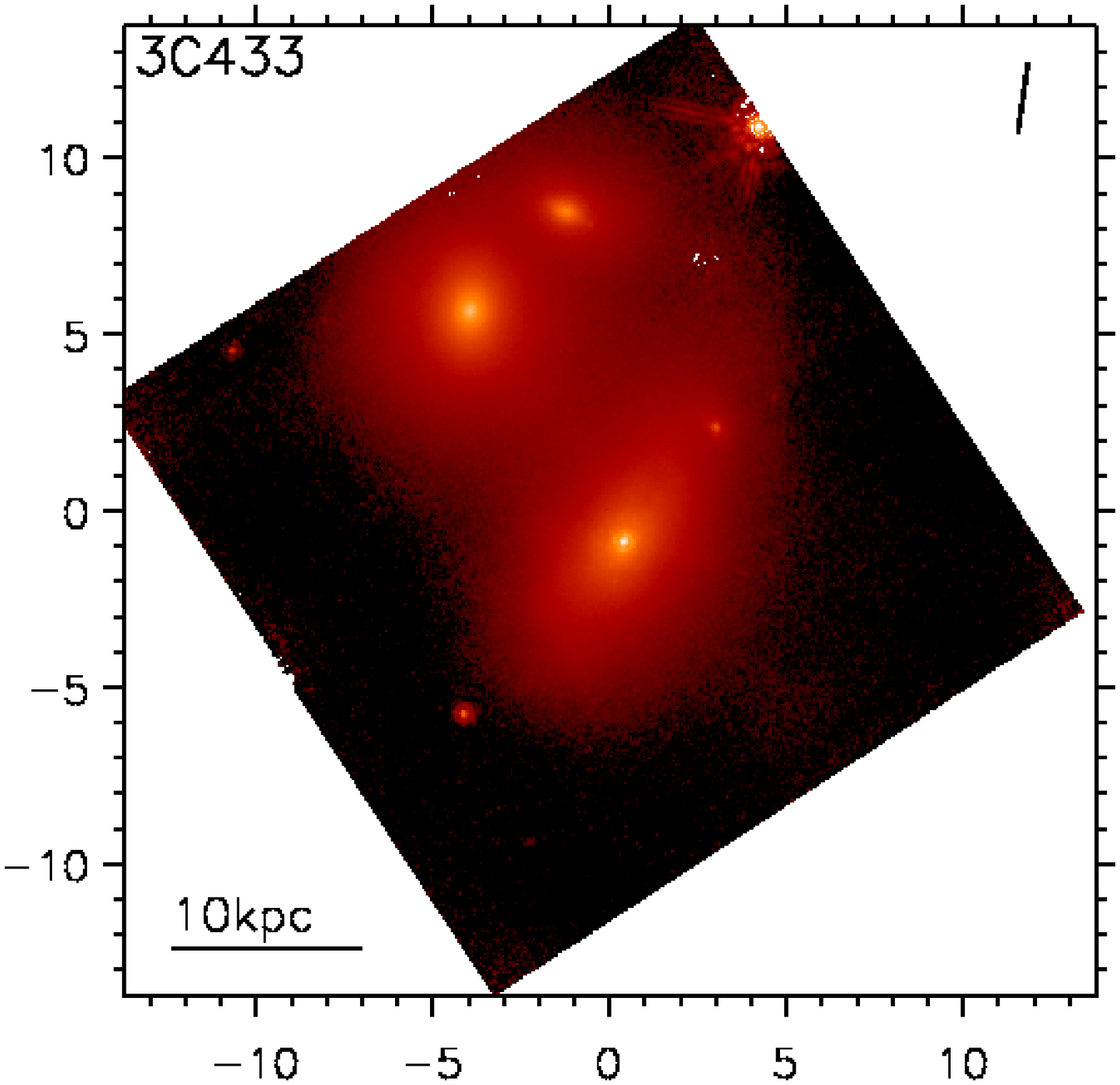}
\caption{HST/NICMOS F160W image of 3C433}
\end{figure}


\begin{figure}
\plotone{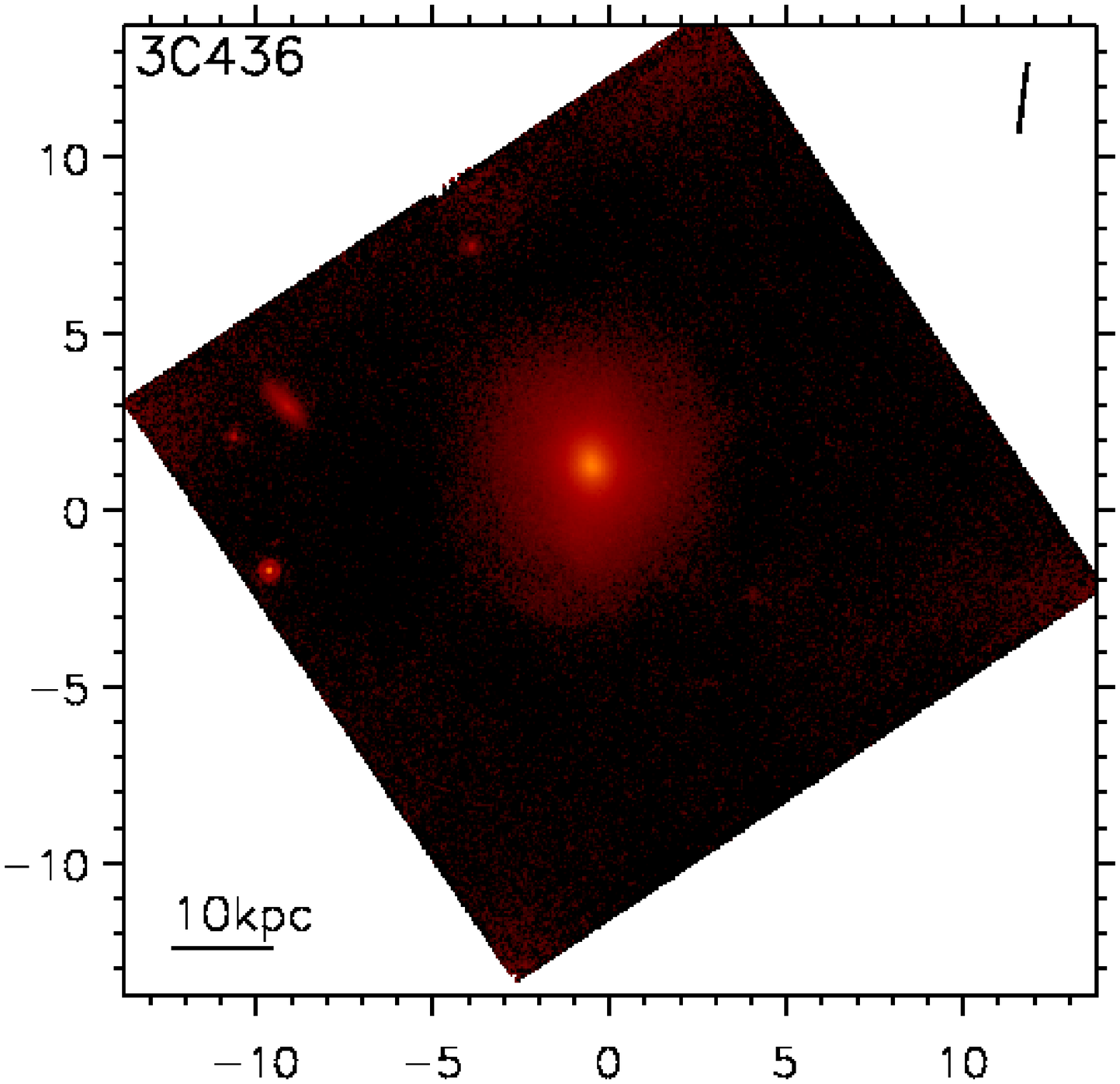}
\caption{HST/NICMOS F160W image of 3C436}
\end{figure}


\begin{figure}
\plotone{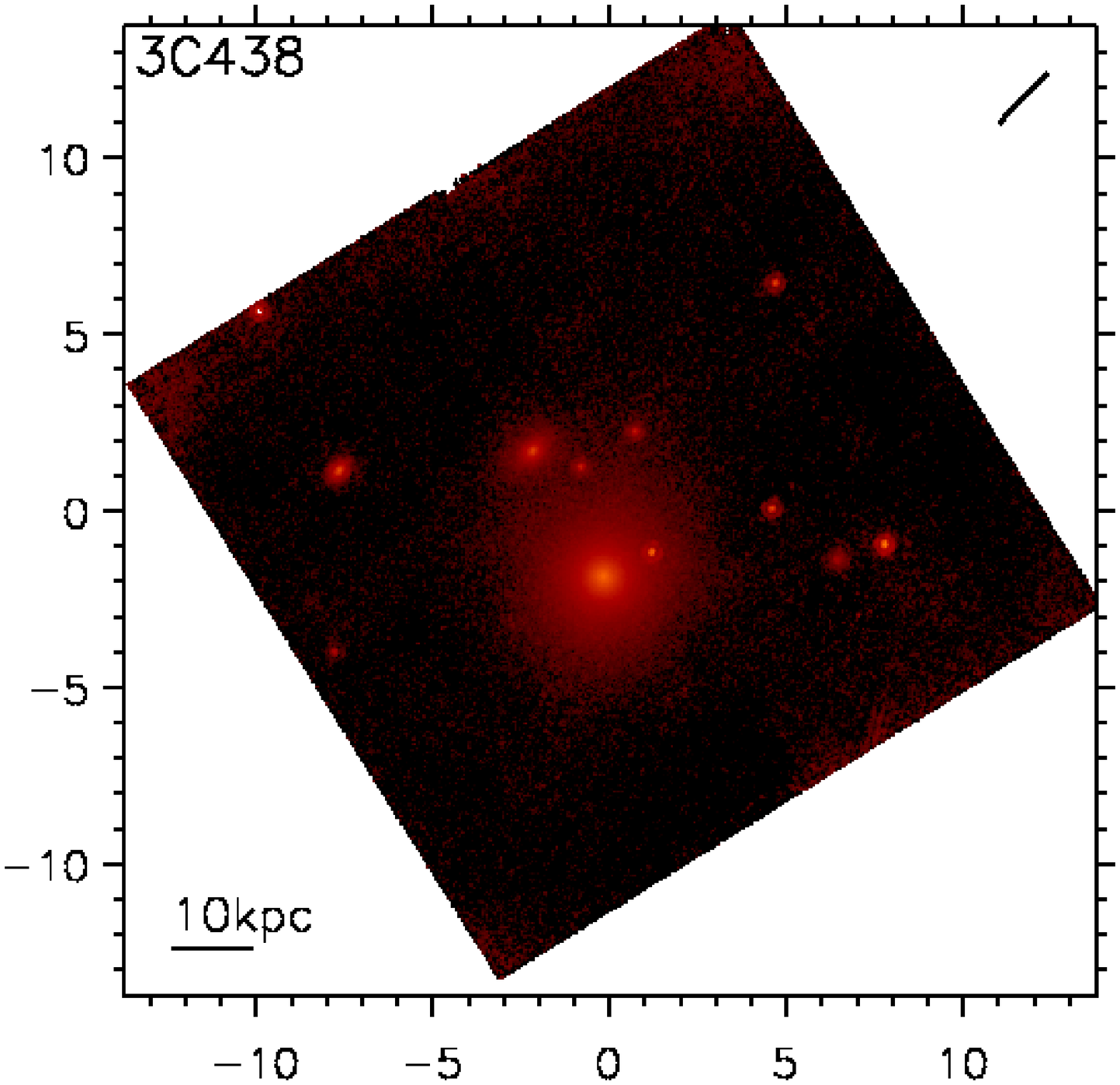}
\caption{HST/NICMOS F160W image of 3C438}
\end{figure}

\clearpage


\begin{figure}
\plotone{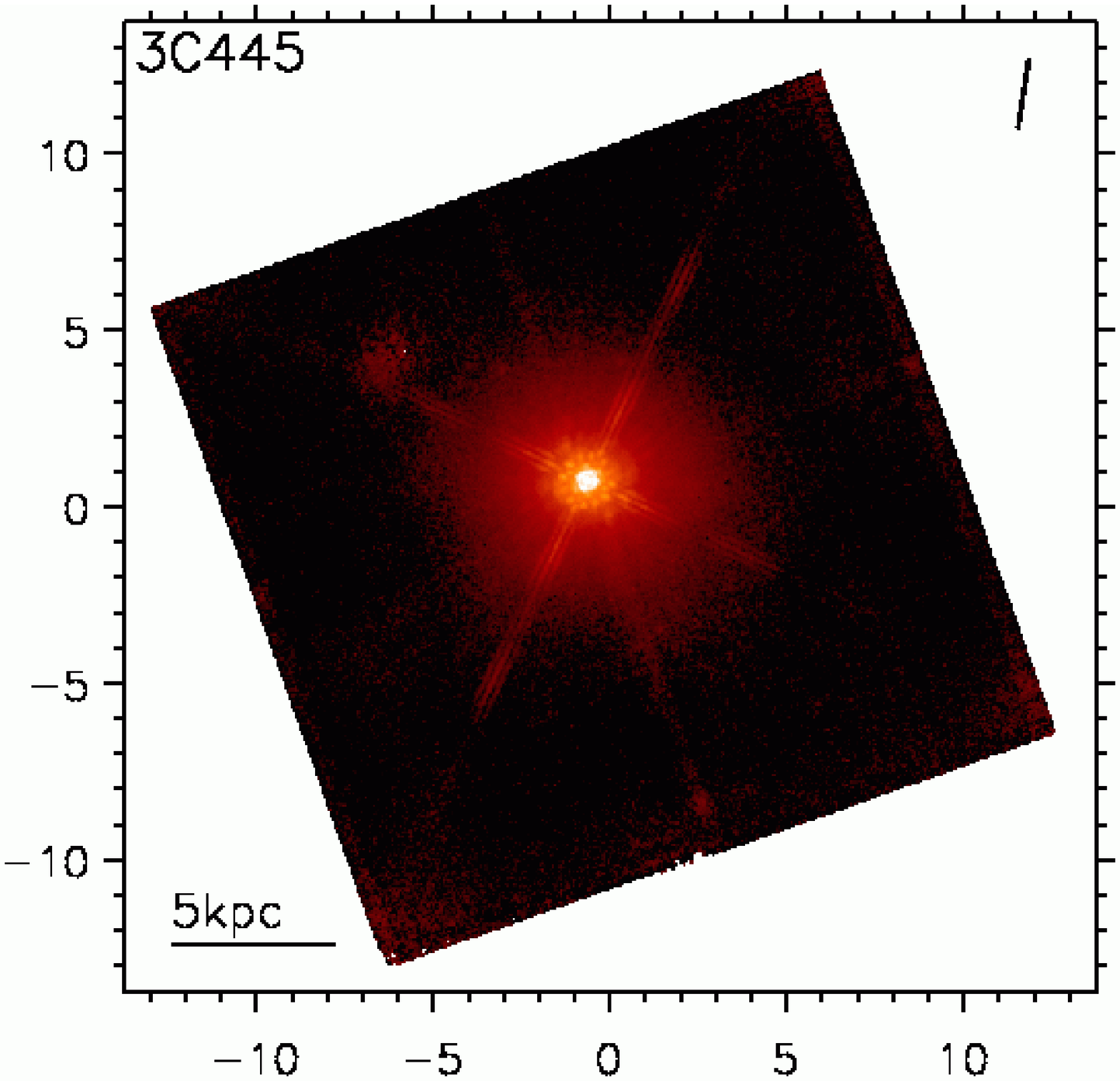}
\caption{HST/NICMOS F160W image of 3C445}
\end{figure}


\begin{figure}
\plotone{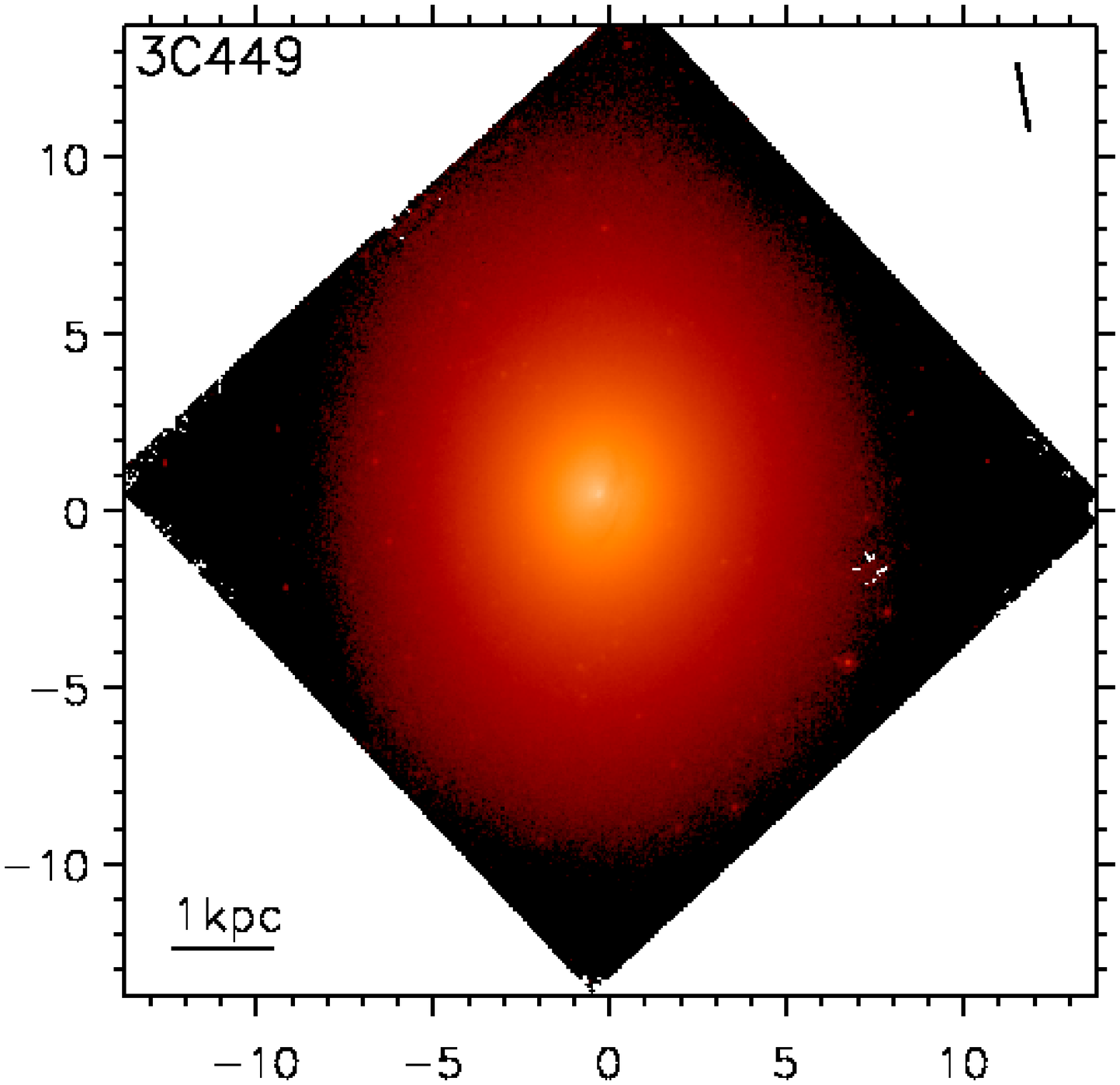}
\caption{HST/NICMOS F160W image of 3C449}
\end{figure}


\begin{figure}
\plotone{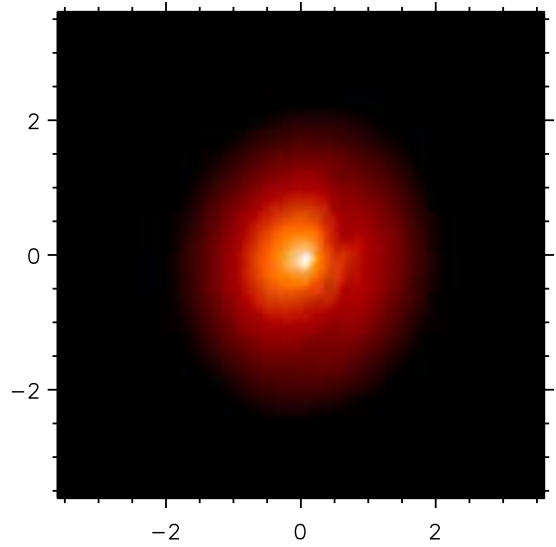}
\caption{Details of the dust lane wrapping around 3C449}
\end{figure}


\begin{figure}
\plotone{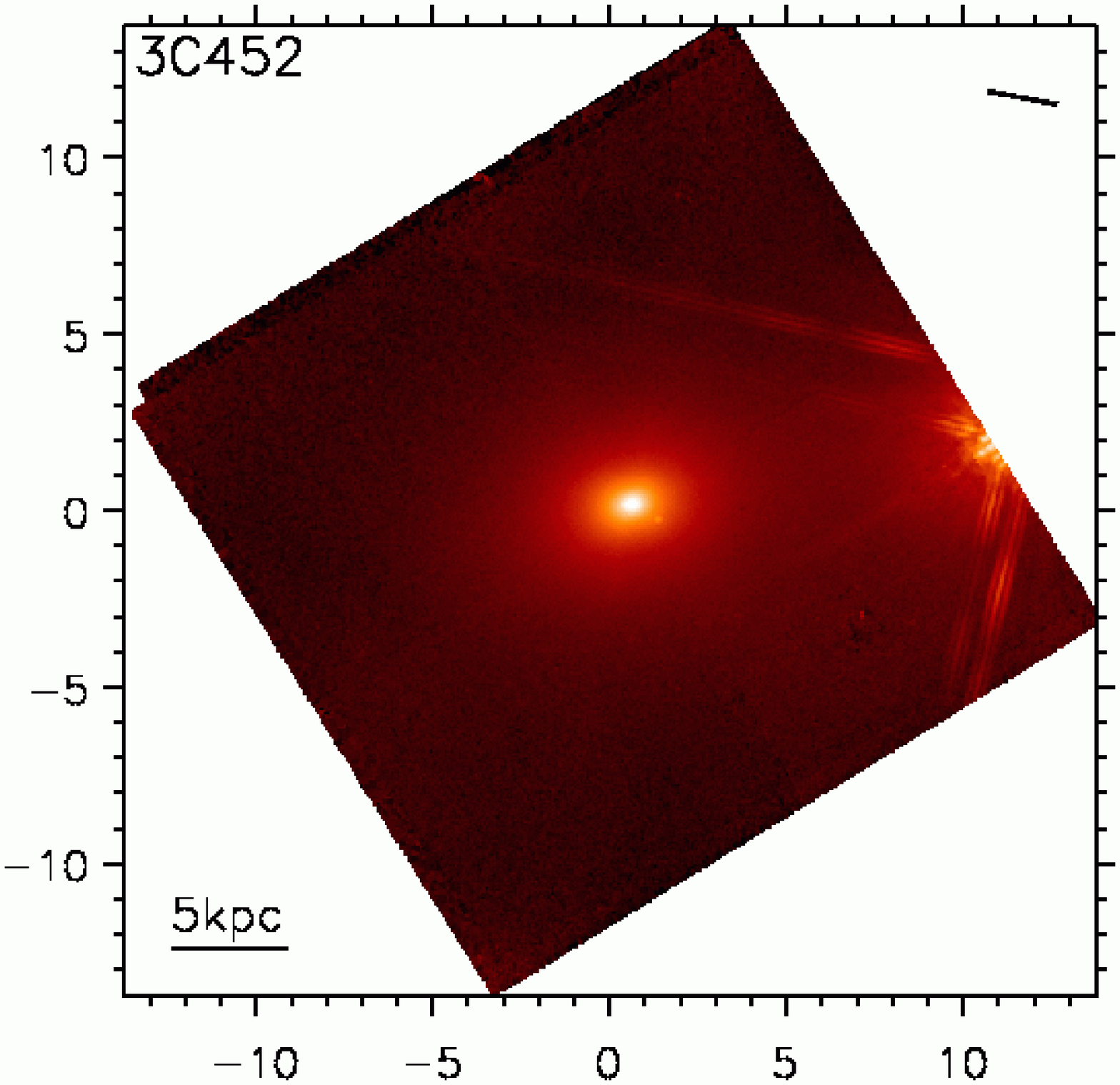}
\caption{HST/NICMOS F160W image of 3C452}
\end{figure}

 \clearpage


\begin{figure}
\plotone{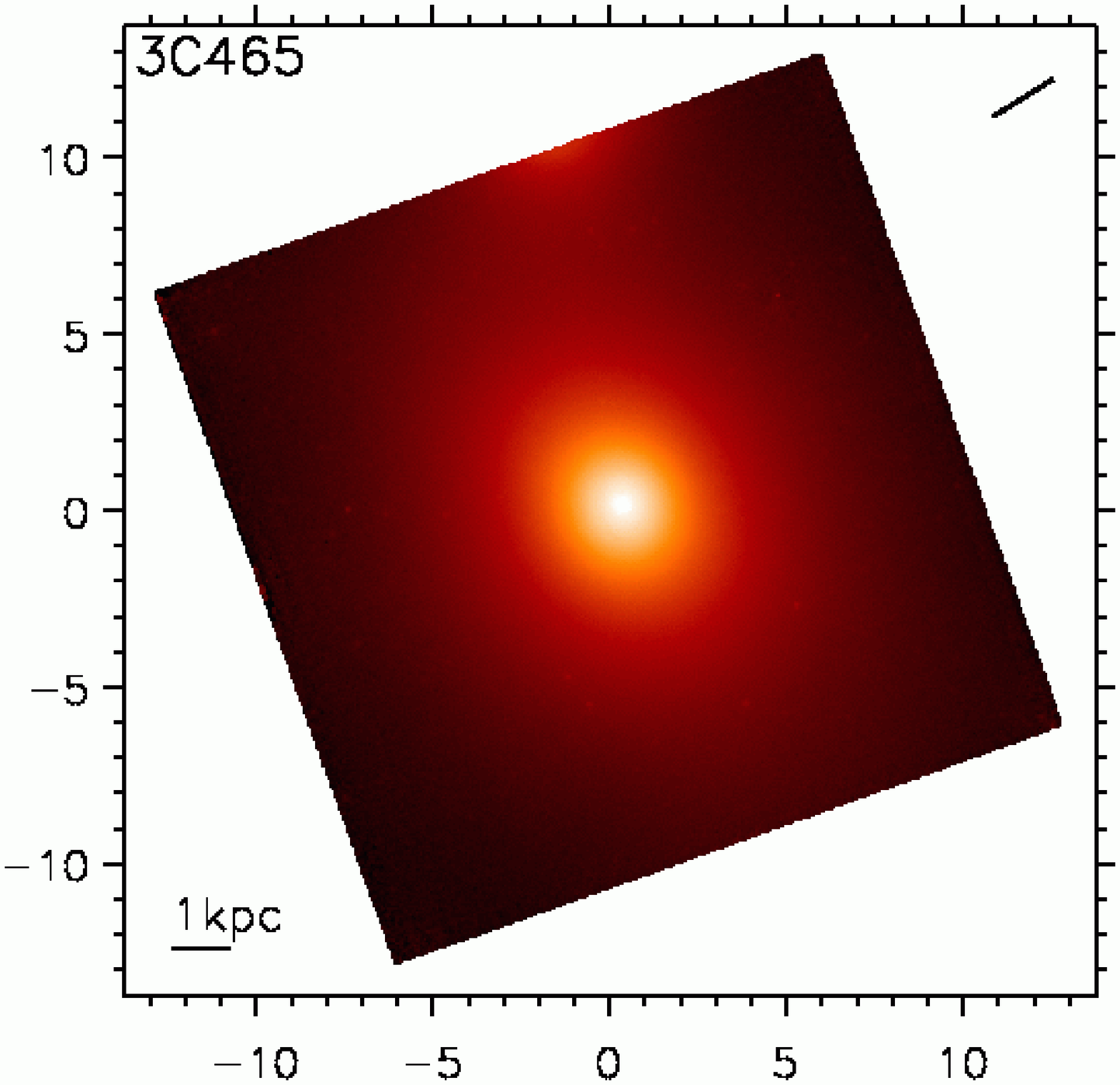}
\caption{HST/NICMOS F160W image of 3C465}
\end{figure}
 


\begin{thebibliography}{}

\bibitem[Allen (2002)]{all02} 
        Allen, M. G., et al. 2002, ApJS, 139, 411
\bibitem[Baum(1988)]{bau88} 
        Baum, S. A., Heckman, T., Bridle, A., van Breugel, W., \& Miley, G. 1988, ApJS, 68, 643
\bibitem[Baum(1996)]{bau96} Baum,  S. A., O'Dea,  C.  P., de  Koff, S.,
        Sparks,  W., Hayes,  J.  J.  E.,  Livio,  M., \&  Golombek, D.   1996, ApJ,465, L5
\bibitem[Bennett (1962)]{ben62}
        Bennett, A. S. 1962a, Mem. RAS, 68, 163 
\bibitem[Bennett (1962)]{bet62}     
        Bennett, A.S. 1962b, MNRAS, 125, 75 
\bibitem[Blundell (1996)]{blu96} 
        Blundell, K. M. 1996, MNRAS, 283, 538     
\bibitem[Bushouse (2000)]{bus00} 
        Bushouse,  H., Dickinson,  M.,  van der  Marel,  R. P.   2000,
        Astronomical Data Analysis Software and Systems IX, ASP conference series, Vol. 216 p. 531
\bibitem[Capetti (2000)]{cap00} 
        Capetti, A., de Ruiter, H.R., Fanti, R., Morganti, R., Parma, P., \& Ulrich, M.-H. 2000,  A\&A, 362, 871    
\bibitem[Chiaberge (1999)]{chi99}     
        Chiaberge, M., Capetti, A., Celotti, A. 1999, A\&A, 349, 77
\bibitem[Chiaberge (2002)]{chi02}     
        Chiaberge, M., Capetti, A., Celotti, A. 2002, A\&A, 394, 791
\bibitem[Chiaberge (2005)]{chi05}     
        Chiaberge, M., et al. 2005, ApJ, 629, 100
\bibitem[deKoff(1996)]{koff96} 
        de Koff, S., Baum, S. A., Sparks, W. B., Biretta, J., Golombek, D., Macchetto, F., 
        McCarthy., P., \& Miley, G. 1996, ApJS, 107, 621
\bibitem[deKoff(2000)]{koff00} 
        de Koff, S., et al. 2000, ApJS, 129, 33
\bibitem[de Vries (1997)]{vri97} 
        de Vries, W.H., et al. 1997, ApJS, 110, 191
\bibitem[Dickinson(2002)]{dic02} 
        Dickinson, M.E.,  et  al. 2002,  HST
        NICMOS data handbook v5.0, ed. B. Mobasher, Baltimore, STScI
\bibitem[Fanaroff(1974)]{fan74} 
        Fanaroff, B. L. \& Riley, J. M. 1974, MNRAS, 167, 31
\bibitem[Floyd(2006)]{flo06}
        Floyd, D. J. E.,  et  al. 2006, astro-ph/0602021  
\bibitem[Fruchter(2002)]{fru02}
        Fruchter, A.S. \& Hook, R. N. 2002, PASP, 114, 144
\bibitem[Hardcastle(1999)]{har99}  
        Hardcastle, M.J. \& Worrall, D.M. 1999, MNRAS, 309, 969
\bibitem[Hardcastle(2003)]{har03}  
        Hardcastle, M.J. 2003, MNRAS, 339, 360
\bibitem[Heckman(1984)]{hec84}
        Heckman, T. M., van Breugel, W. J. M., Miley, G. K. 1984, ApJ, 286, 509
\bibitem[Jedrzejewski(1987)]{jed87}
        Jedrzejewski, R. 1987, MNRAS, 226, 747
\bibitem[Martel(1999)]{mar99} 
        Martel, A., et al. 1999, ApJS, 122, 81
\bibitem[McCarthy(1995)]{mcc95}  
        McCarthy, P. J., Spinrad, H., van Breugel, W. 1995, ApJS, 99, 27
\bibitem[McCarthy(1997)]{mcc97}  
        McCarthy, P. J., Miley, G. K., de Koff, S., Baum, S. A., Sparks, W. B.,
        Golombek, D., Biretta, J., \& Macchetto, F. 1997, ApJS, 112, 415
\bibitem[McLure(1999)]{mc99}  
        McLure, R.J., et al. 1999, MNRAS, 308, 377
\bibitem[Miley(1981)]{mil81}  
        Miley, G. K., Heckman, T. M., Butcher, H. R, van Breugel, W. 1981, ApJ, 247, L5
\bibitem[Noll,  K.,  et  al.  (2004)]{nol04}
        Noll,  K.,  et  al. 2004, NICMOS  Instrument  Handbook", Version 7.0, (Baltimore: STScI)
\bibitem[O'dea(2001)]{dea01}
        O'Dea, C.P., et al. 2001, AJ, 121,1915
\bibitem[Prestage(1988)]{pre88}
        Prestage, R.M. \& Peacock, J.A. 1988, MNRAS, 230, 131
\bibitem[Roche(2000)]{roc00}
        Roche, N., \& Eales, S.A. 2000, MNRAS, 317, 120
\bibitem[Sadun(1993)]{sad93}
        Sadun, A. C., \& Hayes, J. J. E. 1993, PASP, 105, 379 
\bibitem[Spinrad(1985)]{spi85}
        Spinrad, H., Djorgovski, S., Marr, J., \& Aguilar, L. 1985, PASP, 97, 932 
\bibitem[Tremblay(2005)]{tre05}
        Tremblay, G. R. et al. 2005, astro-ph/0510650
\bibitem[Breugel(1985)]{bre85}
        van Breugel, W., Miley, G., Heckman, T., Butcher, H., Bridle, A. 1985, ApJ, 290, 496
\bibitem[Zirbel(1998)]{zir98}
        Zirbel, E. L. \& Baum, S.A. 1998, ApJS, 114, 177



\end{thebibliography}
\end{document}